\documentclass[twoside,11pt] {article}

\usepackage{amsmath}
\usepackage{graphicx}
\usepackage{amssymb}
\begin{document}

\title{Exactly-solvable problems for two-dimensional excitons}

\author{D.G.W.~Parfitt and M.E.~Portnoi} 
%\address{School of Physics, University of
%Exeter, Stocker Road, Exeter EX4 4QL, United Kingdom}
\date{School of Physics, University of Exeter,\\ 
Stocker Road, Exeter EX4 4QL, United Kingdom}

\maketitle
\pagestyle{myheadings} 
\thispagestyle{plain}         
\markboth{D.G.W.~Parfitt and M.E.~Portnoi}
{Exactly-solvable problems for two-dimensional excitons} 
\setcounter{page}{1}

\begin{abstract}
Several problems in mathematical physics relating to excitons in two 
dimensions are considered. First, a fascinating numerical result from 
a theoretical treatment of screened excitons stimulates a re-evaluation 
of the familiar two-dimensional hydrogen atom. Formulating the latter 
problem in momentum space leads to a new integral relation in terms of 
special functions, and fresh insights into the dynamical symmetry of 
the system are also obtained. A discussion of an alternative potential 
to model screened excitons is given, and the variable phase method is 
used to compare bound-state energies and scattering phase shifts for 
this potential with those obtained using the two-dimensional analogue 
of the Yukawa potential. The second problem relates to excitons in a 
quantising magnetic field in the fractional quantum Hall regime. 
An exciton against the background of an incompressible quantum liquid 
is modelled as a few-particle neutral composite consisting of a 
positively-charged hole and several quasielectrons with fractional 
negative charge. A complete set of exciton basis functions is derived, 
and these functions are classified using a result from the theory of 
partitions. Some exact results are obtained for this complex 
few-particle problem.
\end{abstract}

\section*{Introduction}

Dimensionality plays an important role in condensed matter physics,
with many fascinating new phenomena emerging as the number of
dimensions is reduced. With the introduction of improved semiconductor 
growth techniques over the past few decades, physical realisation of 
these systems has become possible. 

One of these systems is a quantum well, fabricated by sandwiching a layer
of semiconductor between two layers of another material with
a higher bandgap.
The motion of electrons and holes is `frozen' in the growth direction,
and in modelling the system we may consider the electrons and holes as 
being essentially confined to a plane. 
As in bulk semiconductors, the Coulomb interaction between electrons 
and holes leads to bound states, known as \emph{excitons}, which are 
extremely important for the optical properties of the quantum well.

The introduction of a strong magnetic field in the growth direction
leads to further interesting phenomena in two-dimensional systems.
At sufficiently low temperatures, many-electron correlations become 
important and new states of matter are formed. One of these states of
matter is an incompressible quantum liquid, whose excitations carry
fractional charge and obey fractional statistics.
Spectroscopic techniques can be used to probe the properties of such
systems, and excitonic effects are again of the utmost importance.

This Chapter will focus on a diverse range of problems in mathematical
physics which have arisen as a result of our work on the theory of
excitons in semiconductor nanostructures.
The areas covered include the theory of special functions, dynamical
symmetries in quantum mechanics, symmetric polynomials, the theory of 
partitions and counting of bound states in short-range potentials.

In Section \ref{S:2dexpart} we discuss a fascinating numerical result 
that arisies from a theoretical treatment of screened excitons. 
This stimulates a re-evaluation of the familiar two-dimensional hydrogen atom.
Formulating the latter problem in momentum space leads to a new integral 
relation in terms of special functions, and fresh insights into the 
dynamical symmetry of the problem are also obtained. 
A discussion of an alternative potential to model screened excitons is 
given, and  bound-state energies and scattering phase shifts for this 
potential are compared with those 
obtained using the two-dimensional analogue of the Yukawa potential.

In Section \ref{S:anyonpart} we consider an exciton consisting of a 
valence hole and several fractionally-charged quasielectrons (anyons). 
A complete set of basis functions is obtained for this `anyon exciton'
and these functions are classified using a result from the theory of 
partitions.
Expressions are derived for the overlap and interaction matrix elements 
of four- and six-particle systems, which describe an exciton against
the background of incompressible quantum liquids with filling factors 
1/3, 2/3 and 2/5.
Some exact results are obtained for the general $(N+1)$-particle case,
including the binding energy for an exciton with zero in-plane momentum 
and zero internal angular momentum.

\section{Two-dimensional exciton: screened and unscreened}\label{S:2dexpart}

\subsection{Motivation}

Modelling of a screened Wannier-Mott exciton in two dimensions using the 
Stern-Howard potential \cite{Stern1967} has revealed an intriguing 
numerical relation \cite{Portnoi1997,Portnoi1999}.
In the search for an analytical derivation of this result, we address the 
simpler problem of an unscreened two-dimensional (2D) exciton, which has 
a well-known solution in real space.
It has been claimed that the aforementioned numerical relation is not 
strictly exact, and an alternative approach to the problem has been proposed 
\cite{Tanguy2001}. This has led to a new potential for modelling screened 
excitons, which is also considered here.

\subsection{Screened 2D Wannier-Mott exciton}

Illumination of a semiconductor with near-bandgap light leads to
the creation of electron-hole bound states known as excitons.
The study of excitons in condensed matter physics has a long
history \cite{Rashba1982}. Broadly speaking, excitons can be divided
into two types: hydrogen atom-like Wannier-Mott excitons and more
tightly-bound Frenkel excitons. In the present work we shall only be
concerned with the first of these.

Advances in semiconductor growth techniques have led to the creation
of almost ideal 2D systems. In such systems, confinement leads to a 
higher exciton binding energy than in the bulk, which makes 2D excitons 
more experimentally accessible.
A semiconductor quantum well is a quasi-two-dimensional 
system, in which photoexcited electrons and holes are essentially confined 
to a plane as their motion is `frozen' in the growth direction. 
Excitons are extremely important for the optical properties 
of the quantum well. 
If enough electrons and holes are excited, the Coulomb attraction between 
particles may be weakened - this phenomenon is known as \emph{screening}.
The effect of screening on an exciton may be approximated by considering
the electron and hole as interacting via the so-called 
Stern-Howard potential \cite{Stern1967}, which has the following form in 
real space:
\vspace{3mm}
\begin{equation}\label{Stern}
V_s(\rho)=-\frac{2}{\rho}\left\{1-\frac{\pi}{2}q_s\rho\left[
\mathbf{H}_0(q_s\rho)-N_0(q_s\rho)\right]\right\}=-\frac{2}{\rho}
\left\{1-q_s\rho\, S_{0,0}(q_s\rho)\right\},
\vspace{3mm}
\end{equation} 
where $q_s$ is known as the 2D screening wavenumber, and $\mathbf{H}_0$, 
$N_0$ and $S_{0,0}$ are the Struve, Neumann and Lommel functions, 
respectively. 
Note that all lengths and energies are measured in units of 
three-dimensional (3D) exciton Bohr radius and excitonic Rydberg,
respectively.
In these units, the radial equation for the relative electron-hole motion 
is given by
\vspace{3mm}
\begin{equation}\label{radial2DSchr}
-\frac{d^2
R(\rho)}{d\rho^2}-\frac{1}{\rho}\frac{dR(\rho)}{d\rho}+\left(V_s(\rho)+
\frac{m^2}{\rho^2}\right)R(\rho)=ER(\rho).
\vspace{3mm}
\end{equation}
This is the radial part of the 2D Schr\"{o}dinger equation,
which is discussed in more detail in Sections \ref{S:2Dha} and \ref{S:radial}.
In Eq.~\eqref{radial2DSchr}, $m$ is the azimuthal quantum number, which
takes integer values.

Numerical calculations \cite{Portnoi1997,Portnoi1999} using 
Eq.~\eqref{Stern} have shown that with increasing screening, bound states 
in this potential disappear at integer values of the inverse screening 
wavenumber $1/q_s$.
This `empirical' observation can be formulated as the following
Sturm-Liouville problem:
\vspace{3mm}
\begin{equation}\label{Sturm}
\left(\hat{A}-\frac{m^2}{x^2}\right)F(x)+\lambda V(x)F(x)=0,
\vspace{3mm}
\end{equation}
where
\vspace{3mm}
\begin{equation}\label{Aoper}
\hat{A}=\frac{1}{x}\frac{d}{dx}x\frac{d}{dx},
\vspace{3mm}
\end{equation}
$\lambda=2/q_s$, and we have made the substitution $x=q_s\rho$.
The zero right-hand side of Eq.~\eqref{Sturm} reflects the fact that
we restrict our interest to zero-energy bound states.
The function $V(x)$ is defined as 
\vspace{3mm}
\begin{equation}\label{newV}
V(x)=\frac{1}{x}-S_{0,0}(x),
\vspace{3mm}
\end{equation}
and it has the following properties:
\vspace{3mm}
\begin{equation}\label{Vprops}
\int_0^\infty V(x)\,x\,dx=1\quad\mbox{and}\quad V(x)=
\hat{A}\left\{\frac{1}{x}-V(x)\right\}.
\vspace{3mm}
\end{equation}

The key result of \cite{Portnoi1997} is that the solution $F$ of
Eq.~\eqref{Sturm} is square-integrable, i.e.
\vspace{3mm}
\begin{equation}\label{squint}
\int_0^\infty |F(x)|^2\,x\,dx=1,
\vspace{3mm}
\end{equation}
only if
\vspace{3mm}
\begin{equation}\label{lamrel}
\lambda=(2|m|+\nu)(2|m|+\nu+1),
\vspace{3mm}
\end{equation}
where $\nu=0, 1, 2,\ldots$ is the number of non-zero nodes in the radial 
wavefunction. 

Notably, the potential defined in Eq.~\eqref{Stern} is simplified 
considerably if we move to Fourier space:
\begin{equation}\label{SternF}
V_s(q)=-\frac{4\pi}{\left(q+q_s\right)}.
\vspace{3mm}
\end{equation}
Eq.~\eqref{SternF} is the 2D analogue of the familiar Yukawa potential
(a standard textbook derivation for both the 2D and 3D cases can be found 
in \cite{Haug1998}, for example).

The most compact way to reformulate the above Sturm-Liouville problem
is in terms of a homogeneous integral equation:
\vspace{3mm}
\begin{equation}\label{intyse}
q^2\Phi(\mathbf{q})=\frac{\lambda}{2\pi}\int\frac{\Phi(\mathbf{q'})}
{|\mathbf{q}-\mathbf{q'}|+1}\,d^2q'.
\vspace{3mm}
\end{equation}
It follows that $\Phi(\mathbf{q})$ is square-integrable in 2D Fourier
space when $\lambda$ satisfies Eq.~\eqref{lamrel}. 
It should be emphasised that this result is of a numerical nature and 
has not been proved analytically. 

In an attempt to find an analytical derivation of Eq.~\eqref{lamrel}
we address the simpler problem of an \emph{unscreened} 2D exciton.
This system is analogous to the 2D hydrogen atom, with the proton
replaced by a valence hole, so we now focus our attention on such
a system.

\subsection{Hydrogen atom and accidental degeneracy}

The in-plane motion of an electron and hole can be described by a 
two-dimensional Schr\"{o}dinger equation for a single particle with a 
reduced mass. 
As in three dimensions, the 2D hydrogen atom exhibits `accidental'
degeneracy of the bound-state energy levels. 

Before proceeding it is necessary to discuss accidental degeneracy 
in more detail. The phenomenon of accidental degeneracy is well known 
from the theory of the 3D hydrogen atom. 
There are three quantum numbers $\lbrace n,l,m\rbrace$, although the energy 
only depends on the principal quantum number $n$. 

Degeneracy in $m$ is of geometrical origin, and is due to the invariance
of the system under spatial rotations. From Noether's theorem 
\cite{Noether1918}, this symmetry implies the existence of a conserved
quantity, which is the angular momentum $\hat{\mathbf{L}}$. 
Indeed, quantum mechanics tells us that whenever there is a conserved 
quantity whose components cannot be measured simultaneously, then the 
energy eigenvalues will be degenerate \cite{Landau1977}.

The origin of the degeneracy in $l$, however, is not so obvious.
In fact, this degeneracy does not have a geometrical origin, but arises
from a hidden \emph{dynamical} symmetry of the system.
The conserved quantity associated with this accidental degeneracy is
the \emph{Runge-Lenz vector} $\hat{\mathbf{A}}$, the components of which
cannot be measured simultaneously.
The existence of the Runge-Lenz vector is not so well known because when 
the Schr\"{o}dinger equation is separated in spherical polar coordinates, 
the natural choice for a complete set of commuting operators is 
$\lbrace\hat{H},\hat{L}_z,\hat{L}^2\rbrace$, where $\hat{H}$ is the
Hamiltonian. 
However, the presence of the Runge-Lenz vector is evident if the 
Schr\"{o}dinger equation is separated in parabolic coordinates, as the 
natural choice for a complete set of commuting operators is then 
$\lbrace\hat{H},\hat{L}_z,\hat{A}_z\rbrace$. 
In general, accidental degeneracy appears when the Schr\"{o}dinger equation 
can be separated in more than one coordinate system (for a detailed review 
see \cite{Kalnins1976}).

In two dimensions there are two quantum numbers $\lbrace n,m\rbrace$. 
However, the energy eigenvalues only depend on $n$, so they are
degenerate in the quantum number $m$. In this case the angular momentum 
only has a $z$-component (where $z$ is in the direction perpendicular
to the plane of motion), and so it can be known precisely. 
The degeneracy in $m$ is therefore due to the existence of a 2D 
Runge-Lenz vector and a corresponding dynamical symmetry of the system.

\subsection{Historical background}

\begin{figure}
\begin{center}
\includegraphics[width=12cm,keepaspectratio]{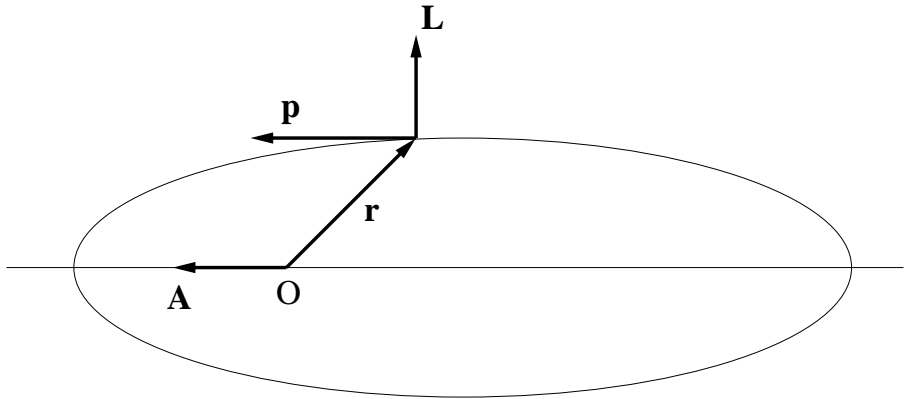}
\end{center}
\caption{Classical Runge-Lenz vector {\bf A}.}
\vspace{1cm}
\label{runge}
\end{figure}

The Runge-Lenz vector has its origin in classical mechanics 
(see Fig.~\ref{runge}), and its discovery has been variously attributed
to Laplace, Runge, Lenz and Hamilton (see \cite{Goldstein1980}
and references therein).
A result known as Bertrand's theorem states that the only central forces 
which result in closed orbits are the inverse square law (Kepler problem)
and Hooke's law \cite{Goldstein1980,Bertrand1873}.  
The closure of orbits is due to the conservation of an additional 
quantity (apart from the energy and angular momentum); in the case of the 
Kepler problem this quantity is the Runge-Lenz vector
\vspace{3mm}
\begin{equation}
\mathbf{A}=\frac{1}{\mu}\left(\mathbf{p}\times\mathbf{L}\right)-
\kappa\frac{\mathbf{r}}{r},
\vspace{3mm}
\end{equation}
where $\mu$ is the reduced mass, $\mathbf{p}$ is the momentum, 
$\mathbf{L}$ is the angular momentum, $\kappa$ is a positive constant 
and $\mathbf{r}$ is the radius vector of the orbit.

A quantum mechanical analogue of the Runge-Lenz vector was first suggested 
by Pauli \cite{Pauli1926}, by analogy with classical mechanics. 
The main difference is that it is now an operator:
\vspace{3mm}
\begin{equation}
\hat{\mathbf{A}}=\frac{1}{2\mu}\left(\hat{\mathbf{p}}\times\hat{\mathbf{L}}
-\hat{\mathbf{L}}\times\hat{\mathbf{p}}\right)-\kappa
\frac{\hat{\mathbf{r}}}{r},
\vspace{3mm}
\end{equation}
where the form of the first term on the right-hand side
ensures that $\hat{\mathbf{A}}$ is Hermitian.
Pauli also made the key conclusion that the existence of a Runge-Lenz 
vector constant in time implies that the hydrogen atom exhibits an 
additional degeneracy to that due to conservation of angular momentum. 
He then used $\hat{\mathbf{A}}$ to derive the correct energy levels 
for the hydrogen atom.

The most important paper concerning the symmetry of the hydrogen atom was 
published by Fock in 1935 \cite{Fock1935}. 
His first step was to consider the Schr\"{o}dinger equation in momentum 
space, which resulted in an integral equation.
Considering only negative energy (bound-state) solutions, he then projected 
the 3D momentum space onto the surface of a four-dimensional (4D) 
hypersphere using a stereographic projection. 
After a suitable transformation of the wavefunction, the resulting 
Schr\"{o}dinger equation is seen to be invariant under rotations in 
4D Euclidean space. 
The kernel is just the Green's function of Laplace's equation in four 
dimensions. 
Application of Green's theorem yields the correct energy eigenvalues, 
as well as eigenfunctions in terms of hyperspherical harmonics. 
Fock deduced that the dynamical symmetry of the hydrogen atom is described 
by the 4D rotation group SO(4), which contains the geometrical 
symmetry SO(3) as a subgroup. 
He related this hidden symmetry to the observed degeneracy of the energy 
eigenvalues.

Shortly afterwards, Bargmann \cite{Bargmann1936} made the connection 
between Pauli's quantum mechanical Runge-Lenz vector and Fock's discovery 
of invariance under rotations in 4D space. 
The real-space Schr\"{o}dinger equation was also separated in parabolic 
coordinates, which demonstrated more explicitly the conservation of the 
Runge-Lenz vector.

A major step forward was taken by Alliluev \cite{Alliluev1958}, 
who extended Fock's method of stereographic projection to the case of 
$d$ dimensions $(d\geqslant 2)$. 
He derived general expressions for the eigenfunctions and eigenvalues
of the problem, and clarified some of Fock's arguments relating to the 
symmetry group for the $d$-dimensional case. 
Alliluev also considered the 2D harmonic oscillator, the symmetry 
group of which is isomorphic to the 3D rotation group SO(3). 
This provided an interesting parallel to the hydrogen atom case.

The first explicit use of Fock's method for the 2D hydrogen 
atom was made by Shibuya and Wulfman \cite{Shibuya1965}. 
In this case, the eigenfunctions are just the usual spherical harmonics 
in three dimensions. 
The authors also included a graphical illustration of the degeneracy, 
in which two eigenfunctions which look very different on a plane are 
shown to be equivalent when projected onto a sphere. 
(It should be noted that the solution of the real-space Schr\"{o}dinger 
equation in two dimensions was first provided in 
\cite{Flugge1952}, although the derivation was later repeated 
in \cite{Zaslow1967}.)

In the 1960s there was a revival of interest in accidental degeneracy 
as it was hoped that it could be applied to the theory of elementary 
particles. 
It attracted the attention of such famous names as Gell-Mann, Schwinger
and Ne'eman.
In a wide-ranging study \cite{Bander1966a,Bander1966b}, Bander and Itzykson 
provided a comprehensive review of work thus far. 
The first part gives a detailed explanation of the earlier work of Fock 
and Alliluev, the details of which are difficult to glean from the 
original works. 
Two systems of generalised spherical harmonics are then considered on the 
4D hypersphere: the standard hyperspherical harmonics, 
and an alternative in terms of Wigner $D$-functions. 
These are related to separation of variables is spherical polar and 
parabolic coordinates, respectively. 
For the 3D hydrogen atom, the Runge-Lenz vector in real 
space is shown to correspond to the infinitesimal generator of rotations 
on the 4D hypersphere. 
A detailed group-theoretical analysis of the hydrogen atom is also given. 
In the second part, the analysis is extended to scattering states, 
where the momentum space is now projected onto a higher-dimensional 
hyperboloid.

The next few decades saw little progress, probably because the important 
questions had already been addressed so thoroughly. 
However, improvements in semiconductor growth techniques enabled the 
manufacture of effectively 2D structures, which soon led 
to a resurgence of interest in the 2D hydrogen atom.

In a series of papers \cite{Yang1991a,Yang1991b,Guo1991}, 
the 2D Runge-Lenz vector was defined for the first time, 
and real-space solutions of the Schr\"{o}dinger equation were applied 
to problems of atomic physics in two dimensions, including dipole matrix 
elements, the Stark effect, photon transition rates and fine/hyperfine 
structure. 
The relativistic case was also considered.

Returning to higher dimensions, the use of hydrogenoid orbitals as basis 
sets for problems in quantum chemistry led to a re-examination of the 
separability of the $d$-dimensional Schr\"{o}dinger equation in different 
coordinate systems \cite{Aquilanti1997}. A novel approach was
taken by Nouri \cite{Nouri1999}, who considered the $d$-dimensional Coulomb 
problem in quantum phase space. This led to an expression for the 
Wigner distribution function, from which it is possible to calculate 
some physical and chemical properties for the hydrogen atom.
The dynamical symmetry of the hydrogen atom has also been related to 
Kac-Moody algebras, rather than the usual Lie algebras 
\cite{Daboul1993,Daboul1998}. 
This provides fertile ground for mathematical investigation and will 
no doubt lead to a new direction for research.

Recent studies \cite{Dahl1997,Kamath2002} have shown that the 
classical Runge-Lenz vector has a deeper physical origin, being tied to 
the generator of Lorentz transformations for the relativistic two-body 
problem; the non-relativistic Kepler problem is then just a zero-order 
description of the more general case. 
In principle, it should be possible to undertake a similar analysis of 
the quantum mechanical problem based on the Dirac equation.

Fock's method for the 3D hydrogen atom has been applied by
Muljarov \emph{et al.}~\cite{Muljarov1999,Muljarov2000} to the problem 
of an exciton with anisotropic mass and dielectric constant, 
providing an interesting alternative to a variational approach 
\cite{Schindlmayr1997}.
The anisotropy is considered as a perturbation, 
and the perturbed wavefunction is expanded in terms of the 
hyperspherical harmonics.
Brillouin-Wigner perturbation theory is then used to calculate the
energy levels and eigenfunctions of the perturbed system.

\subsection{Fock's method}

We now present an outline of Fock's approach to the 
3D hydrogen atom \cite{Fock1935}, which led to important
insights into the dynamical symmetry of the problem. We start from the 
Schr\"{o}dinger equation in real space
\vspace{3mm}
\begin{equation}
\left(-\nabla^2-\frac{2}{r}\right)\Psi(\mathbf{r})=E\Psi(\mathbf{r}),
\vspace{3mm}
\end{equation}
with distances and energies measured in terms of the 3D Bohr radius 
$a_0=\hbar^2/\mu e^2$ and Rydberg $Ry=\mu e^4/2\hbar^2$, where $\mu$ is
the reduced mass of the system.

Introducing the pair of Fourier transforms between real space and
momentum space
\begin{align}
\Phi(\mathbf{p})&=\int\Psi(\mathbf{r})e^{i\mathbf{p}\cdot
\mathbf{r}}\,d\mathbf{r}, \\
\Psi(\mathbf{r})&=\frac{1}{(2\pi)^3}\int\Phi(\mathbf{p})e^{-i\mathbf{p}
\cdot\mathbf{r}}\,d\mathbf{p},
\end{align}
and noting that for bound states the energy $E=-p_0^2$ will be negative,
we may write the Schr\"{o}dinger equation as an integral equation
\vspace{3mm}
\begin{equation}\label{inteqn4D1}
\left(p^2+p_0^2\right)\Phi(\mathbf{p})=\frac{1}{\pi^2}
\int\frac{\Phi(\mathbf{p'})\,d\mathbf{p'}}{|\mathbf{p}-\mathbf{p'}|^2}.
\vspace{3mm}
\end{equation}

The 3D momentum space is now projected onto the surface
of a 4D unit hypersphere centred at the origin, and the
in-plane momentum is scaled by $p_0$. A point on the hypersphere is
specified by three angles, $\theta$, $\phi$ and $\alpha$, where
$\alpha$ is the angle between the position vector of the point and the
fourth axis introduced in the 4D space.
The transformation is given explicitly by
\begin{align}
u_1&=\sin\alpha\sin\theta\cos\phi=\frac{2p_0p_x}{p^2+{p_0}^2}, \\
u_2&=\sin\alpha\sin\theta\sin\phi=\frac{2p_0p_y}{p^2+{p_0}^2}, \\
u_3&=\sin\alpha\cos\theta=\frac{2p_0p_z}{p^2+{p_0}^2}, \\
u_4&=\cos\alpha=\frac{p^2-{p_0}^2}{p^2+{p_0}^2}.
\end{align}
An element of surface area on the unit hypersphere is given by
\begin{equation}
d\Omega=\sin^2\alpha\sin\theta\,d\alpha\,d\theta\,d\phi=
\left(\frac{2p_0}{p^2+{p_0}^2}\right)^3\,d\mathbf{p},
\end{equation}
and the distance between two points transforms as:
\begin{equation}
|\mathbf{u}-\mathbf{u'}|^2=\frac{(2p_0)^2}{(p^2+{p_0}^2)
(p'^2+{p_0}^2)}|\mathbf{p}-\mathbf{p'}|^2.
\end{equation}
If the wavefunction on the sphere is expressed as
\begin{equation}
\chi(\mathbf{u})=\frac{1}{\sqrt{p_0}}\left(\frac{p^2+{p_0}^2}
{2p_0}\right)^2\Phi(\mathbf{p}),
\end{equation}
then Eq.~\eqref{inteqn4D1} reduces to the simple form:
\begin{equation}\label{inteqn4D2}
\chi(\mathbf{u})=\frac{1}{2\pi^2
p_0}\int\frac{\chi(\mathbf{u'})\,d\Omega'}{|\mathbf{u}-\mathbf{u'}|^2}.
\vspace{3mm}
\end{equation}

\begin{figure}
\begin{center}
\includegraphics[width=7cm,keepaspectratio]{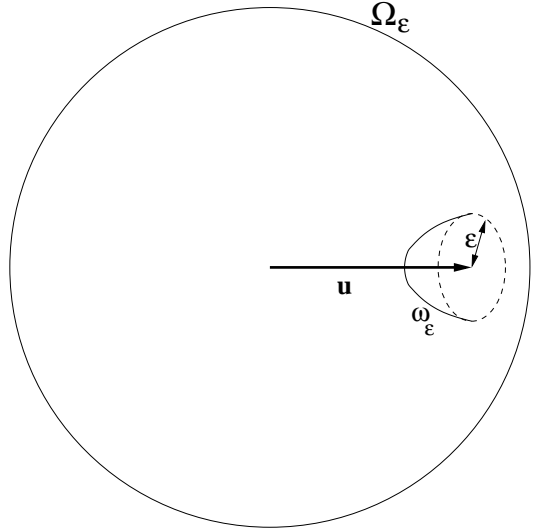}
\end{center}
\caption{Surface of integration in Green's formula.}
\vspace{1cm}
\label{epsilon}
\end{figure}

We now recognise that the kernel of the integral equation 
\eqref{inteqn4D2} is just the Green's function for the Laplacian
operator in four dimensions:
\begin{equation}
G=\frac{1}{|\mathbf{u}-\mathbf{u'}|^2},
\end{equation}
which satisfies
\begin{equation}
\nabla^2 G=-4\pi^2\delta(\mathbf{u}-\mathbf{u'}).
\end{equation}
We also introduce a function $\psi$ which is a solution of
Laplace's equation, i.e. it is a harmonic function.
We now wish to apply Green's theorem for $G$ and $\psi$
to the volume bounded by the unit hypersphere, letting
$\mathbf{\hat{n}}$ denote the unit vector from the origin to
the north pole of the hypersphere:
\vspace{3mm}
\begin{equation}\label{grr}
\int_{\tau}\left[\psi(\mathbf{u'})\nabla^2 G-G\nabla^2\psi(\mathbf{u'})
\right]\,d\tau=\int_{\Omega}\left[\psi(\mathbf{u'})\frac{\partial G}
{\partial n}-G\frac{\partial\psi(\mathbf{u'})}{\partial n}\right]
\,d\Omega.
\vspace{3mm}
\end{equation}
However, this theorem can only be applied when both $G$ and $\psi$
are harmonic inside the hypersphere and on its surface, whereas
$G$ has a singularity at $\mathbf{u}=\mathbf{u'}$.
To overcome this problem we define our surface of integration to
be the surface of the unit hypersphere with a small `hemisphere'
of radius $\varepsilon$ removed around the point $\mathbf{u}$
(see Fig.~\ref{epsilon}):
\begin{equation}
\Omega_\varepsilon\equiv\{|\mathbf{u'}|^2=1,\,|\mathbf{u}-\mathbf{u'}|^2
\geqslant\varepsilon\}\cup\{|\mathbf{u'}|^2\leqslant 1,\,
|\mathbf{u}-\mathbf{u'}|^2=\varepsilon\}.
\end{equation}
As $\varepsilon\rightarrow 0$ we regain the original unit hypersphere.

Applying Green's theorem to this new surface we obtain
\vspace{3mm}
\begin{equation}
0=\int_{\Omega_\varepsilon}\left[\psi(\mathbf{u'})\frac{\partial G}
{\partial n}-G\frac{\partial\psi(\mathbf{u'})}{\partial n}\right]
\,d\Omega_\varepsilon,
\vspace{3mm}
\end{equation}
because both $G$ and $\psi$ are harmonic throughout this new region.
Now, the integral on the right-hand side is made up of two parts:
the first part will tend to the usual surface integral over the 
original unit hypersphere as $\varepsilon\rightarrow 0$, whereas
the second part over the small `hemisphere' must be considered in
more detail.

For the first term, denoting the small hemispherical surface by
$\omega_\varepsilon$:
\vspace{3mm}
\begin{equation}
\int_{\omega_\varepsilon}G\frac{\partial\psi(\mathbf{u'})}{\partial n}
\,d\omega_\varepsilon\sim\int_{r=\varepsilon}\frac{1}{r^2}
\frac{\partial\psi}{\partial r}r^3\,dS=O(\varepsilon),
\vspace{3mm}
\end{equation}
where $dS$ is an element of solid angle, and so this term vanishes
as $\varepsilon\rightarrow 0$. For the second term we have
\vspace{3mm}
\begin{equation}
\int_{\omega_\varepsilon}\psi(\mathbf{u'})\frac{\partial G}{\partial n}
\,d\omega_\varepsilon=-\int_{r=\varepsilon}\psi(\mathbf{u'})\frac{\partial}
{\partial r}\left(\frac{1}{r^2}\right)r^3\,dS\rightarrow
2\pi^2\psi(\mathbf{u})\quad\mbox{as}\quad\varepsilon\rightarrow 0.
\vspace{3mm}
\end{equation}
Substituting back into Green's formula \eqref{grr} and rearranging we 
obtain
\vspace{3mm}
\begin{equation}\label{Greenfinal}
\psi(\mathbf{u})=\frac{1}{2\pi^2}\int_\Omega\left[G
\frac{\partial\psi(\mathbf{u'})}{\partial r}-\psi(\mathbf{u'})
\frac{\partial G}{\partial n}\right]\,d\Omega.
\vspace{3mm}
\end{equation}

We now write our solution to Laplace's equation in the form
\begin{equation}
\psi=r^\lambda\chi,\qquad\lambda=0,1,2,\ldots\; .
\end{equation}
Substituting for $\psi$ in Eq.~\eqref{Greenfinal} and restricting
$\mathbf{u}$ and $\mathbf{v}$ to the surface of the unit hypersphere,
so that
\begin{equation}
\frac{\partial}{\partial n}\frac{1}{|\mathbf{u}-\mathbf{u'}|^2}
=-\frac{1}{|\mathbf{u}-\mathbf{u'}|^2},
\end{equation}
we obtain
\begin{equation}
\chi(\mathbf{u})=\frac{\lambda +1}{2\pi^2}\int\frac{\chi(\mathbf{u'})
d\Omega}{|\mathbf{u}-\mathbf{u'}|^2}.
\vspace{3mm}
\end{equation}
Comparing this with Eq.~\eqref{inteqn4D2} gives
\begin{equation}
\frac{1}{2\pi^2p_0}=\frac{\lambda+1}{2\pi^2}.
\end{equation}
Substituting for $p_0$ and associating $\lambda+1$ with the principal
quantum number $n$, we find that
\begin{equation}
E_n=-\frac{1}{n^2},\qquad n=1,2,3,\ldots\; .
\vspace{3mm}
\end{equation}

The corresponding eigenfunctions $\chi(\mathbf{u})$ are just the 
solutions of Laplace's equation in four dimensions, 
i.e. the hyperspherical harmonics.
These may be expressed alternatively as
\begin{equation}
\chi(\mathbf{u})=(-2i)^l l!\sqrt{\frac{2n(n-l-1)!}{\pi(n+l)!}}\,
\sin^l\alpha\:C_{n-l-1}^{l+1}(\cos\alpha)\,Y_l^m(\theta,\phi),
\vspace{3mm}
\end{equation}
where $C_\nu^\mu$ are the Gegenbauer polynomials \cite{Gradshteyn2000} 
and $Y_l^m$ are the conventional spherical harmonics.

\subsection{Two-dimensional hydrogen atom}\label{S:2Dha}

The relative in-plane motion of an electron and hole, with effective
masses $m_e$ and $m_h$, respectively, may be treated as that of a
single particle with reduced mass $\mu=m_em_h/(m_e+m_h)$ and energy
$E$, moving in a Coulomb potential $V(\rho)$. The wavefunction of the
particle satisfies the stationary Schr\"{o}dinger equation
\vspace{3mm}
\begin{equation}\label{wannier}
\hat{H}\Psi(\boldsymbol{\rho})=\left[-\frac{1}{\rho}\frac{\partial}
{\partial\rho}\left(\rho\frac{\partial}{\partial\rho}\right)-\frac{1}
{\rho^2}\frac{\partial^2}{\partial\phi^2}+V(\rho)\right]
\Psi(\boldsymbol{\rho})=E\Psi(\boldsymbol{\rho}),
\vspace{3mm}
\end{equation}
where $(\rho,\phi)$ are plane polar coordinates.  Note that excitonic
Rydberg units are used throughout this Chapter, which leads to a
potential of the form $V(\rho)=-2/\rho$.

The eigenfunctions of Eq.~\eqref{wannier} are derived in Section
\ref{S:radial}. It is well known that the bound state energy levels
are of the form \cite{Flugge1952}:
\vspace{3mm}
\begin{equation}\label{elevel}
E=-\frac{1}{\left(n+1/2\right)^2},\qquad n=0,1,2,\ldots,
\vspace{3mm}
\end{equation}
where $n$ is the principal quantum number. Notably, Eq.~\eqref{elevel} does
not contain explicitly the azimuthal quantum number $m$, which enters
the radial equation (see Section \ref{S:radial},
Eq.~\eqref{radialeqn}). Each energy level is $(2n+1)$-fold 
degenerate, the so-called accidental degeneracy.
The difference between the factor $-1/(n+1/2)^2$ in the 2D case and 
$-1/n^2$ in the 3D case has been interpreted as arising from the 
topological peculiarity of 2D space \cite{Bateman1992}:
considering the origin as a singular point, paths that enclose the origin
and those that do not cannot be continuously deformed into one another.

It is convenient to introduce a vector operator corresponding to the
$z$-projection of the angular momentum
$\hat{\mathbf{L}}_z=\mathbf{e}_z\hat{L}_z$, where $\mathbf{e}_z$ is a
unit vector normal to the plane of motion of the electron and hole. We
now introduce the 2D analogue of the quantum-mechanical
Runge-Lenz vector as the dimensionless operator
\begin{equation}\label{dimless}
\hat{\mathbf{A}}=(\hat{\mathbf{q}}\times\hat{\mathbf{L}}_z-
\hat{\mathbf{L}}_z\times\hat{\mathbf{q}})-\frac{2}{\rho}\boldsymbol{\rho},
\end{equation}
where $\hat{\mathbf{q}}=-i\nabla$ is the momentum operator. Note that
$\hat{\mathbf{A}}$ lies in the plane and has Cartesian components
$\hat{A}_x$ and $\hat{A}_y$.

$\hat{L}_z$, $\hat{A}_x$ and $\hat{A}_y$ represent conserved
quantities and therefore commute with the Hamiltonian:
\begin{equation}
[\,\hat{H},\,\hat{L}_z\,]=[\,\hat{H},\,\hat{A}_x\,]=[\,\hat{H},
\,\hat{A}_y\,]=0.
\end{equation}
They also satisfy the following commutation relations:
\begin{align}
[\,\hat{L}_z,\,\hat{A}_x\,]&=i\hat{A}_y, \\
[\,\hat{L}_z,\,\hat{A}_y\,]&=-i\hat{A}_x, \\
[\,\hat{A}_x,\,\hat{A}_y\,]&=-4i\hat{L}_z\hat{H}.
\end{align}

The existence of the non-commuting operators $\hat{A}_x$ and
$\hat{A}_y$, representing conserved physical quantities, implies that
the Runge-Lenz vector is related to the accidental degeneracy of the
energy levels in two dimensions \cite{Landau1977}. We now present a simple
interpretation of the hidden symmetry underlying this degeneracy.

For eigenfunctions of the Hamiltonian we can replace $\hat{H}$ by the
energy $E$, and defining
\begin{equation}
\hat{\mathbf{A}}'=\frac{\hat{\mathbf{A}}}{2\sqrt{-E}},
\end{equation}
we obtain the new commutation relations:
\begin{align}
[\,\hat{L}_z,\,\hat{A}_x'\,]&=i\hat{A}_y', \\
[\,\hat{L}_z,\,\hat{A}_y'\,]&=-i\hat{A}_x', \\
[\,\hat{A}_x',\,\hat{A}_y'\,]&=i\hat{L}_z.
\end{align}

If we now construct a 3D vector operator
\begin{equation}
\hat{\mathbf{J}}=\hat{\mathbf{A}}'+\hat{\mathbf{L}}_z,
\end{equation}
then the components of $\hat{\mathbf{J}}$ satisfy the commutation
rules of ordinary angular momentum:
\begin{equation}
[\,\hat{J}_j,\,\hat{J}_k\,]=i\epsilon_{jkl}\hat{J}_l,
\end{equation}
where $\epsilon_{jkl}$ is the Levi-Civita symbol. 
Indeed, the components of $\hat{\mathbf{J}}$ are isomorphic to the
generators of the 3D rotation group SO(3).

Noting that
$\hat{\mathbf{A}}'\cdot\hat{\mathbf{L}}_z=\hat{\mathbf{L}}_z\cdot
\hat{\mathbf{A}}'=0$,
we have
\begin{equation}\label{jsq}
\hat{\mathbf{J}}^2={(\hat{\mathbf{A}}'+\hat{\mathbf{L}}_z)}{}^2=
{\hat{\mathbf{A}}
'}{}^2+\hat{\mathbf{L}}_z^2,
\end{equation}
where the operator $\hat{\mathbf{J}}^2$ has eigenvalues $j(j+1)$ and
commutes with the Hamiltonian. The operator $\hat{\mathbf{J}}^2$ is
known as a \emph{Casimir operator} of the group consisting of the
components of $\hat{\mathbf{J}}$. From Racah's theorem 
\cite{Racah1951}, we find that for a Lie group of rank $l$, 
there exists a set of $l$ Casimir operators which are functions 
of the generators and commute with every operator of the group. 
The present group is of rank 1 as none of the generators commutes 
with any other, which means that $\hat{\mathbf{J}}^2$ is unique.

We now make use of a special expression relating $\hat{\mathbf{A}}^2$
and $\hat{\mathbf{L}}_z^2$, the derivation of which is given in
Section \ref{S:asquared}:
\begin{equation}\label{asqeqn}
\hat{\mathbf{A}}^2=\hat{H}(4\hat{\mathbf{L}}_z^2+1)+4.
\end{equation}
Substituting in Eq.~\eqref{jsq} and again replacing $\hat{H}$ with
$E$, we obtain
\begin{equation}\label{jsqu}
\hat{\mathbf{J}}^2=-\frac{1}{4E}\left[E(4\hat{\mathbf{L}}_z^2+1)+4\right]+
\hat{\mathbf{L}}_z^2.
\end{equation}
Because $[\,\hat{H},\,\hat{\mathbf{J}}^2\,]=0$, an eigenfunction of
the Hamiltonian will also be an eigenfunction of $\hat{\mathbf{J}}^2$.
Operating with both sides of Eq.~\eqref{jsqu} on an eigenfunction of
the Hamiltonian, we obtain for the eigenvalues of $\hat{\mathbf{J}}^2$:
\begin{equation}
j(j+1)=-\left(\frac{1}{4}+\frac{1}{E}\right).
\end{equation}
Rearranging, and identifying $j$ with the principal quantum number
$n$, we obtain the correct expression for the energy eigenvalues:
\begin{equation}
E=-\frac{1}{\left(n+1/2\right)^2},\qquad n=0,1,2,\ldots\; .
\vspace{3mm}
\end{equation}

Note that the $z$-component of $\hat{\mathbf{J}}$ is simply
$\hat{L}_z$. Recalling that the eigenvalues of $\hat{L}_z$ are denoted by $m$,
there are $(2j+1)$ values of $m$ for a given $j$. However, as $j=n$,
we see that there are $(2n+1)$ values of $m$ for a given energy, which
corresponds to the observed $(2n+1)$-fold degeneracy.

\subsection{Fock's method in two dimensions}\label{S:sec3}

\subsubsection{Stereographic projection}\label{SS:Fock}

The method of Fock \cite{Fock1935}, in which a 3D momentum
space is projected onto the surface of a 4D hypersphere,
may be applied to our 2D problem \cite{Torres1998,Parfitt2002a}. 
We begin by defining a pair of 2D Fourier transforms between 
real space and momentum space:
\begin{align}
\Phi(\mathbf{q})&=\int\Psi(\boldsymbol{\rho})e^{i\mathbf{q}\cdot
\boldsymbol{\rho}}\,d\boldsymbol{\rho},
\label{f1}\\
\Psi(\boldsymbol{\rho})&=\frac{1}{(2\pi)^2}\int\Phi(\mathbf{q})e^{-i\mathbf{q}
\cdot\boldsymbol{\rho}}\,d\mathbf{q}.
\label{f2}
\end{align}
We shall restrict our interest to bound states, and hence the energy
$E=-q_0^2$ will be negative.

Substitution of Eq.~\eqref{f2} in Eq.~\eqref{wannier} yields the
following integral equation for $\Phi(\mathbf{q})$:
\begin{equation}\label{intSE}
\left(q^2+q_0^2\right)\Phi(\mathbf{q})=\frac{1}{\pi}
\int\frac{\Phi(\mathbf{q'})\,d\mathbf{q'}}{|\mathbf{q}-\mathbf{q'}|}.
\end{equation}

\begin{figure}
\begin{center}
\includegraphics[width=12cm,keepaspectratio]{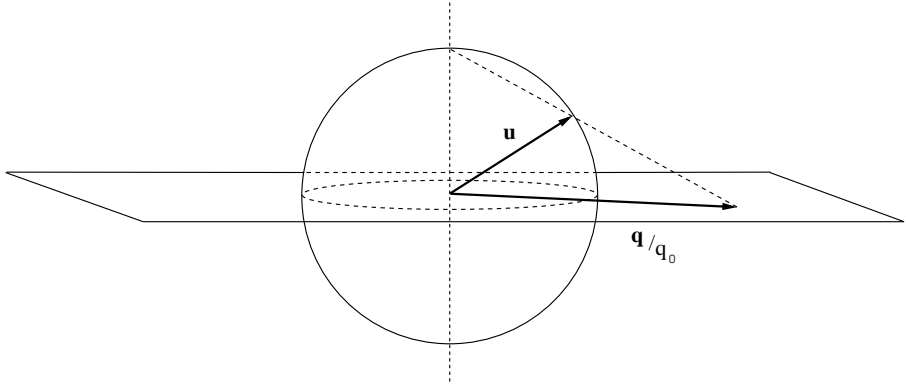}
\end{center}
\caption{Projection of 2D momentum space onto unit sphere.}
\vspace{1cm}
\label{projection}
\end{figure}

The 2D momentum space is now projected onto the surface
of a 3D unit sphere centred at the origin, and so it is
natural to scale the in-plane momentum by $q_0$ (see Fig.~\ref{projection}, 
adapted from our paper \cite{Parfitt2003a}). 
Each point on a unit
sphere is completely defined by two polar angles, $\theta$ and $\phi$,
and the Cartesian coordinates of a point on the unit sphere are given
by
\begin{align}
u_x&=\sin\theta\cos\phi=\frac{2q_0q_x}{q^2+{q_0}^2},\label{c1} \\
u_y&=\sin\theta\sin\phi=\frac{2q_0q_y}{q^2+{q_0}^2},\label{c2} \\
u_z&=\cos\theta=\frac{q^2-{q_0}^2}{q^2+{q_0}^2}.\label{c3}
\end{align}

An element of surface area on the unit sphere is given by
\vspace{3mm}
\begin{equation}
d\Omega=\sin\theta\,d\theta\,d\phi=\left(\frac{2q_0}{q^2+{q_0}^2}\right)^2
\,d\mathbf{q},
\vspace{3mm}
\end{equation}
and the distance between two points transforms as:
\vspace{3mm}
\begin{equation}
|\mathbf{u}-\mathbf{u'}|=\frac{2q_0}{(q^2+{q_0}^2)^{1/2}
(q'^2+{q_0}^2)^{1/2}}|\mathbf{q}-\mathbf{q'}|.
\vspace{3mm}
\end{equation}

If the wavefunction on the sphere is expressed as
\vspace{3mm}
\begin{equation}\label{trans}
\chi(\mathbf{u})=\frac{1}{\sqrt{q_0}}\left(\frac{q^2+{q_0}^2}
{2q_0}\right)^{3/2}\Phi(\mathbf{q}),
\vspace{3mm}
\end{equation}
then Eq.~\eqref{intSE} reduces to the simple form:
\begin{equation}\label{int}
\chi(\mathbf{u})=\frac{1}{2\pi
q_0}\int\frac{\chi(\mathbf{u'})\,d\Omega'}{|\mathbf{u}-\mathbf{u'}|}.
\vspace{3mm}
\end{equation}

\subsubsection{Expansion in spherical harmonics}

Any function on a sphere can be expressed in terms of spherical
harmonics, so for $\chi(\mathbf{u})$ we have
\begin{equation}\label{sph1}
\chi(\mathbf{u})=\sum_{l=0}^{\infty}\sum_{m=-l}^{l}A_{lm}Y_l^m(\theta,\phi),
\vspace{3mm}
\end{equation}
where $Y_l^m(\theta,\phi)$ are basically defined as in \cite{Mathews1970}:
\begin{equation}
Y_{l}^{m}(\theta,\phi)=c_{lm}\sqrt{\frac{2l+1}{4\pi}\frac{(l-|m|)!}
{(l+|m|)!}}\;P_l^{|m|}(\cos\theta)e^{im\phi},
\end{equation}
where $P_n^{|m|}(\cos\theta)$ is an associated Legendre function as
defined in \cite{Gradshteyn2000}. The constant $c_{lm}$ is an arbitrary 
`phase factor'. 
As long as $|c_{lm}|^2=1$ we are free to choose $c_{lm}$, and
for reasons which will become clear in Section \ref{SS:ints} we set
\begin{equation}\label{phase}
c_{lm}=(-i)^{|m|}.
\end{equation}

The kernel of the integral in Eq.~\eqref{int} can also be
expanded in this basis as \cite{Arfken1985}:
\vspace{1mm}
\begin{equation}\label{sph2}
\frac{1}{|\mathbf{u}-\mathbf{u'}|}=\sum_{\lambda=0}^{\infty}
\sum_{\mu=-\lambda}^{\lambda}\frac{4\pi}{2\lambda+1}
Y_\lambda^\mu(\theta,\phi)Y_\lambda^{\mu*}(\theta ',\phi ').
\vspace{3mm}
\end{equation}

Substituting Eqs.~\eqref{sph1} and \eqref{sph2} into Eq.~\eqref{int}
we have
\vspace{3mm}
\begin{multline}\label{long}
\sum_{l=0}^{\infty}\sum_{m=-l}^{l}A_{lm}Y_l^m(\theta,\phi)=\frac{2}{q_0}
\sum_{l_1=0}^{\infty}\sum_{l_2=0}^{\infty}\sum_{m_1=-l_1}^{l_1}
\sum_{m_2=-l_2}^{l_2} \\
\times\int\frac{1}{2l_2+1}A_{l_1m_1}Y_{l_1}^{m_1}(\theta',\phi')
Y_{l_2}^{m_2}(\theta,\phi)Y_{l_2}^{m_2*}(\theta',\phi')\,d\Omega'.
\vspace{3mm}
\end{multline}

We now make use of the orthogonality property of spherical harmonics
to reduce Eq.~\eqref{long} to the following:
\vspace{3mm}
\begin{equation}\label{short}
\sum_{l=0}^{\infty}\sum_{m=-l}^{l}A_{lm}Y_l^m(\theta,\phi)=\frac{2}{q_0}
\sum_{l_1=0}^{\infty}\sum_{m_1=-l_1}^{l_1}\frac{1}{2l_1+1}A_{l_1m_1}
Y_{l_1}^{m_1}(\theta,\phi).
\vspace{3mm}
\end{equation}
Multiplying both sides of Eq.~\eqref{short} by
$Y_n^{m'*}(\theta,\phi)$ and integrating over $d\Omega$ gives
\vspace{3mm}
\begin{equation}
A_{nm'}=\frac{2}{q_0(2n+1)}A_{nm'},
\vspace{3mm}
\end{equation}
where we have again used the orthogonality relation for spherical
harmonics. The final step is to rearrange for $q_0$ and identify the
index $n$ with the principal quantum number. This enables us to find
an expression for the energy in excitonic Rydbergs:
\begin{equation}
E=-q_0^2=-\frac{1}{\left(n+1/2\right)^2},\qquad n=0,1,2,\ldots\; .
\vspace{3mm}
\end{equation}
This is seen to be identical to Eq. \eqref{elevel}.

For a particular value of $n$, the general solution of Eq.~\eqref{int}
can be expressed as
\begin{equation}\label{chin}
\chi_n(\mathbf{u})=\sum_{m=-n}^{n}A_{nm}Y_{n}^{m}(\theta,\phi).
\vspace{3mm}
\end{equation}
Each of the functions entering the sum in Eq.~\eqref{chin}
satisfies Eq.~\eqref{int} separately.
So, for each value of $n$ we have $(2n+1)$ linearly-independent
solutions, and this explains the observed $(2n+1)$-fold degeneracy.

We are free to choose any linear combination of spherical harmonics for our
eigenfunctions, but for convenience we simply choose
\begin{equation}\label{eig}
\chi_{nm}(\mathbf{u})=A_{nm}Y_{n}^{m}(\theta,\phi).
\end{equation}
If we also require our eigenfunctions to be normalised as follows:
\vspace{3mm}
\begin{equation}
\frac{1}{(2\pi)^2}\int|\chi(\mathbf{u})|^2\,d\Omega=\frac{1}{(2\pi)^2}
\int\frac{q^2+q_0^2}{2q_0^2}|\Phi(\mathbf{q})|^2\,d\mathbf{q}=
\int|\Psi(\boldsymbol{\rho})|^2\,d\boldsymbol{\rho}=1,
\vspace{3mm}
\end{equation}
then Eq.~\eqref{eig} reduces to
\begin{equation}
\chi_{nm}(\mathbf{u})=2\pi Y_{n}^{m}(\theta,\phi).
\end{equation}

Applying the transformation in Eq.~\eqref{trans}, we can obtain an
explicit expression for the orthonormal eigenfunctions of Eq.~\eqref{intSE}:
\vspace{3mm}
\begin{equation}\label{phiq}
\Phi_{nm}(\mathbf{q})=c_{nm}\sqrt{2\pi\frac{(n-|m|)!}{(n+|m|)!}}
\left(\frac{2q_0}{q^2+q_0^2}\right)^{3/2}P_n^{|m|}(\cos\theta)e^{im\phi},
\vspace{3mm}
\end{equation}
where we have used the fact that $q_0=(n+1/2)^{-1}$, and $\theta$ and 
$\phi$ are defined by Eqs.~\eqref{c1}--\eqref{c3}.

\subsubsection{New integral relation}\label{SS:ints}

To obtain the real-space eigenfunctions $\Psi(\boldsymbol{\rho})$ we make
an inverse Fourier transform:
\begin{equation}\label{phitrans}
\Psi(\boldsymbol{\rho})=\frac{1}{(2\pi)^2}\int\Phi(\mathbf{q})
e^{-i\mathbf{q}\cdot\boldsymbol{\rho}}\,d\mathbf{q}=\frac{1}{(2\pi)^2}
\int_0^{2\pi}\int_0^{\infty}\Phi(\mathbf{q})e^{-iq\rho\cos\phi'}
\,q\,dq\,d\phi',
\vspace{3mm}
\end{equation}
where $\phi'$ is the azimuthal angle between the vectors
$\boldsymbol{\rho}$ and $\mathbf{q}$. However, if we now substitute
Eq.~\eqref{phiq} into this expression we have to be careful with our
notation. The angle labelled $\phi$ in Eq.~\eqref{phiq} is actually
related to $\phi'$ via
\begin{equation}\label{phirel}
\phi=\phi'+\phi_\rho,
\end{equation}
where $\phi_\rho$ is the azimuthal angle of the vector
$\boldsymbol{\rho}$, which can be treated as constant for the purposes
of our integration.

Taking this into account, the substitution of Eq.~\eqref{phiq} into
Eq.~\eqref{phitrans} yields:
\vspace{3mm}
\begin{multline}\label{phi2}
\Psi(\boldsymbol{\rho})=\frac{c_{nm}}{(2\pi)^{3/2}}
\sqrt{\frac{(n-|m|)!}{(n+|m|)!}}\,e^{im\phi_\rho} \\
\times\int_0^{2\pi}\int_0^{\infty}\left(\frac{2q_0}{q^2+q_0^2}
\right)^{3/2}
P_n^{|m|}(\cos\theta)e^{i(m\phi'-q\rho\cos\phi')}\,q\,dq\,d\phi'.
\vspace{3mm}
\end{multline}

From Eq.~\eqref{c3} we obtain
\begin{equation}\label{theta}
P_n^{|m|}(\cos\theta)=P_n^{|m|}\left(\frac{q^2-{q_0}^2}{q^2+{q_0}^2}\right),
\vspace{3mm}
\end{equation}
and we use the following form of Bessel's integral \cite{Arfken1985}:
\vspace{3mm}
\begin{equation}\label{bes}
\int_0^{2\pi}e^{i(m\phi'-q\rho\cos\phi')}\,d\phi'=2\pi(-i)^m
J_m(q\rho),
\vspace{3mm}
\end{equation}
where $J_m(q\rho)$ is a Bessel function of the first kind of order
$m$.

Substituting Eqs.~\eqref{theta} and \eqref{bes} into
Eq.~\eqref{phi2} leads to
\vspace{3mm}
\begin{multline}\label{rho1}
\Psi(\boldsymbol{\rho})=\frac{c_{nm}(-i)^m}{\sqrt{2\pi}}
\sqrt{\frac{(n-|m|)!}{(n+|m|)!}}\,e^{im\phi_\rho} \\
\times\int_0^{\infty}\left(\frac{2q_0}{q^2+q_0^2}\right)^{3/2}P_n^{|m|}
\left(\frac{q^2-{q_0}^2}{q^2+{q_0}^2}\right)J_m(q\rho)\,q\,dq.
\vspace{3mm}
\end{multline}

We now make a change of variables, $x=q_0\rho$ and $y=q^2/q_0^2$, so
that Eq.~\eqref{rho1} becomes
\begin{multline}\label{rad}
\Psi(\boldsymbol{\rho})=c_{nm}(-1)^{n+m}(-i)^m\sqrt{\frac{q_0(n-|m|)!}
{\pi(n+|m|)!}}\,e^{im\phi_\rho} \\
\times\int_0^{\infty}P_n^{|m|}
\left(\frac{1-y}{1+y}\right)\frac{J_m(x\sqrt{y})}{(1+y)^{3/2}}\,dy,
\end{multline}
where we have used the fact that \cite{Arfken1985}:
\begin{equation}
P_n^{|m|}\left(\frac{y-1}{y+1}\right)=(-1)^{n+m}P_n^{|m|}
\left(\frac{1-y}{1+y}\right).
\vspace{3mm}
\end{equation}

If we now equate the expression for $\Psi(\boldsymbol{\rho})$ in 
Eq.~\eqref{rad} with that derived in Section \ref{S:radial}, we obtain
the following:
\vspace{3mm}
\begin{equation}\label{newint1}
c_{nm}(-1)^{n+m}(-i)^m\int_0^{\infty}P_n^{|m|}\left(\frac{1-y}{1+y}\right)
\frac{J_m(x\sqrt{y})}{(1+y)^{3/2}}\,dy=
\frac{(2x)^{|m|}e^{-x}}{n+1/2}L_{n-|m|}^{2|m|}(2x).
\vspace{3mm}
\end{equation}
The value of $c_{nm}$ chosen earlier in Eq.~\eqref{phase} ensures that
both sides of Eq.~\eqref{newint1} are numerically equal.
For $m\geqslant 0$ the relation simplifies to
\begin{multline}\label{newint2}
\int_0^{\infty}P_n^{m}\left(\frac{1-y}{1+y}\right)\frac{J_m(x\sqrt{y}
)}{(1+y)^{3/2}}\,dy=\frac{(-1)^n(2x)^{m}e^{-x}}{n+1/2}L_{n-m}^{2m}(2x), \\
n,m=0,1,2,\ldots;\:m\leqslant n.
\end{multline}

As far as we can ascertain, this integral relation between special
functions has not previously been tabulated. For $n,m=0$ we recover
the known integral relation \cite{Gradshteyn2000}:
\begin{equation}
\int_0^{\infty}\frac{J_0(x\sqrt{y})}{(1+y)^{3/2}}\,dy=2e^{-x}.
\vspace{3mm}
\end{equation}

\subsection{Dynamical symmetry}\label{S:sec4}

\subsubsection{Infinitesimal generators}

Consider now a vector $\mathbf{u}$ from the origin to a point on the
3D unit sphere defined in Section \ref{SS:Fock}. 
If this vector is rotated through an infinitesimal angle $\alpha$ in 
the $(u_xu_z)$ plane, we have a new vector
\begin{equation}\label{defn}
\mathbf{u}'=\mathbf{u}+\mathbf{\delta u},
\end{equation}
where the components of $\mathbf{u}$ are given in
Eqs.~\eqref{c1}--\eqref{c3}, and
\begin{equation}
\delta\mathbf{u}=\alpha\,\mathbf{e}_y\times\mathbf{u}.
\end{equation}
This rotation on the sphere corresponds to a change in the 2D
momentum from $\mathbf{q}$ to $\mathbf{q'}$.
The Cartesian components of Eq.~\eqref{defn} are then found to be
\begin{align}
u_{x}^{\prime}=\frac{2q_0 q_x^{\prime}}{q'{}^2+q_0^2}&=
\frac{2q_0 q_x}{q^2+q_0^2}+\alpha\frac{q^2-q_0^2}{q^2+q_0^2}, \\ 
u_{y}^{\prime}=\frac{2q_0 q_y^{\prime}}{q'{}^2+q_0^2}&=
\frac{2q_0 q_y}{q^2+q_0^2}, \\
u_{z}^{\prime}=\frac{q'{}^2-q_0^2}{q'{}^2+q_0^2}&=
\frac{q^2-q_0^2}{q^2+q_0^2}-\alpha\frac{2q_0 q_x}{q^2+q_0^2},
\end{align}
where $q^2=q_x^2+q_y^2$.

After some manipulation we can also find the components of
$\delta\mathbf{q}=\mathbf{q}'-\mathbf{q}$:
\begin{align}\label{deltas}
\delta q_x&=\alpha\frac{q^2-q_0^2-2q_x^2}{2q_0}, \\
\delta q_y&=-\alpha\frac{q_x q_y}{q_0}.
\end{align}

The corresponding change in $\Phi(\mathbf{q})$ is given by
\vspace{3mm}
\begin{equation}
\delta\Phi(\mathbf{q})=\frac{\alpha}{(q^2+q_0^2)^{3/2}}
\left(\frac{q^2-q_0^2-2q_x^2}{2q_0}\frac{\partial}{\partial
q_x}-\frac{q_x q_y}{q_0}\frac{\partial}{\partial
q_y}\right)\left[(q^2+q_0^2)^{3/2}\Phi(\mathbf{q})\right].
\vspace{3mm}
\end{equation}
We can write this as
\begin{equation}
\delta\Phi(\mathbf{q})=-\frac{i}{2q_0}\alpha\hat{\mathcal{A}}_x
\Phi(\mathbf{q}),
\vspace{3mm}
\end{equation}
where the infinitesimal generator is given by
\vspace{3mm}
\begin{equation}
\hat{\mathcal{A}}_x=\frac{i}{(q^2+q_0^2)^{3/2}}\left((q^2-q_0^2-2q_x^2)
\frac{\partial}{\partial
q_x}-2q_xq_y\frac{\partial}{\partial q_y}\right)(q^2+q_0^2)^{3/2}.
\vspace{3mm}
\end{equation}

We now make use of the following operator expression in the momentum 
representation:
\begin{equation}
\hat{\boldsymbol{\rho}}=\mathbf{e}_x\hat{x}+\mathbf{e}_y\hat{y}=
i\nabla_q,
\end{equation}
and the commutation relation
\begin{equation}
[\,\hat{\boldsymbol{\rho}},\,f(\mathbf{q})\,]=i\nabla_q f,
\end{equation}
to derive a more compact expression for $\hat{\mathcal{A}}_x$:
\begin{equation}\label{axe1}
\hat{\mathcal{A}}_x=(q^2-q_0^2)\hat{x}-2q_x(\mathbf{q}\cdot
\hat{\boldsymbol{\rho}})-3iq_x.
\end{equation}
By considering an infinitesimal rotation in the $(u_yu_z)$ plane we
can obtain a similar expression for $\hat{\mathcal{A}}_y$:
\begin{equation}\label{axe2}
\hat{\mathcal{A}}_y=(q^2-q_0^2)\hat{y}-2q_y(\mathbf{q}\cdot
\hat{\boldsymbol{\rho}})-3iq_y.
\end{equation}

These expressions operate on a particular energy eigenfunction with
eigenvalue $-q_0^2$. If we move the constant $-q_0^2$ to the right of
the coordinate operators in Eqs.~\eqref{axe1} and \eqref{axe2}, and
replace it with the Hamiltonian in momentum space, $\hat{\mathcal{H}}$:
\begin{align}
\hat{\mathcal{A}}_x&=q^2\hat{x}+\hat{x}\hat{\mathcal{H}}-2q_x(\mathbf{q}
\cdot\hat{\boldsymbol{\rho}})-3iq_x,
\label{ax} \\
\hat{\mathcal{A}}_y&=q^2\hat{y}+\hat{y}\hat{\mathcal{H}}-2q_y(\mathbf{q}
\cdot\hat{\boldsymbol{\rho}})-3iq_y,
\label{ay}
\end{align}
then $\hat{\mathcal{A}}_x$ and $\hat{\mathcal{A}}_y$ can operate on
any linear combination of eigenfunctions.

\subsubsection{Relation to Runge-Lenz vector}

Recall the definition of the 2D Runge-Lenz vector in real space:
\vspace{3mm}
\begin{equation}\label{rl2}
\hat{\mathbf{A}}=(\hat{\mathbf{q}}\times\hat{\mathbf{L}}_z-
\hat{\mathbf{L}}_z\times\hat{\mathbf{q}})-\frac{2}{\rho}\boldsymbol{\rho}.
\vspace{3mm}
\end{equation}
Using $\hat{\mathbf{L}}_z=\boldsymbol{\rho}\times\hat{\mathbf{q}}$,
and the following identity for the triple product of three vectors:
\begin{equation}
\mathbf{a}\times(\mathbf{b}\times\mathbf{c})=(\mathbf{a}
\cdot\mathbf{c})\mathbf{b}-(\mathbf{a}\cdot\mathbf{b})\mathbf{c},
\end{equation}
we can apply the commutation relation
$[\,\boldsymbol{\rho},\,\hat{\mathbf{q}}\,]=i$ to rewrite Eq.~\eqref{rl2}
in the form:
\begin{equation}\label{rl3}
\hat{\mathbf{A}}=\hat{\mathbf{q}}^2\boldsymbol{\rho}+\boldsymbol{\rho}
\left(\hat{\mathbf{q}}^2-\frac{2}{\rho}\right)-2\hat{\mathbf{q}}
(\hat{\mathbf{q}}\cdot\boldsymbol{\rho})-3i\hat{\mathbf{q}}.
\vspace{3mm}
\end{equation}

If we now return to the expression for the real-space Hamiltonian in 
Eq.~\eqref{wannier}, it is apparent that we may substitute
\begin{equation}\label{ham2}
\hat{\mathbf{q}}^2-\frac{2}{\rho}=\hat{H},
\end{equation}
in Eq.~\eqref{rl3} to yield
\begin{equation}
\hat{\mathbf{A}}=\hat{\mathbf{q}}^2\boldsymbol{\rho}+\boldsymbol{\rho}
\hat{H}-2\hat{\mathbf{q}}(\hat{\mathbf{q}}\cdot\boldsymbol{\rho})-3i
\hat{\mathbf{q}}.
\end{equation}
Comparing this with Eqs.~\eqref{ax} and \eqref{ay}, it is evident that
the two components of the Runge-Lenz vector in real space correspond
to the generators of infinitesimal rotations in the $(u_xu_z)$ and
$(u_yu_z)$ planes.

\subsection{Postscript}

Although projection into momentum space has proved useful for the 
unscreened case, it offers no further insights into the analytical 
derivation of Eq.~\eqref{lamrel}. An alternative approach to this problem 
has recently been taken by Tanguy \cite{Tanguy2001}. He claims, using the
following argument, that the numerical result in Eq.~\eqref{lamrel} for 
the critical screening lengths is not strictly exact.

In a semi-classical approximation \cite{Berry1973,Yi1994}, the 2D 
formulation of energy quantisation for zero-energy bound states in
the Stern-Howard potential \eqref{Stern} may be expressed as
\vspace{3mm}
\begin{equation}
\int_0^\infty\sqrt{\frac{2}{\rho}\left\{1-\frac{\pi}{2}q_s\rho\left[
\mathbf{H}_0(q_s\rho)-N_0(q_s\rho)\right]\right\}}\,d\rho=
\left(\nu+\frac{1}{2}\right)\pi.
\vspace{3mm}
\end{equation}
Rearranging, the following expression for $\lambda=2/q_s$ is obtained in 
the semi-classical approximation:
\begin{equation}\label{lamsc}
\lambda_{sc}=\left(\frac{\pi}{2\mathcal{I}}\right)^2
\left(\nu+\frac{1}{2}\right)^2,
\end{equation}
where
\vspace{3mm}
\begin{equation}\label{calI}
\mathcal{I}=\int_0^\infty\sqrt{1-\frac{\pi}{2}u^2\left[
\mathbf{H}_0(u^2)-N_0(u^2)\right]}\,du.
\vspace{3mm}
\end{equation}
and $u=\sqrt{q_s\rho}$. We would expect Eq.~\eqref{lamsc} to be accurate 
for $\nu\gg 1$.

Tanguy asserts that for Eq.~\eqref{lamrel} to be true we would 
expect $2\mathcal{I}=\pi$, so as to obtain the correct leading order
dependence of $\lambda_{sc}$ on $\nu^2$. He evaluates the integral 
\eqref{calI} numerically and obtains
\begin{equation}
2\mathcal{I}=3.14057414,
\end{equation}
from which he concludes that $\lambda_{sc}$ differs from $\lambda$,
and so Eq.~\eqref{lamrel} is not exact.
Using the fact that the potential \eqref{Stern} behaves as $-2/\rho$ 
at small $\rho$ and as $-2/\left(q_s^2\rho^3\right)$ as 
$\rho\rightarrow\infty$, Tanguy proposes another potential 
\begin{equation}\label{Tanguy}
V(\rho)=-\frac{2}{\rho\left(1+q_s\rho\right)^2},
\vspace{3mm}
\end{equation}
which has the same asymptotic behaviour as potential \eqref{Stern}, 
and has zero-energy bound states when $\lambda=2/q_s$ satisfies 
Eq.~\eqref{lamrel}. 
The corresponding zero-energy radial wavefunctions are of the form
\vspace{3mm}
\begin{equation}
\phi\propto\rho^{|m|}\;{}_2F_1\left(-2|m|-\nu,2|m|+\nu+1;2|m|+1;
\frac{\rho}{\rho+\frac{(2|m|+\nu)(2|m|+\nu+1)}{2}}\right),
\vspace{3mm}
\end{equation}
where ${}_2F_1$ is a hypergeometric function.

\begin{figure}
\begin{center}
\includegraphics[width=8cm,keepaspectratio]{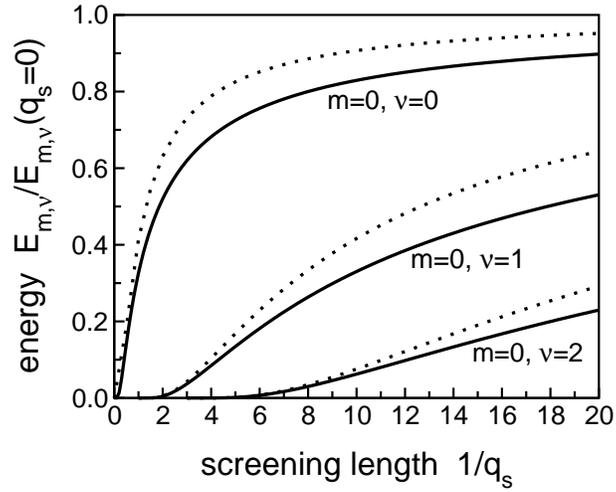}
\end{center}
\caption{Bound-state energies $E_{m,\nu}$ as a function of inverse 
screening wavenumber $1/q_s$ for the first three $m=0$ states in 
Stern-Howard (solid lines) and Tanguy (dotted lines) potentials.
All values are normalised by the corresponding energies for the
unscreened potential, $E_{m,\nu}(q_s=0)=-(m+\nu+1/2)^{-2}$.}
\vspace{1cm}
\label{bsenergy}
\end{figure}

\begin{figure}
\begin{center}
\includegraphics[width=8cm,keepaspectratio]{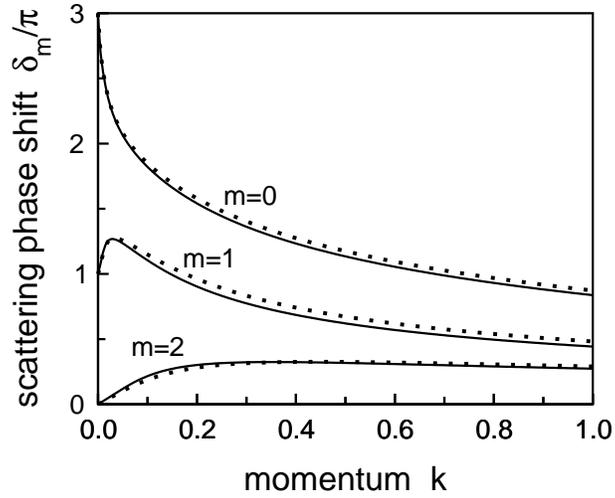}
\end{center}
\caption{Scattering phase shifts $\delta_m$ as a function of the 
in-plane momentum $k$ for $1/q_s=5$. Solid lines: Stern-Howard potential;
dotted lines: Tanguy potential.}
\vspace{1cm}
\label{phshift}
\end{figure}

The potential in Eq.~\eqref{Tanguy} is clearly much simpler than the
Stern-Howard potential \eqref{Stern}, and could certainly be used  
to model the screening of 2D excitons. 
To provide a comparison between the two potentials we use a 2D modification 
of the variable phase method (see Section \ref{S:vphase} for the details)
to calculate the bound-state energies and scattering phase shifts for 
Tanguy's potential \eqref{Tanguy}, 
and compare these with similar calculations for the Stern-Howard potential. 
The results of our calculations \cite{Parfitt2003b} are shown in 
Figs.~\ref{bsenergy} and \ref{phshift}.
Notably, for a wide range of values of $q_s$ there is at least ten per cent 
difference between numerically calculated binding energies for the two 
potentials, despite the fact that both potentials support the same number 
of bound states for the same value of $q_s$. 
The difference in the scattering phase shifts is less pronounced.


\subsection{Conclusions}

We have shown that the accidental degeneracy in the energy eigenvalues
of the 2D Kepler problem may be explained by the existence of a planar 
analogue of the familiar 3D Runge-Lenz vector.  
By moving into momentum space and making a stereographic projection 
onto a 3D sphere, a new integral relation in terms of 
special functions has been obtained, which to our knowledge has not 
previously been tabulated.
This approach has also yielded the momentum-space eigenfunctions
for bound states, which will prove useful in future applications.
We have also demonstrated explicitly that 
the components of the 2D Runge-Lenz vector in real space 
are intimately related to infinitesimal rotations in 3D
momentum space.
Although no further insight has been gained into the screened exciton
problem, this is more than compensated for by the unexpected beauty of 
the unscreened case. 

With regard to the numerical expression \eqref{lamrel} for critical 
screening lengths, it is still possible that it remains exactly true for 
both the Stern-Howard and Tanguy potentials.
Tanguy's results are inconclusive: the semi-classical approximation 
remains an approximation, and it is possible that rounding errors in the
numerical evaluation of the integral \eqref{calI} lead to its deviation 
from $\pi$. 
Indeed, it is interesting that the difference between the two potentials 
changes sign at $q_s\sim 1$, which allows for the fact that 
Eq.~\eqref{lamrel} may be simultaneously true for both potentials.

The obvious extension of the 2D hydrogenic problem is to consider 
scattering states as well as bound states, in order to obtain a complete 
set of basis functions.
However, rather than projection onto a sphere, this would require
the 2D momentum space to be projected onto a two-sheeted hyperboloid.
This has been done for the 3D hydrogen atom 
\cite{Bander1966b}, but the eigenfunctions are found to consist of two 
solutions, arising from the switching between sheets of the hyperboloid. 
Because of this complexity it is unlikely that the momentum-space 
eigenfunctions will produce any useful integral relations when compared 
with real-space solutions.

\section{Anyon exciton: exact solutions}\label{S:anyonpart}

\subsection{Motivation}

Intrinsic photoluminescence experiments have revealed interesting
features in the optical spectra of incompressible quantum liquids 
(IQLs) at filling factors $\nu=1/3$, $2/3$ and $2/5$
\cite{Goldberg1988,Heiman1988,Turberfield1990}.
The anyon exciton model (AEM) was developed to provide a model of
an exciton consisting of a valence hole and three quasielectrons
(anyons) against the background of $\nu=1/3$ and $2/3$ IQLs
\cite{Portnoi1996}, and has provided major insights into the role of 
electron-hole separation in determining their optical spectra.
Recent developments in experimental techniques for studying such
systems \cite{Plentz1996,Ashkinadze2002a} have rekindled our interest 
in developing a more general form of the model which may be applied 
to higher fractions.
To this end, we seek a generalised form of the AEM for a 
neutral exciton made up of a valence hole and an arbitrary number 
$N$ of anyons.
The particular case of $N=5$ allows us to consider an exciton against
the background of an IQL with filling factor $\nu=2/5$.

\subsection{Quantum Hall effects}

The quantum Hall effect (QHE) was discovered in 1980 by von Klitzing 
\emph{et al.}~\cite{Klitzing1980}, who studied an effectively 
2D electron system at high magnetic field and low temperature. 
The Hall conductivity $\sigma_{xy}$ exhibits plateaus at integer multiples 
of $e^2/h$ (corresponding to an integer number of filled Landau levels), 
and is accompanied by a vanishing diagonal conductivity $\sigma_{xx}$. 
This quantisation is independent of the shape of the specimen or presence 
of impurities and is accurate to one part in ten million, 
leading to its use as a standard of electrical resistance.

Another significant breakthrough was made in 1982 by Tsui \emph{et al.}
\cite{Tsui1982}, who investigated higher-mobility semiconductor samples 
at increased magnetic fields and lower temperatures.
Plateaus in $\sigma_{xy}$ were observed at fractional values of $e^2/h$,
accompanied by minima in $\sigma_{xx}$. It soon became apparent that
this phenomenon, labelled the \emph{fractional quantum Hall effect} (FQHE),
could not be explained in terms of a single-electron formulation;
any theory would need to take into account the strongly-correlated 
behaviour of many electrons.
(For a comprehensive review of the integer and fractional quantum Hall
effects see \cite{Prange1987,Chakraborty1995}.)

In 1983, Laughlin showed \cite{Laughlin1983b} that the stability of the
system at fractional filling factor (number of filled Landau levels)
is due to the condensation of the electron system into an incompressible
quantum liquid (IQL). 
By analogy with Jastrow wavefunctions in the theory of ${}^4$He, Laughlin 
proposed a many-electron variational wavefunction for filling factor 
$\nu=1/m$ ($m$ odd) to describe the ground state of the IQL:
\begin{equation}
\psi=\prod_{j<k}(z_j-z_k)^m\prod_l\exp\left(-|z_l|^2/4\right),
\vspace{3mm}
\end{equation}
where $z_i$ are the complex electron coordinates scaled in units of
magnetic length, and $m$ is an odd integer.
He also noted the existence of an energy gap between the ground 
and excited states of the IQL, and predicted that excitations would
be quasiparticles with fractional charge -- quasielectrons and
quasiholes. The first evidence for the existence of energy gaps was
soon provided by magnetotransport experiments \cite{Chang1983} and 
spectroscopic methods \cite{Kukushkin1986,Buhmann1990}, and
magnetotransport measurements also provided a signature of charge
fractionalisation \cite{Clark1988,Simmons1989,Dorozhkin1995}.

An alternative to Laughlin's method was provided by Haldane 
\cite{Haldane1983}, who considered a system based on a spherical geometry
in which the electrons are confined to the surface of a sphere with a magnetic
monopole at the centre.
Numerical diagonalisation of systems with up to eight electrons for
$\nu=1/3$ in a spherical geometry \cite{Haldane1985}
confirmed the accuracy of Laughlin's wavefunction in describing the 
ground state, as well as the presence of a finite energy gap between 
ground and excited states.
This work was later extended to $\nu=1/5$ with similar results
\cite{Fano1986}.
The use of Haldane's spherical geometry in numerical models of
an exciton against the background of an IQL is considered in
Section \ref{S:theory}.

Further progress on the nature of the quasiparticle excitations was
made by Halperin \cite{Halperin1984}, who developed pseudo-wavefunctions
to describe low-lying excited states in terms of a small number of 
quasiparticles added to the IQL ground state.
These pseudo-wavefunctions are of the form
\begin{equation}\label{hpseudo}
\Psi=P(z_k)\prod_{j<k}\left(z_j-z_k\right)^\alpha\prod_{l=1}^{N}
\exp\left\{-|z_l|^2/(4m)\right\},
\end{equation}
where $z_i$ are the complex coordinates of the $N$ quasiparticles,
$\alpha$ is a statistical factor and $P(z_i)$ is a symmetric 
polynomial in the variables $z_i$.
Halperin then used this method to construct the whole hierarchy of 
fractional states.
From the quantisation rules that determine the allowed quasiparticle
spacings he also determined that the quasiparticles are anyons
obeying fractional statistics, i.e. the wavefunction changes by a 
complex phase factor when two particles are interchanged. (For more 
details on anyons and fractional statistics see 
\cite{Wilczek1990,Wilczek1991}.)

In addition to charged quasiparticles, the IQL also supports dispersive
neutral excitations. The lowest branch of these neutral excitations is 
labelled a \emph{magnetoroton} \cite{Girvin1985,Girvin1986}, and can be 
pictured as a quasielectron-quasihole pair. 
Strong evidence of the existence of these excitations has been provided 
by inelastic light scattering (Raman scattering) experiments 
\cite{Pinczuk1993}. 

Returning to Laughlin's wavefunction, a picture began to emerge of magnetic
`vortices' generated by flux quanta, where the position of each vortex
corresponds to a zero of the many-body wavefunction. The $\nu=1/3$ FQHE,
for example, could be interpreted as the attachment of three flux quanta
to each electron, forming \emph{composite bosons} \cite{Kivelson1992}.
Stability of the system at this filling factor then corresponds to a
Bose condensate of these composite particles.

The failure of this model to account for higher-order states with 
odd denominator, as well as even-denominator states, led to the
development by Jain of the \emph{composite fermion model} 
\cite{Jain1989,Jain1990}.
Vortices are again attached to electrons, but the fractional quantum Hall
effect becomes an integer quantum Hall effect for the new composite
particles. For example, in the $\nu=1/3$ state each electron is
considered bound to two flux quanta, such that the composite particle
feels an effective field of one flux quantum per particle.
These composites can then be considered as fermions filling one
Landau level. The composite fermion model is generally accepted as
providing the best current description of higher-order FQHE states
(although not everyone subscribes to this view \cite{Dyakonov2001}), 
as well as a method for generating many-particle wavefunctions and 
their excitations. 

One further development that should be mentioned is the recent direct
observation of fractionally-charged quasiparticles in `shot noise'
measurements \cite{Picciotto1997,Saminadayar1997}. 
This noise arises from a current of individual quasielectrons with
charge $-e/3$ tunnelling through a narrow barrier 
(see \cite{Collins1997} for more details).
Whereas similar claims \cite{Goldman1995} have been questioned in the 
past, the results do look extremely promising, and a great deal of work 
is being carried out to verify these observations. 

\subsection{Photoluminescence experiments}

Although the earliest experimental work on the FQHE was based on 
magnetotransport experiments, these only allowed energies near the 
Fermi level to be probed.
The introduction of spectroscopic techniques was a significant step
as it enabled the whole spectrum of energies to be studied via
contactless measurements.
The ideal system for such optical experiments is a GaAs-AlGaAs
heterostructure. Donors are placed in the layer of higher-bandgap material
(AlGaAs) and the electrons associated with these donors move towards the
region of lower-bandgap material (GaAs). This process, known as modulation
doping, creates a high-mobility two-dimensional electron gas (2DEG) in the
layer of GaAs, which may be used as the basis for experimental study of
the IQL.
However, the electric field that is used to create the 2DEG also repels
photoexcited holes, which necessitates adding a second barrier to 
restrict these holes. 
The physical system can be modelled by two parallel planes separated
by a distance of $h$ magnetic lengths. It has been found that when 
$h=0$ the system possesses a hidden symmetry and is insensitive to electron 
correlations \cite{MacDonald1992,Apalkov1991b} (This is analogous to the 
familiar Kohn theorem, which states that the excitation spectrum of a 
translationally-invariant system is unaffected by electron-electron
interactions \cite{Kohn1961}.) 
As a result, the frequencies of all
allowed transitions coincide exactly with those of an empty crystal.
For this reason, charge asymmetric systems (with $h\neq 0$) are the
most interesting for experimental observation.

The phenomenon of photoluminescence also requires some explanation.
Illumination of the semiconductor heterostructure produces non-equilibrium 
electrons and holes, which may then recombine radiatively via two different 
mechanisms. 
In \emph{extrinsic} photoluminescence, electrons recombine with holes
localised on acceptor dopants, introduced via $\delta$-doping, whereas
in \emph{intrinsic} photoluminescence electrons recombine with `free' holes.

Experiments to study extrinsic emission provided the first 
optical measurement of FQHE energy gaps \cite{Kukushkin1986,Buhmann1990}, 
and a corresponding theory has been developed with good success 
\cite{Apalkov1991a}.
Of most interest to the present work, however, are experiments on 
intrinsic PL.
The first such experiments were by Heiman \emph{et al.} 
\cite{Goldberg1988,Heiman1988}, who studied high-mobility GaAs-AlGaAs 
multiple quantum wells at varying filling factor.
In this real system the hidden symmetry is broken, thus allowing electron
correlation effects to influence the optical spectra.
Several intrinsic emission lines were observed in the photoluminescence 
spectrum for $\nu<1$, and the most significant feature was a doublet near 
$\nu=2/3$ which had a similar temperature dependence to the 
magnetoresistance associated with the IQL in transport experiments.
The authors later studied a higher-mobility single quantum well,
which led to the observation of a similar doublet at filling factors
$\nu=1/3$ and $2/5$. 
The results for $\nu=2/3$ were also reported by Turberfield 
\emph{et al.} \cite{Turberfield1990}.

The only experimental work to study the effects of electron-hole 
separation on intrinsic photoluminescence was that by 
Plentz \emph{et al.} \cite{Plentz1996}.
Their apparatus allowed the 2DEG density to be changed by using
applied gate voltages. This enabled the effective electron-hole separation
(in units of magnetic length) to be varied, while keeping the filling 
factor constant (for more experimental details see \cite{Plentz1997}).
The PL spectra obtained for $\nu=1/3$ show a pronounced doublet
structure. With increasing electron-hole separation, the energy splitting
of the doublet decreases and the higher-energy peak dominates.
Refinements to this technique have recently been reported by
another group \cite{Yusa2001,Yusa2002}, and a similar method,
which uses variations in the light intensity to change the 2DEG
density, has also been developed \cite{Ashkinadze2002a,Ashkinadze2002b}.

These experimental methods have a great deal of potential for further 
investigation of spatially-separated systems. Although little work
has been done in this area, it is hoped that the present study and
subsequent work will rekindle interest in this direction.

\subsection{Theoretical studies}\label{S:theory}

Any attempt to model intrinsic PL must take into account 
low-energy excitonic bound states (due to electron-hole Coulomb attraction) 
against the background of an IQL. The first such work was by Apalkov 
\emph{et al.} 
\cite{Apalkov1992a,Apalkov1992b,Apalkov1995a,Apalkov1995b,Apalkov1995c},
who numerically diagonalised systems comprising a small number of electrons 
together with a valence hole on the surface of a Haldane sphere. 
The electron-hole interaction was modified to take into account the 
layer separation $h$, which allowed both symmetric and asymmetric 
systems to be studied.
Energy spectra were calculated for a range of values of $h$ and exciton
branches were classified in terms of the exciton internal angular 
momentum $L$.
For $h\leqslant 1.5l$, where $l$ is the magnetic length, there was a 
single exciton branch with $L=0$, whereas for $h>1.5l$ new branches appeared, 
with higher values of $L$ reaching down to become the ground state.
The branches were labelled as either \emph{tight excitons} or 
\emph{anyon excitons}. For small $h$, the $L=0$ tight exciton forms the
ground state, and as $h$ is increased, anyon exciton branches form
a succession of ground states and the tight excitons move to higher
energies.

Emission spectra were also calculated, and these also showed a strong
dependence on $h$. For $h\lesssim 0.5l$ a single peak was observed, 
whereas for $h\gtrsim 0.5l$ a second lower-energy peak appears.
A tentative interpretation of these results says that the upper peak
is from direct transitions for exciton momentum $k=0$, whereas the
lower peak corresponds to magnetoroton-assisted transitions with
finite $k$. The double-peak structure shows a qualitative agreement
with experiment \cite{Heiman1988,Plentz1997}.
The results of Apalkov \emph{et al.} were supported by a similar
numerical study by Zang and Birman \cite{Zang1995}, who also produced
PL spectra for different $h$ and obtained the familiar doublet structure.

In parallel with finite-size calculations, an analytical model of an
exciton against the background of an IQL was formulated by Portnoi
and Rashba 
\cite{Portnoi1996,Rashba1993,Portnoi1994,Portnoi1995a,Portnoi1995b}, 
termed the \emph{anyon exciton model} (AEM).
This model is applicable for electron-hole separation $h\gtrsim 2l$,
and thus complements the numerical approach. Indeed, remarkable 
agreement with finite-size results is obtained for $h\approx 2l$.
Initially, the simplest model of a hole with two semions (quasielectrons
with charge $-e/2$) was considered,
but this was soon extended to a more-realistic four-particle exciton 
consisting of a hole and three quasielectrons with charge $-e/3$.
The results of the AEM showed a multiple-branch energy spectrum, and a 
full set of exciton basis functions were obtained. More details of
the formulation and results of the AEM can be found below.

\subsection{Review of anyon exciton model}

\subsubsection{Formulation}\label{SS:formulation}

We present here a brief overview of the anyon exciton model (AEM)
applied to the case of a four-particle exciton \cite{Portnoi1996}.
The model considers a neutral exciton consisting of a valence
hole with charge $+e$ and three quasielectrons (anyons) with
charge $-e/3$ and statistical factor $\alpha$, subject to a strong 
perpendicular magnetic field $\mathbf{H}$.
The hole and anyons move in two different planes separated by a
distance of $h$ magnetic lengths. (Note that from now on we measure
all distances in terms of the magnetic length $l=(c\hbar/eH)^{1/2}$
unless otherwise specified.)

This four-particle exciton will have a total of four degrees of
freedom if all the particles are in the lowest Landau level. 
As the exciton is neutral we may assign it an in-plane
momentum $\mathbf{k}$, which absorbs two of these degrees of
freedom. The remaining degrees of freedom result in internal
quantum numbers and a multiple-branch energy spectrum.
Indeed, the two internal quantum numbers are assigned the labels
$L$ and $M$, and enumerate exciton branches. 
Note that for $\mathbf{k}=0$, $L$ is related to the projection of
the angular momentum $L_z$.

To simplify the description of the exciton, the hole and anyons
are considered as moving in the same plane, and the separation
$h$ can be introduced later as it does not appear in the exciton
wavefunction. It is also convenient to move from the in-plane hole 
and anyon coordinates $\mathbf{r}_h$ and $\mathbf{r}_i$ to the new 
2D coordinates
\vspace{3mm}
\begin{equation}\label{2dcts}
\mathbf{R}=\frac{1}{2}\left(\mathbf{r}_h+\frac{1}{3}\sum_{i=1}^3\:
\mathbf{r}_i\right),\quad\boldsymbol{\rho}=\frac{1}{3}\sum_{i=1}^3\:
\mathbf{r}_i-\mathbf{r}_h,\quad\mathbf{r}_{jl}=\mathbf{r}_j-\mathbf{r}_l,
\vspace{3mm}
\end{equation}
where $jl=12$, $23$ and $31$, together with the complex coordinates
$z_{jl}=x_{jl}+iy_{jl}$. In Eq.~\eqref{2dcts}, $\mathbf{R}$ is the 
coordinate of the geometrical centre of charge of the exciton, 
$\boldsymbol{\rho}$ is the position of the centre of negative charge 
relative to the hole, and $\mathbf{r}_{jl}$ are 
\emph{difference coordinates} between pairs of anyons.
The following constraint should also be noted:
\begin{equation}
\mathbf{r}_{12}+\mathbf{r}_{23}+\mathbf{r}_{31}=0,
\end{equation}
which greatly simplifies the subsequent formulation.

The motion of an anyon exciton is described by a Halperin
pseudo-wavefunction \cite{Halperin1984}:
\begin{multline}
\Psi (\mathbf{R},\boldsymbol{\rho},\{\bar{z}_{jl}\} )=\frac{1}
{\sqrt{2\pi A}}\exp\left\{i\mathbf{k}\cdot\mathbf{R}+\frac{i}{2}
\hat{\mathbf{z}}\cdot[\boldsymbol{\rho}\times\mathbf{R}]-
\frac{1}{4}(\boldsymbol{\rho}-\mathbf{d})^2\right\} \\
\times P_L(\ldots ,\bar{z_{jl}},\ldots)\prod_{jl}(\bar{z}_{jl})
^\alpha\exp\left\{-|z_{jl}|^2/36\right\},
\end{multline}
where $A$ is the area of the sample and $P_L$ is a homogeneous
polynomial of degree $L$ in the difference coordinates.
Note that the statistical factor $\alpha=0$ for bosons,
$\alpha=1$ for fermions, $\alpha=-1/3$ for quasielectrons in a $\nu=1/3$
IQL and $\alpha=+1/3$ for quasielectrons in a $\nu=2/3$ IQL.

The properties of the polynomial $P_L$ are of the utmost importance.
By definition it is symmetric in the anyon coordinates $\bar{z}_j$,
but the permutation rules with respect to the difference coordinates 
$\bar{z}_{jl}$ are not obvious. Indeed, it turns out that if $L$ is even,
$P_L$ is symmetric in $\bar{z}_{jl}$, whereas if $L$ is odd, the
polynomial is antisymmetric in $\bar{z}_{jl}$.
For $L=3$, the simplest case, this polynomial is a 
\emph{Vandermonde determinant}:
\vspace{3mm}
\begin{equation}\label{vand}
W\left(\bar{z}_{12},\bar{z}_{23},\bar{z}_{31}\right)=
\left|\begin{array}{ccc}
1 & 1 & 1 \\
\bar{z}_{12} & \bar{z}_{23} & \bar{z}_{31} \\
{\bar{z}_{12}}^2 & {\bar{z}_{23}}^2 & {\bar{z}_{31}}^2  
\end{array}\right|
=(\bar{z}_{12}-\bar{z}_{23})(\bar{z}_{23}-\bar{z}_{31})
(\bar{z}_{31}-\bar{z}_{12}).
\end{equation}

The linearly-independent $L$-even polynomials can be chosen as
\begin{equation}
P_{L,M}({\bar{z}_{12}},{\bar{z}_{23}},{\bar{z}_{31}})=
{\bar{z}_{12}}^{L-4M}{\bar{z}_{23}}^{2M}{\bar{z}_{31}}^{2M}
+{\bar{z}_{23}}^{L-4M}{\bar{z}_{31}}^{2M}{\bar{z}_{12}}^{2M}
+{\bar{z}_{31}}^{L-4M}{\bar{z}_{12}}^{2M}{\bar{z}_{23}}^{2M},
\end{equation}
where the quantum number $M$ enumerates different polynomials 
with the same degree $L$.
The total number of $L$-even polynomials with a given $L$ is
therefore $[L/6]+1$, where the brackets $[\; ]$ denote the integer part. 
The $L$-odd polynomials are constructed as follows:
\begin{equation}
P_{3,0}=W,\quad P_{L,M}=WP_{L-3,M},
\end{equation}
and there are $[(L-3)/6]+1$ of these polynomials for a given $L$.

Given an expression for the exciton basis functions
$\Psi_i$ in terms of the quantum numbers $\mathbf{k}$, $L$ and $M$,
any exciton wavefunction may be expressed in terms of this basis:
\begin{equation}
\Phi=\sum_i\chi_i\Psi_i.
\vspace{3mm}
\end{equation}
However, the basis functions are not orthogonal in $M$,
so the Schr\"{o}dinger equation is written as follows:
\begin{equation}
\hat{H}\boldsymbol{\chi}=\varepsilon\hat{B}\boldsymbol{\chi},
\end{equation}
where $\hat{B}$ is the overlap matrix of scalar products and
$\varepsilon$ is an energy eigenvalue.
The evaluation of the matrix elements $\hat{H}$ and $\hat{B}$
is rather complicated, and the reader is referred to 
\cite{Portnoi1996} for further details.
These matrix elements allow the calculation of the energy spectra and 
electron density for the exciton, and these results are outlined below.

\subsubsection{Results}\label{SS:results}

For energy eigenvalue calculations at $k=0$,
a comparison of the energy for statistical factor $\alpha=-1/3$
and $\alpha=0$ shows good agreement for $h>2$ 
(see Fig.~1 in \cite{Portnoi1996}), and hence a boson approximation 
is made to simplify further calculations.
For $k\neq 0$, the energy spectra must be calculated
numerically, and it is found that in the region $h>2$ 
the ground state of the system has non-zero angular momentum $L=3m$, 
where $m\geqslant 2$ is an integer.
Note also that the wavefunctions of these ground states contain the 
polynomials $P_{6M,M}$ and $WP_{6M,M}$. 
These satisfy the hard-core constraint, in that they go to zero if 
any two of the anyon coordinates are the same. This is very 
encouraging as hard-core functions are the accepted standard for
a proper description of anyons \cite{Wu1984a,Wu1984b}.
 
\begin{figure}
\begin{center}
\includegraphics[width=12cm,keepaspectratio]{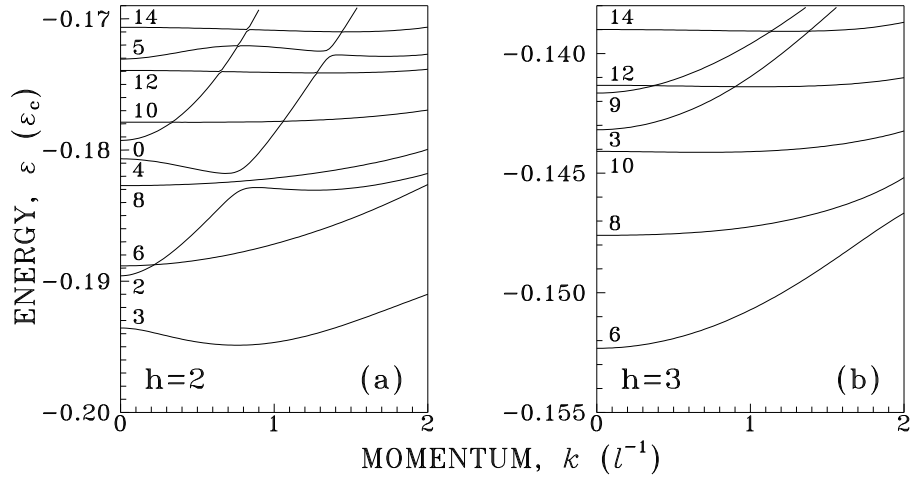}
\end{center}
\caption{Anyon exciton dispersion $\varepsilon (k)$ for two values
of electron-hole separation $h$. Numbers show $L$ values;
$h$ is in units of magnetic length $l$ and $\varepsilon$ is in units
of the Coulomb energy $\varepsilon_c=e^2/\epsilon l$.}
\vspace{1cm}
\label{energy}
\end{figure}

Energy spectra for $h=2$ and $3$ are shown in Fig.~\ref{energy}
\cite{Portnoi1996}.
For $h=2$, the minimum of the spectrum occurs for non-zero
$k$, which means that the exciton in the lowest
energy state possesses a non-zero dipole moment.
The negative dispersion in Fig.~\ref{energy}(a) arises because of 
the mutual repulsion of $L=2$ and $L=3$ branches.
For higher values of $h$ the ground state is always at $k=0$,
as in Fig.~\ref{energy}(b), and the lowest-branch dispersion is always
positive.

\begin{figure}
\begin{center}
\includegraphics[width=13cm,keepaspectratio]{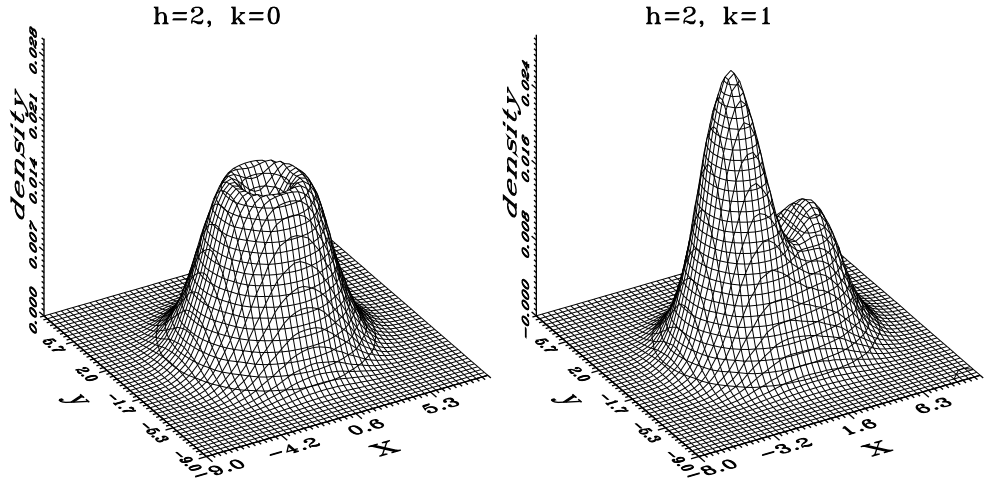}
\end{center}
\caption{Electron density distribution in an anyon exciton for different 
values of the exciton in-plane momentum $k$. The distance $h$ between the 
hole and the incompressible electron liquid is equal to two magnetic lengths. 
The hole is at the origin; the $x$-axis is chosen along the exciton dipole 
moment.}
\vspace{1cm}
\label{density}
\end{figure}

Calculations of the electron density around the hole also lead
to remarkable results (see Fig.~\ref{density} \cite{Portnoi1996}). 
For $k=0$ there is a crater-like dip in the density around the origin; 
this is an obvious signature of charge fractionalisation as a normal 
magnetoexciton always has a maximum at this point.
The distribution for non-zero $k$ clearly shows two anyons close 
to the hole and a single anyon at a slight distance from it, 
providing further clear evidence of charge fractionalisation in
anyon excitons.

\subsection{Exciton basis functions}

\subsubsection{Derivation}

The concept of exciton basis functions is now generalised 
\cite{Parfitt2003c} to an exciton consisting of a valence hole with 
charge $+e$ and $N$ anyons with charge $-e/N$ and  statistical factor $\alpha$.
The hole and anyons reside in two different layers, separated by a distance
of $h$ magnetic lengths, and are subject to a magnetic field 
$\mathbf{H}=H\hat{\mathbf{z}}$ perpendicular to their planes of confinement.
Unless otherwise stated, we shall assume that the hole and the
quasielectrons are in their corresponding lowest Landau levels.
To simplify the description of the exciton, the hole and anyons may be
considered as moving in the same plane with coordinates $\mathbf{r}_h$
and $\mathbf{r}_j$, respectively, and the the layer separation $h$ can
be introduced later when considering the anyon-hole interaction. 

An exciton consisting of a hole and $N$ anyons, all in the lowest
Landau level, will have a total of $N+1$ degrees of freedom. 
Assigning the exciton an in-plane momentum $\mathbf{k}$ absorbs two 
of these degrees of freedom, so for $N\geqslant 2$ the exciton will 
have $N-1$ internal degrees of freedom.
At this stage we consider the particles as non-interacting and introduce
interactions later as necessary.
We can therefore write the $(N+1)$-particle Hamiltonian as
\vspace{3mm}
\begin{equation}\label{hamil1}
\hat{H}_0=\frac{1}{2m_h}\left({\hat{\mathbf{p}}}_h-\frac{q_h
\mathbf{A}}{c}\right)^2+\sum_{j=1}^N\:\frac{1}{2m_a}\left(
{\hat{\mathbf{p}}}_j-\frac{q_a\mathbf{A}}{c}\right)^2,
\vspace{3mm}
\end{equation}
where $q_h=+e$ and $q_a=-e/N$ are the hole and anyon charges,
respectively.

Choosing the symmetric gauge 
\begin{equation}
\mathbf{A}=\frac{1}{2}\left[\mathbf{H}\times\mathbf{r}\right],
\end{equation}
and scaling all distances with the magnetic length $l=(c\hbar/eH)^{1/2}$,
we obtain
\vspace{3mm}
\begin{equation}\label{hamil3}
\hat{H}_0=\frac{1}{2m_h}\left(\frac{1}{i}\nabla_h-\left[\hat{\mathbf{z}}
\times\mathbf{r}_h\right]\right)^2+\sum_{j=1}^N\:\frac{1}{2m_a}
\left(\frac{1}{i}\nabla_j+\frac{1}{N}\left[\hat{\mathbf{z}}
\times\mathbf{r}_j\right]\right)^2,
\vspace{3mm}
\end{equation}
where, as usual, $e$, $\hbar$ and $c$ are assumed equal to unity.

\begin{figure}
\begin{center}
\includegraphics[width=13cm,keepaspectratio]{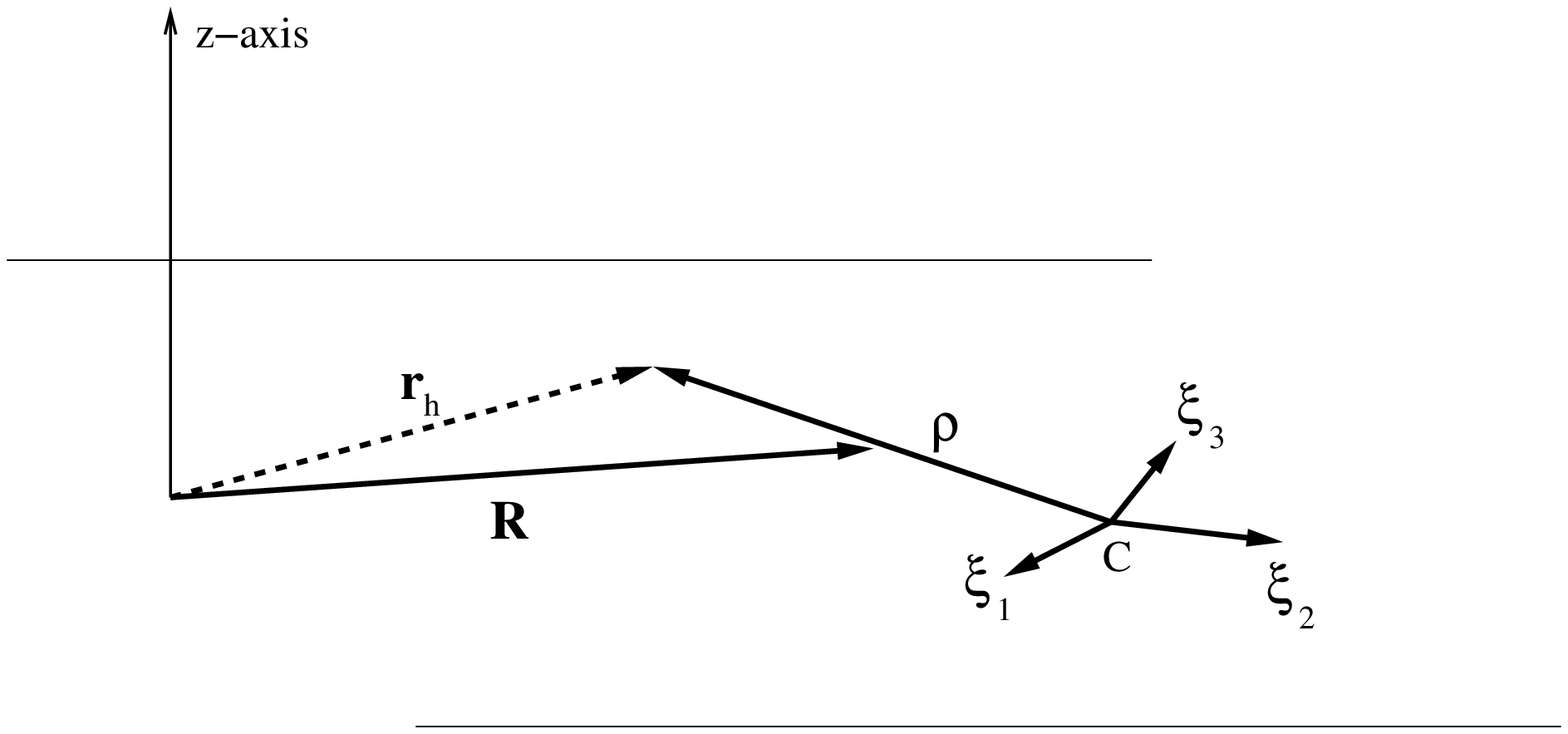}
\end{center}
\caption{Two-dimensional coordinate system for hole and $N$ anyons 
($C$ indicates the centre of negative charge). The hole is considered as
being in the same plane as the anyons, with layer separation introduced
as required.}
\vspace{1cm}
\label{coords}
\end{figure}

We now introduce the following new coordinates (see Fig.~\ref{coords}):
\vspace{3mm}
\begin{equation}
\mathbf{R}=\frac{1}{2}\left(\mathbf{r}_h+\frac{1}{N}\sum_{j=1}^N\:
\mathbf{r}_j\right),\quad\boldsymbol{\rho}=\mathbf{r}_h-\frac{1}{N}
\sum_{j=1}^N\:\mathbf{r}_j,\quad\boldsymbol{\xi}_j=\mathbf{r}_j
-\frac{1}{N}\sum_{l=1}^N\:\mathbf{r}_l,
\vspace{3mm}
\end{equation}
together with the complex coordinates $\zeta_j=\xi_{xj}+i\xi_{yj}$.
Note also the following constraint on these coordinates:
\begin{equation}\label{constraint}
\sum_{j=1}^N\:\boldsymbol{\xi}_j=\sum_{j=1}^N\:\zeta_j=0.
\vspace{3mm}
\end{equation}
The fact that the variables $\boldsymbol{\xi}_j$ are not independent
means that derivatives with respect to these variables are not defined.
However, it is possible to introduce the derivatives 
$\nabla_{\boldsymbol{\xi}_j}$
which treat the variables $\boldsymbol{\xi}_j$ as if they were
independent \cite{Bolton1994}.

Let $S$ be the $2N$-dimensional space of physical coordinates 
$\mathbf{r}_j$, and $S_\xi$ be the $2N$-dimensional space of the
\emph{independent} coordinates $\boldsymbol{\xi}_j$.
Due to the constraint above, physical motion only takes place within
a submanifold $M_\xi\subset S_\xi$, so $\xi$ maps the physical space
$S$ to the submanifold $M_\xi$, i.e. $\xi: S\rightarrow M_\xi$.
An arbitrary complex function $f$ can always be expressed explicitly 
as a function of the coordinates $\boldsymbol{\xi}_j$, and so we can
write that $f$ maps $M_\xi$ to $\mathbb{C}$, i.e. 
$f: M_\xi\rightarrow\mathbb{C}$. If we now keep the expression for
$f$ fixed, we can extend its definition over the whole space $S_{\xi}$
by removing the constraint, and it is now correct to talk about
its derivatives $\nabla_{\boldsymbol{\xi}_j}
f(\ldots,\boldsymbol{\xi}_j,\ldots)$.

\begin{figure}
\begin{center}
\includegraphics[width=12cm,keepaspectratio]{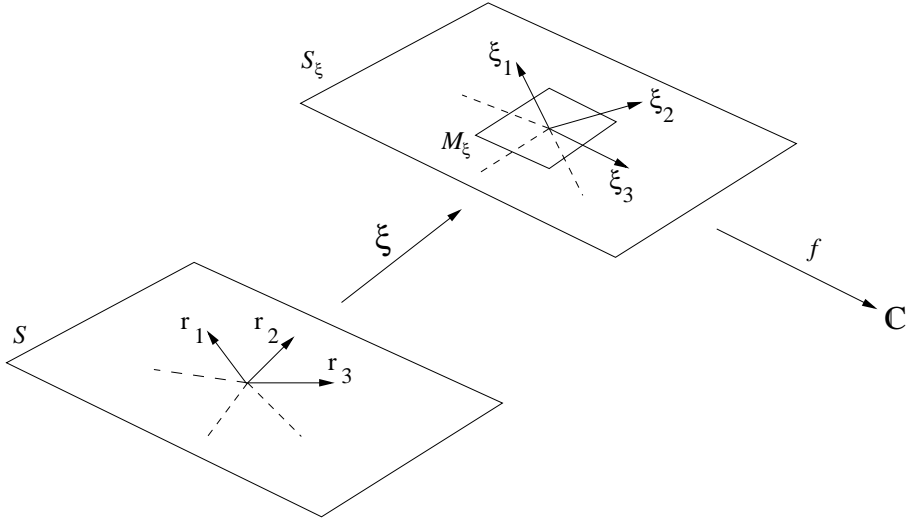}
\end{center}
\caption{Wavefunction of relative motion expressed as a composition of
two mappings $S\stackrel{\xi}
{\rightarrow}M_\xi\subset S_\xi\stackrel{f}{\rightarrow}\mathbb{C}$.}
\vspace{1cm}
\label{manifold}
\end{figure}

We may now consider a composition of the two maps $S\stackrel{\xi}
{\rightarrow}M_\xi\subset S_\xi\stackrel{f}{\rightarrow}\mathbb{C}$
(see Fig.~\ref{manifold}).
Wavefunctions of the kind used to describe motion with respect to 
the centre of charge of the anyon subsystem are compositions of
$f$ and $\xi$, such that
\begin{equation}
(f\circ\xi)(\mathbf{r}_1,\ldots ,\mathbf{r}_N)=
f(\ldots,\boldsymbol{\xi}_j(\mathbf{r}_1,\ldots ,\mathbf{r}_N),
\ldots).
\end{equation}
We do not distinguish between $f$ and $f\circ\xi$ in what follows,
but rather write both as $f$.

Using the chain rule, we may now rewrite the derivatives of $f$
with respect to $\mathbf{r}_j$ as
\begin{align}\label{xideriv}
\nabla_j f&=\sum_{l=1}^N\:\nabla_{\boldsymbol{\xi}_l}
f(\ldots ,\boldsymbol{\xi}_l ,\ldots )\nabla_j\xi_l \notag \\
&=\nabla_{\boldsymbol{\xi}_j}f(\ldots ,\boldsymbol{\xi}_j ,\ldots )
-\frac{1}{N}\sum_{l=1}^N\:\nabla_{\boldsymbol{\xi}_l}f(\ldots ,
\boldsymbol{\xi}_l ,\ldots ).
\vspace{3mm}
\end{align}

We may now express the old coordinates in terms of the new
ones as follows:
\begin{equation}\label{newvars2}
\mathbf{r}_h=\mathbf{R}+\frac{\boldsymbol{\rho}}{2},\quad
\mathbf{r}_j=\mathbf{R}-\frac{\boldsymbol{\rho}}{2}+
\boldsymbol{\xi}_j,
\end{equation}
and the derivatives with respect to the old coordinates are therefore
\begin{align}
\nabla_h&=\frac{1}{2}\nabla_{\mathbf{R}}+
\nabla_{\boldsymbol{\rho}}, \\
\nabla_j&=\frac{1}{2N}\nabla_{\mathbf{R}}-\frac{1}{N}
\nabla_{\boldsymbol{\rho}}+\nabla_{\boldsymbol{\xi}_j}
-\frac{1}{N}\sum_{l=1}^N\:\nabla_{\boldsymbol{\xi}_l}. \label{endone}
\vspace{3mm}
\end{align}
It is now possible to write the Hamiltonian \eqref{hamil3} in terms of 
the new variables. 
Using Eqs.~\eqref{xideriv}--\eqref{endone} we obtain
\begin{equation}
\hat{H}_0=\hat{H}_{exc}+\hat{H}_{\xi},
\end{equation}
where
\begin{multline}
\hat{H}_{exc}=\frac{1}{2m_h}\left\{\frac{1}{2i}\nabla_{\mathbf{R}}
+\frac{1}{i}\nabla_{\boldsymbol{\rho}}-\frac{1}{2}\left[\hat{\mathbf{z}}
\times\left(\mathbf{R}-\frac{\boldsymbol{\rho}}{2}\right)\right]\right\}^2 \\
+\frac{1}{2Nm_a}\left\{\frac{1}{2i}\nabla_{\mathbf{R}}
-\frac{1}{i}\nabla_{\boldsymbol{\rho}}+\frac{1}{2}\left[\hat{\mathbf{z}}
\times\left(\mathbf{R}+\frac{\boldsymbol{\rho}}{2}\right)\right]\right\}^2,
\end{multline}
and 
\vspace{3mm}
\begin{equation}\label{Hxi}
\hat{H}_{\xi}=\frac{1}{2m_a}\sum_{j=1}^N\left\{\frac{1}{i}
\nabla_{\boldsymbol{\xi}_j}
-\frac{1}{Ni}\sum_{l=1}^N\nabla_{\boldsymbol{\xi}_l}+\frac{1}{2N}
\left[\hat{\mathbf{z}}\times\boldsymbol{\xi}_j\right]\right\}^2.
\vspace{3mm}
\end{equation}
Note that Eq.~\eqref{Hxi} was obtained by applying constraint 
\eqref{constraint}.

The first part $\hat{H}_{exc}$ is similar to that for a standard
diamagnetic exciton \cite{Gorkov1968,Lerner1980}, with the electron mass
replaced with $N$ anyon masses. The eigenfunctions of this operator can be 
written straightforwardly as
\begin{equation}\label{wfpsi}
\Psi_{exc}(\mathbf{R} ,\boldsymbol{\rho})=\exp\left\{i\mathbf{k}
\cdot\mathbf{R}+i\hat{\mathbf{z}}\cdot\left[\mathbf{R}\times\boldsymbol{\rho}
\right]/2\right\}
\Phi(\boldsymbol{\rho}),
\end{equation}
where $\Phi(\boldsymbol{\rho})$ satisfies the equation
\vspace{3mm}
\begin{equation}\label{funcyone}
\left(\frac{1}{2m_h}\left\{\frac{1}{i}\nabla_{\boldsymbol{\rho}}
-\frac{\left[\hat{\mathbf{z}}
\times\left(\boldsymbol{\rho}-\mathbf{d}\right)\right]}{2}\right\}^2
+\frac{1}{2Nm_a}\left\{\frac{1}{i}\nabla_{\boldsymbol{\rho}}
+\frac{\left[\hat{\mathbf{z}}\times\left(\boldsymbol{\rho}-\mathbf{d}
\right)\right]}{2}\right\}^2\right)\Phi=E_{exc}\Phi.
\vspace{3mm}
\end{equation}
Here, $\mathbf{d}=\mathbf{k}\times\hat{\mathbf{z}}$ is the exciton 
dipole moment. Note that Eq.~\eqref{wfpsi} does not contain the 
mass-dependent phase factor that appears in the wavefunction in
\cite{Lerner1980}. This is due to our choice of the centre of
charge of the exciton for our coordinate $\mathbf{R}$, rather than
the centre of mass chosen in \cite{Gorkov1968,Lerner1980}.
Thus, the quantum number $\mathbf{k}$ entering Eq.~\eqref{wfpsi}
represents the momentum of the geometrical centre of the exciton.

The solutions of Eq.~\eqref{funcyone} are
\begin{equation}\label{eigfun}
\Phi_{nm}=|\boldsymbol{\rho}|^{|m|}L_n^{|m|}\left\{(\boldsymbol{\rho}
-\mathbf{d})^2/2\right\}\exp\left\{im\phi-(\boldsymbol{\rho}
-\mathbf{d})^2/4\right\},
\end{equation}
where $\phi$ is the azimuthal angle of the vector $(\boldsymbol{\rho}
-\mathbf{d})$ and $L_n^m(z)$ are the associated Laguerre polynomials.
The corresponding eigenvalues are 
\vspace{3mm}
\begin{equation}\label{eigval}
E_{exc}=\frac{1}{m_h}\left(n+\frac{|m|-m+1}{2}\right)
+\frac{1}{Nm_a}\left(n+\frac{|m|+m+1}{2}\right).
\vspace{3mm}
\end{equation}

Note that in Eqs.~\eqref{eigfun} and \eqref{eigval} we no longer
restricted ourselves to the lowest Landau level for the hole and
quasielectrons.
For the ground-state ($n,m=0$), Eq.~\eqref{wfpsi} reduces to
\begin{equation}
\Psi_{exc}(\mathbf{R} ,\boldsymbol{\rho})=\exp\left\{i\mathbf{k}\cdot
\mathbf{R}+i\hat{\mathbf{z}}\cdot\left[\mathbf{R}\times\boldsymbol{\rho}
\right]/2-(\boldsymbol{\rho}-
\mathbf{d})^2/4\right\},
\end{equation}
which depends on the quantum number $\mathbf{k}$ alone.

We now move to consider the term $\hat{H}_{\xi}$ in Eq.~\eqref{Hxi}, 
which may be divided into two parts:
\vspace{3mm}
\begin{equation}
\hat{H}_{1\xi}=\frac{1}{2m_{a}}\left(1-\frac{1}{N}\right)
\sum_{j=1}^N\left(\frac{1}{i}\nabla_{\boldsymbol{\xi}_j}+\frac{1}{2N}
\left[\hat{\mathbf{z}}\times{\boldsymbol{\xi}}_j\right]\right)^2,
\vspace{3mm}
\end{equation}
and
\vspace{3mm}
\begin{multline}
\hat{H}_{2\xi}=\frac{1}{2Nm_a}\sum_{j=1}^N\Bigg\{-
\nabla_{\boldsymbol{\xi}_j}^2+\frac{1}{Ni}\left[\hat{\mathbf{z}}\times
{\boldsymbol{\xi}}_j\right]\cdot\left(\nabla_{\boldsymbol{\xi}_j}-
\sum_{l=1}^N\nabla_{\boldsymbol{\xi}_l}\right)^{ } \\
+\frac{1}{N^2}|{\boldsymbol{\xi}_j}|^2
+2\nabla_{\boldsymbol{\xi}_j}\cdot
\left(\sum_{l=1}^N\nabla_{\boldsymbol{\xi}_l}\right)
-\frac{1}{N}\left(\sum_{l=1}^N\nabla_{\boldsymbol{\xi}_l}\right)^2
\Bigg\}.
\end{multline}
\vspace{1mm}

The eigenfunctions of the ground state of $\hat{H}_{1\xi}$ can then
be written in terms of complex coordinates as
\begin{equation}
\Phi_{\xi}=\prod_j\left(\bar{\zeta}_j\right)^{\beta_j}\exp\left\{
-|\zeta_j|^2/4N\right\},
\vspace{3mm}
\end{equation}
with the corresponding eigenvalues
\begin{equation}
E_{\xi}=\frac{1}{2m_{a}}\left(1-\frac{1}{N}\right).
\vspace{3mm}
\end{equation}
If we now apply constraint \eqref{constraint}, we find that the
function $\Phi_{\xi}$ is also an eigenfunction of $\hat{H}_{2\xi}$,
with a corresponding eigenvalue of zero, i.e.
\vspace{3mm}
\begin{equation}
\hat{H}_{2\xi}\Phi_{\xi}\propto\left(\sum_{j=1}^N
{\boldsymbol{\xi}}_j\right)^2\Phi_{\xi}=0.
\vspace{3mm}
\end{equation}
A calculation of the total energy of the ground state gives
\vspace{3mm}
\begin{equation}
E=E_{exc}+E_{\xi}=\frac{1}{2m_h}+\frac{1}{2m_{a}}=\frac{1}{2}(\omega_h
+N\omega_a),
\vspace{3mm}
\end{equation}
which is indeed the ground-state energy of the original $(N+1)$-particle
Hamiltonian in Eq.~\eqref{hamil3}.

The most general form for the ground-state eigenfunction of $\hat{H}_0$
is
\begin{multline}
\Psi\left(\mathbf{R},\boldsymbol{\rho},\{\bar{\zeta}_i\}\right)=
\exp\left\{i\mathbf{k}\cdot\mathbf{R}+i\hat{\mathbf{z}}\cdot
\left[\mathbf{R}\times\boldsymbol{\rho}
\right]/2-
(\boldsymbol{\rho}-\mathbf{d})^2/4\right\} \\
\times\quad F(\ldots ,\bar{\zeta}_i,\ldots)\prod_p\exp\left\{
-|\zeta_p|^2/4N\right\},
\end{multline}
where $F$ is a function of the complex conjugates $\bar{\zeta}_i$ alone.
We choose $F$ as follows to satisfy the interchange rules for anyons:
\vspace{3mm}
\begin{equation}
F(\ldots ,\bar{\zeta}_i,\ldots)=P_L(\ldots ,\bar{\zeta}_i,\ldots)
\prod_{j<l}(\bar{\zeta}_j-\bar{\zeta}_l)^\alpha,
\vspace{3mm}
\end{equation}
where $P_L$ is a symmetric polynomial of degree $L$ in the variables
$\bar{\zeta}_i$. Note that for $\mathbf{k}=0$ the problem has rotational
symmetry about the $z$-axis and the degree of the symmetric polynomial
$L$ is related to the exciton angular momentum 
\mbox{$[L_z=-L-N(N-1)\alpha/2]$}.

We are now in a position to express the anyon exciton basis functions in
the final form
\begin{multline}\label{finalbasis}
\Psi\left(\mathbf{R},\boldsymbol{\rho},\{\zeta_i\}\right)=
\exp\left\{i\mathbf{k}\cdot\mathbf{R}+i\hat{\mathbf{z}}\cdot
\left[\mathbf{R}\times\boldsymbol{\rho}\right]/2-
(\boldsymbol{\rho}-\mathbf{d})^2/4\right\} \\
\times\quad P_L(\ldots ,\bar{\zeta}_i,\ldots)
\prod_{j<l}(\bar{\zeta}_j-\bar{\zeta}_l)^\alpha\prod_p\exp\left\{
-|\zeta_p|^2/4N\right\}.
\end{multline}

\subsubsection{Symmetric polynomials}\label{SS:symmetric}

We now consider the precise structure of the symmetric polynomials
$P_L$ which appear in Eq.~\eqref{finalbasis}.
To determine the symmetric basis polynomials of order $L$, we apply the 
\emph{fundamental theorem of symmetric polynomials} \cite{Birkhoff1977},
which states that any symmetric polynomial in $N$ variables, 
$P(x_1,\ldots ,x_N)$, can be uniquely expressed in terms of the 
elementary symmetric polynomials
\vspace{3mm}
\begin{equation}\label{spolyno}
\sigma_1=\sum_i x_i,\;\sigma_2=\sum_{i<j}x_i x_j,\;
\sigma_3=\sum_{i<j<k}x_i x_j x_k,
\;\ldots\; ,\;\sigma_N=x_1 x_2\cdots x_N.
\vspace{3mm}
\end{equation}
It is apparent that all linearly-independent symmetric 
basis polynomials of a particular degree may be enumerated 
by considering the possible products of the elementary symmetric 
polynomials \eqref{spolyno}, as the total degree $L$ is the sum of the 
degrees of the constituent polynomials.
For example, the possible symmetric polynomials of degree four are 
constructed from $\sigma_1^4$, $\sigma_1^2\sigma_2$, $\sigma_1\sigma_3$, 
$\sigma_2^2$ and $\sigma_4$.

There are two further problems. The first is to determine the number
of possible products of elementary symmetric polynomials for
a total degree $L$. The second is to determine the structure of
these products.

To calculate the number of linearly-independent symmetric basis polynomials 
of degree $L$ which may be constructed from the elementary symmetric 
polynomials $\sigma_1,\ldots ,\sigma_N$, we use a result from the theory of 
partitions \cite{Slomson1991}.

The number of ways of partitioning a number $L$ into parts of
size $1,2,\ldots ,P$ is given by the coefficient of $x^L$ in the
expansion of
\vspace{3mm}
\begin{equation}\label{genprod}
\frac{1}{1-x}\cdot\frac{1}{1-x^2}\cdots\frac{1}{1-x^P}
=\prod_{k=1}^P\; \frac{1}{1-x^k},
\vspace{3mm}
\end{equation}
where each term in the product is known as a \emph{generating function}.
The number of ways of partitioning increases rapidly with $L$
and does not follow any pattern.
Furthermore, the different products of elementary symmetric
polynomials for a particular value of $L$ must be determined
by hand. An approximate formula for calculating the number of 
ways of partitioning a number does exist, the so-called
Hardy-Ramanujan formula \cite{Hardy1918}, but it is not applicable
to the present case as we have no polynomial of order one due
to constraint \eqref{constraint}. This constraint significantly
reduces the number of possible symmetric polynomials by removing
the first factor in the product \eqref{genprod}.
For example, for $N=3$ the number of polynomials of degree $L$
is equal to the integer part of $L/6+1$ for even $L$ and
the integer part of $(L-3)/6+1$ for odd $L$, which corresponds 
to the result obtained in \cite{Portnoi1996}.

\subsubsection{Discussion}

The key difference between the current general formulation and that
outlined in \cite{Portnoi1996} for the four-particle case $(N=3)$
is the replacement of anyon difference coordinates $\bar{z}_{jl}$
by the new coordinates $\bar{\zeta}_i$. 
The principal advantage of the difference coordinates was that they
simplified the calculation of inter-anyon repulsion matrix elements,
as well as the form of the statistical factor in the exciton wavefunction.
However, they had the disadvantage that the classification of symmetric
polynomials was more difficult, as it was necessary to introduce a
Vandermonde determinant for odd-$L$ polynomials. For example, in the
simplest case $L=3$, even though the number of anyon coordinates is the
same in both formulations, we have 
\mbox{$\sigma_3=(\bar{z}_{12}-\bar{z}_{23})(\bar{z}_{23}-\bar{z}_{31})
(\bar{z}_{31}-\bar{z}_{12})$}
in terms of difference coordinates, whereas we have a simple product 
$\sigma_3=\bar{\zeta}_1\bar{\zeta}_2\bar{\zeta}_3$ in terms of the new 
coordinates.
For a number of anyons greater than three this disadvantage is
crucial, as the classification of polynomials in terms of $\bar{z}_{jl}$
becomes too cumbersome and the number of constraints on these 
coordinates is greater than one.

\subsection{Reformulation of four-particle anyon exciton}

\subsubsection{Formulation}\label{SS:form4}

We now use the above results to formulate the problem
of a four-particle exciton in terms of the new coordinates 
$\bar{\zeta}_i$.
We consider an exciton consisting of a hole and three anyons
with charge $-e/3$, and make a boson approximation so that the
statistical factor $\alpha=0$. As mentioned in Section \ref{SS:results}, 
it is evident from Fig.~1 of \cite{Portnoi1996} that for large values of $h$ 
(required for the AEM to be valid) 
the statistical factor $\alpha$ becomes unimportant, and the results 
for $\alpha=0$ were very similar to those for $\alpha=\pm 1/3$.
We are therefore justified in approximating the anyons as bosons.

The basis functions for a four-particle anyon exciton in a boson
approximation are, from Eq.~\eqref{finalbasis}:
\begin{multline}\label{basis4}
\Psi\left(\mathbf{R},\boldsymbol{\rho},\bar{\zeta}_1,\bar{\zeta}_2,
\bar{\zeta}_3\right)=
\exp\left\{i\mathbf{k}\cdot\mathbf{R}+i\hat{\mathbf{z}}
\cdot\left[\mathbf{R}\times\boldsymbol{\rho}\right]/2-
(\boldsymbol{\rho}-\mathbf{d})^2/4\right\} \\
\times P_{L,M}(\bar{\zeta}_1,\bar{\zeta}_2,\bar{\zeta}_3)
\prod_{p=1}^3\exp\left\{-|\zeta_p|^2/12\right\},
\end{multline}
where the index $M$ enumerates different linearly-independent symmetric
polynomials of degree $L$.

As a result of constraint \eqref{constraint}, 
$\sigma_1=\bar{\zeta}_1+\bar{\zeta}_2+\bar{\zeta}_3=0$, and so
we need to consider only the elementary symmetric polynomials $\sigma_2$ 
and $\sigma_3$ when constructing the symmetric polynomial $P_{L,M}$ .
From Eq.~\eqref{genprod}, the number of possible ways of constructing a 
polynomial of degree $L$ is the coefficient of $x^L$ in the 
expansion
\vspace{3mm}
\begin{equation}
\frac{1}{1-x^2}\cdot\frac{1}{1-x^3}=1+x^2+x^3+x^4+x^5+2x^6+x^7+2x^8+\ldots
\;.
\vspace{3mm}
\end{equation}
Table~\ref{Table1} shows the possible ways of constructing the first 
twelve polynomials $P_L$.

\begin{table}
\caption{Possible ways of constructing a symmetric polynomial $P_L$
from the elementary symmetric polynomials $\sigma_2$ and $\sigma_3$.}
\vspace{0.5cm}
\begin{center}
\begin{tabular}{|c|c|l|} \hline
Order, $L$ & No. of polynomials & Structure \\
\hline
0 & 1 & 1 \\
\hline
1 & 0 & - \\
\hline
2 & 1 & $\sigma_2$ \\
\hline
3 & 1 & $\sigma_3$ \\
\hline
4 & 1 & $\sigma_2^2$ \\
\hline
5 & 1 & $\sigma_2\sigma_3$ \\
\hline
6 & 2 & $\sigma_2^3$, $\sigma_3^2$ \\
\hline
7 & 1 & $\sigma_2^2\sigma_3$ \\
\hline
8 & 2 & $\sigma_2^4$, $\sigma_2\sigma_3^2$ \\
\hline
9 & 2 & $\sigma_2^3\sigma_3$, $\sigma_3^3$ \\
\hline
10 & 2 & $\sigma_2^5$, $\sigma_2^2\sigma_3^2$ \\
\hline
11 & 2 & $\sigma_2^4\sigma_3$, $\sigma_2\sigma_3^3$ \\
\hline
12 & 3 & $\sigma_2^6$, $\sigma_2^3\sigma_3^2$, $\sigma_3^4$ \\
\hline
%13 & 2 & $\sigma_2^5\sigma_3$, $\sigma_2^2\sigma_3^3$ \\
%\hline
%14 & 3 & $\sigma_2^7$, $\sigma_2^4\sigma_3^2$, $\sigma_2\sigma_3^4$ \\
%\hline
%15 & 3 & $\sigma_2^6\sigma_3$, $\sigma_2^3\sigma_3^3$, $\sigma_3^5$ \\
%\hline
\end{tabular}
\label{Table1}
\end{center}
\vspace{1cm}
\end{table}

As in Section \ref{SS:formulation}, a general exciton wavefunction for 
given $\mathbf{k}$ may be expanded in terms of the complete set of basis 
functions \eqref{basis4} as follows:
\begin{equation}
\Phi=\sum_i\chi_i\Psi_i.
\end{equation}
Basis functions with different values of $\mathbf{k}$ are orthogonal. 
We shall later show that functions with different $L$ are also orthogonal. 
However, basis functions with the same value of $L$ but different $M$ 
are not necessarily orthogonal, so that their scalar products 
$\langle L,M|L,M^{\prime}\rangle$ will be non-zero.
We therefore write the Schr\"{o}dinger equation in matrix form as
\begin{equation}\label{mateqn4}
\hat{H}\boldsymbol{\chi}=\varepsilon\hat{B}\boldsymbol{\chi},
\end{equation}
where $\hat{B}$ is the block diagonal matrix of scalar products 
(the overlap matrix) and $\varepsilon$ is an energy eigenvalue.
The size of each block in $\hat{B}$ depends on the number of different
wavefunctions with given $L$. 
In the case of a four-particle exciton (see Table~\ref{Table1}), 
the block corresponding to $L=12$ will be 
of size $3\times 3$. 
Both the overlap matrix $\hat{B}$ and the Hamiltonian matrix $\hat{H}$ 
are diagonal in $\mathbf{k}$.
Note that for $\mathbf{k}=0$ the matrix $\hat{H}$ takes the same
block diagonal form as $\hat{B}$ (as will be shown later), and as a result 
the problem becomes exactly solvable. 

We shall now proceed to evaluate the matrix elements in Eq.~\eqref{mateqn4}.
As $\hat{H}$ and $\hat{B}$ are diagonal in $\mathbf{k}$, in what
follows we consider only the matrix elements diagonal in $\mathbf{k}$.
Since all terms in the polynomial $P_{L,M}$ are of the form of a product
of the coordinates ${\bar{\zeta}}_j$, we use the following functions
in monomials in the matrix element calculations:
\begin{multline}\label{monoms}
\Psi_{\{n\},\mathbf{k}}\left(\mathbf{R},\boldsymbol{\rho},\{\bar{\zeta_i}\}
\right)=C\exp\left\{i\mathbf{k}\cdot\mathbf{R}
+i\hat{\mathbf{z}}\cdot\left[\mathbf{R}\times\boldsymbol{\rho}
\right]/2-(\boldsymbol{\rho}-\mathbf{d})^2/4\right\} \\
\times {\bar{\zeta}}_1^{n_1}{\bar{\zeta}}_2^{n_2}
{\bar{\zeta}}_3^{n_3}\prod_{j=1}^3\exp\left\{-|\zeta_j|^2/12\right\},
\end{multline}
where $\{n\}$ denotes the set of quantum numbers $n_1$, $n_2$ and $n_3$,
and the constant $C$ will be defined in the next Section.
The basis functions in terms of symmetric polynomials of degree $L$
can be obtained as a linear combination of the monomial functions 
\eqref{monoms}, and $n_1+n_2+n_3=L$.

We also take into account constraint \eqref{constraint} via the
following transformation:
\begin{equation}
\delta\left(\boldsymbol{\xi}_1+\boldsymbol{\xi}_2+\boldsymbol{\xi}_3\right)
=\int\frac{d\mathbf{f}}{(2\pi)^2}\,\exp\left\{i\mathbf{f}\cdot
\left(\boldsymbol{\xi}_1+\boldsymbol{\xi}_2+\boldsymbol{\xi}_3\right)\right\}.
\end{equation}

\subsubsection{Overlap matrix elements}\label{SS:over4}

To calculate the overlap matrix $\hat{B}$ we consider
the scalar product of two monomial functions $\Psi_{\{n\},\mathbf{k}}$ 
and $\Psi_{\{n'\},\mathbf{k}}$:
\begin{multline}
\langle\left\{n\right\}|\left\{n'\right\}\rangle=\int d\mathbf{R}
\int d\boldsymbol{\rho}\int\frac{d\mathbf{f}}{(2\pi)^2}
\int d\boldsymbol{\xi}_1 d\boldsymbol{\xi}_2 d\boldsymbol{\xi}_3 \\
\times\bar{\Psi}_{{\{n\},\mathbf{k}}}\Psi_{\{n'\},\mathbf{k}}
\exp\left\{i\mathbf{f}\cdot
\left(\boldsymbol{\xi}_1+\boldsymbol{\xi}_2+\boldsymbol{\xi}_3\right)\right\}.
\end{multline}
Integration over $\mathbf{R}$ and Gaussian integration over 
$\boldsymbol{\rho}$ give a factor of $2\pi A$, where $A$ is the area of
the sample, and this yields
\begin{equation}
\langle\left\{n\right\}|\left\{n'\right\}\rangle=
C^2(2\pi A)\int\frac{d\mathbf{f}}
{(2\pi)^2}\prod_{j=1}^3 M_{n_j n_j^\prime}(\mathbf{f}),
\end{equation}
where
\begin{equation}\label{mnm}
M_{m m^\prime}(\mathbf{f})=\int d\boldsymbol{\xi}\:
\zeta^m{\bar{\zeta}}^{m'}e^{-\xi^2/6+i\mathbf{f}\cdot
\boldsymbol{\xi}}.
\vspace{3mm}
\end{equation}
We now express the complex variable $\zeta$ in polar form as 
$\zeta=\xi e^{i\phi_{\boldsymbol{\xi}}}$,
and let $\phi$ be the angle between $\mathbf{f}$ and $\boldsymbol{\xi}$, 
i.e.
\begin{equation}
\phi=\phi_{\boldsymbol{\xi}}-\phi_\mathbf{f}.
\end{equation}
Eq.~\eqref{mnm} then becomes
\begin{multline}\label{mnm2}
M_{m m^\prime}(\mathbf{f})=e^{i(m-m')\phi_{\mathbf{f}}}
\int_0^{2\pi}d\phi\int_0^{\infty}d\xi\:\zeta^{1+m+m'} \\
\times\exp\left\{i(m-m')\phi+if\xi\cos\phi-\xi^2/6\right\}.
\end{multline}
The integration over $\phi$ can be carried out using Bessel's integral 
\cite{Arfken1985}:
\vspace{3mm}
\begin{equation}
\int_0^{2\pi}d\phi\: e^{\pm i(m-m')\phi+if\xi\cos\phi}
=2\pi i^{|m-m'|}J_{|m-m'|}(f\xi),
\vspace{3mm}
\end{equation}
where $J_m(z)$ is a Bessel function of order $m$.
This simplifies Eq.~\eqref{mnm2} to
\begin{equation}
M_{m m^\prime}(\mathbf{f})=2\pi i^{|m-m'|}e^{i(m-m')
\phi_{\mathbf{f}}}\mathcal{M}_{mm'}(f),
\end{equation}
where we define
\begin{equation}\label{curlym4}
\mathcal{M}_{mm'}(f)=\int_0^{\infty}d\xi\:\zeta^{1+m+m'}
e^{-\xi^2/6}J_{|m-m'|}(f\xi).
\end{equation}
The integral in Eq.~\eqref{curlym4} may be expressed \cite{Gradshteyn2000} 
in terms of a confluent hypergeometric function $\Phi(\beta,\gamma;z)$ as
\begin{multline}\label{chyp}
\mathcal{M}_{mm'}(f)=\frac{2^{(m+m')/2}3^{(m+m')/2+1}
\Gamma\left(\max\{m,m'\}+1\right)}{|m-m'|!}t^{|m-m'|/2}
\\ 
\times\Phi\left(\max\{m,m'\}+1,|m-m'|+1;-t\right),
\end{multline}
where $t=3f^2/2$ and $\max\{m,m'\}$ is the largest of the integers
$m$ and $m'$. Applying the Kummer transformation
\begin{equation}
\Phi(\beta,\gamma;-t)=e^{-t}\Phi(\gamma-\beta,\gamma;t),
\end{equation} 
simplifies Eq.~\eqref{chyp} further.
The fact that $(\gamma-\beta)$ is always a non-positive integer:
\begin{equation}
\gamma-\beta=|m-m'|-\max\{m,m'\},
\end{equation}
means that $\Phi(\gamma-\beta,\gamma;t)$ reduces to a polynomial,
and $\Phi(\beta,\gamma;-t)$ is then just a polynomial in $t$ multiplied 
by $e^{-t}$.

Integrating over $\phi_{\mathbf{f}}$ yields the final form of the overlap 
matrix elements
\vspace{3mm}
\begin{equation}\label{monmat}
\langle\left\{n\right\}|\left\{n'\right\}\rangle=C^2A(2\pi)^3\:\delta_{LL'}
\int_0^{\infty}df\: f\prod_{j=1}^3 i^{|n_j-n'_j|}\mathcal{M}_{n_j n'_j}(f),
\vspace{3mm}
\end{equation}
where $L=n_1+n_2+n_3$.
We define the constant $C$ so that for $L=L^{\prime}=0$
the matrix element $\langle\left\{n\right\}|\left\{n'\right\}\rangle=
\langle 0|0\rangle=1$, which gives
\begin{equation}\label{csq4}
C^2=\frac{1}{3(2\pi)^3A}.
\end{equation}

It should be noted that matrix elements in terms of symmetric polynomials 
can be constructed as a linear combination of the monomial matrix elements 
\eqref{monmat}.

\subsubsection{Anyon-anyon interaction}

The interaction between anyons is of the form
\begin{equation}\label{coul4}
\hat{V}_{aa}=\frac{1}{9}\left\{\frac{1}{\left|{\boldsymbol{\xi}}_1
-{\boldsymbol{\xi}}_2\right|}+\frac{1}{\left|{\boldsymbol{\xi}}_2
-{\boldsymbol{\xi}}_3\right|}+\frac{1}{\left|{\boldsymbol{\xi}}_3
-{\boldsymbol{\xi}}_1\right|}\right\}.
\end{equation}
Only matrix elements for the first term $\hat{V}_{12}$ will be calculated
here as the others follow by analogy.
Taking the Fourier transform of Eq.~\eqref{coul4} yields
\begin{equation}
\hat{V}_{12}\left({\boldsymbol{\xi}}_1,{\boldsymbol{\xi}}_2\right)=
\frac{1}{9}\int\frac{d\mathbf{q}}{2\pi q}
\exp\left\{i\mathbf{q}\cdot\left({\boldsymbol{\xi}}_1
-{\boldsymbol{\xi}}_2\right)\right\},
\end{equation}
which gives a matrix element
\begin{multline}
\langle\left\{n\right\}|\hat{V}_{12}|\left\{n'\right\}\rangle=
\frac{1}{9}\int d\mathbf{R}
\int d\boldsymbol{\rho}\int\frac{d\mathbf{f}}{(2\pi)^2}
\int\frac{d\mathbf{q}}{2\pi q}
\int d\boldsymbol{\xi}_1 d\boldsymbol{\xi}_2 d\boldsymbol{\xi}_3 \\ 
\times\bar{\Psi}_{{\{n\},\mathbf{k}}}\Psi_{\{n'\},\mathbf{k}}
\exp\left\{i\mathbf{f}\cdot
\left(\boldsymbol{\xi}_1+\boldsymbol{\xi}_2+\boldsymbol{\xi}_3\right)\right\}
\exp\left\{i\mathbf{q}\cdot\left({\boldsymbol{\xi}}_1
-{\boldsymbol{\xi}}_2\right)\right\}.
\end{multline}
Following the method of Section \ref{SS:over4} gives
\vspace{3mm}
\begin{align}\label{phiphi}
\langle\left\{n\right\}|\hat{V}_{12}|\left\{n'\right\}\rangle &=
\frac{C^2(2\pi A)}{9}\prod_{j=1}^3 i^{|n_j-n'_j|}
\int d\mathbf{f}\int d\mathbf{q} \notag \\
&\times\exp\left\{i\phi_{+}(n_1-n_1^{\prime})
+i\phi_{-}(n_2-n_2^{\prime})+i\phi_{\mathbf{f}}(n_3-n_3^{\prime})\right\}
\notag \\
&\times\mathcal{M}_{n_1 n'_1}(f_{+})\mathcal{M}_{n_2 n'_2}(f_{-})
\mathcal{M}_{n_3 n'_3}(f),
\end{align}
where $\mathcal{M}_{mm'}$ are as defined in Eq.~\eqref{curlym4}, 
$\mathbf{f}_{\pm}=\mathbf{f}\pm\mathbf{q}$, and $\phi_\pm$ are
the phases of $\mathbf{f}_{\pm}$.

We now eliminate $\phi_\mathbf{f}$ by the change of variables
$\phi=\phi_{\mathbf{f}}-\phi_{\mathbf{q}}$ and
$\psi_{\pm}=\phi_{\pm}-\phi_{\mathbf{f}}$.
Substituting for $\phi_{\mathbf{f}}$, $\phi_{+}$ and $\phi_{-}$ in
Eq.~\eqref{phiphi} and integrating over $\phi_{\mathbf{q}}$,
we obtain
\vspace{3mm}
\begin{align}\label{aafinal4}
\langle\left\{n\right\}|\hat{V}_{12}|\left\{n'\right\}\rangle &=
\frac{C^2A(2\pi)^2}{9}\:\delta_{LL'}\prod_{j=1}^3 i^{|n_j-n'_j|}
\int_0^{\infty}df\: f
\int_0^{\infty}dq \notag \\
&\times\int_0^{2\pi}d\phi\:\exp\left\{i\phi(L-L^{\prime})\right\}
\exp\left\{i\psi_{+}(n_1-n_1^{\prime})+i\psi_{-}(n_2-n_2^{\prime})
\right\} \notag \\
&\times\mathcal{M}_{n_1 n'_1}(f_{+})\mathcal{M}_{n_2 n'_2}(f_{-})
\mathcal{M}_{n_3 n'_3}(f),
\end{align}
where $\psi_{\pm}$ and $f_{\pm}$ can be expressed in terms of the 
variables of integration, $f$, $q$ and $\phi$, as
\begin{equation}
e^{i\psi_{\pm}}=\left(f\pm qe^{-i\phi}\right)/f_{\pm},\qquad
f_{\pm}^2=f^2+q^2\pm 2fq\cos\phi.
\vspace{3mm}
\end{equation}
It is evident from Eq.~\eqref{aafinal4} that the anyon-anyon interaction
matrix has the same block diagonal structure as the overlap matrix.
The integrand in Eq.~\eqref{aafinal4} can be further reduced to a
product of a polynomial in $q$, $f$ and $e^{\pm i\phi}$, and an
exponential factor \mbox{$\exp(-9f^2/2-3q^2)$}.
This integral can therefore be evaluated analytically for any
$\{n\}$ and $\{n'\}$.
For the case of $L=L^{\prime}=0$, using Eq.~\eqref{csq4} for $C^2$,
the anyon-anyon interaction matrix element reduces to
\begin{equation}\label{aa4}
\langle 0|\hat{V}_{aa}|0\rangle=\frac{1}{6}\sqrt{\frac{\pi}{3}}.
\end{equation}

\subsubsection{Anyon-hole interaction}

The anyon-hole interaction is of the form
\vspace{3mm}
\begin{equation}\label{ahcoul}
\hat{V}_{ah}=-\frac{1}{3}\left\{\frac{1}{\left|\mathbf{r}_{1h}\right|}
+\frac{1}{\left|\mathbf{r}_{2h}\right|}+\frac{1}{\left|\mathbf{r}_{3h}
\right|}\right\},
\vspace{3mm}
\end{equation}
where we define
\begin{equation}
\mathbf{r}_{jh}=\boldsymbol{\xi}_j-\boldsymbol{\rho}+h\hat{\mathbf{z}}.
\vspace{3mm}
\end{equation}
Considering only $\hat{V}_{1h}$, we take the Fourier transform:
\vspace{3mm}
\begin{equation}
\hat{V}_{1h}\left(\boldsymbol{\rho},{\boldsymbol{\xi}}_1\right)=
-\frac{1}{3}\int\frac{d\mathbf{q}}{(2\pi)^2}V_{ah}(q)
\exp\left\{i\mathbf{q}\cdot\left({\boldsymbol{\xi}}_1-\boldsymbol{\rho}
\right)\right\},
\vspace{3mm}
\end{equation}
where
\begin{equation}
V_{ah}(q)=\frac{2\pi}{q}e^{-qh}.
\vspace{3mm}
\end{equation}
The matrix elements are then
\begin{multline}
\langle\left\{n\right\}|\hat{V}_{1h}|\left\{n'\right\}\rangle=
-\frac{1}{3}\int d\mathbf{R}
\int d\boldsymbol{\rho}\int\frac{d\mathbf{f}}{(2\pi)^2}
\int\frac{d\mathbf{q}}{(2\pi)^2}V_{ah}(q)
\int d\boldsymbol{\xi}_1 d\boldsymbol{\xi}_2 d\boldsymbol{\xi}_3 \\
\times\bar{\Psi}_{{\{n\},\mathbf{k}}}\Psi_{\{n'\},\mathbf{k}}
\exp\left\{i\mathbf{f}\cdot
\left(\boldsymbol{\xi}_1+\boldsymbol{\xi}_2+\boldsymbol{\xi}_3\right)\right\}
\exp\left\{i\mathbf{q}\cdot\left(
{\boldsymbol{\xi}}_1-\boldsymbol{\rho}\right)\right\}.
\end{multline}
Integration over $\mathbf{R}$ is straightforward and gives a factor
of $A$. However, the integration over $\boldsymbol{\rho}$ is more
complicated:
\vspace{3mm}
\begin{equation}
\int d\boldsymbol{\rho}\;\exp\left\{-(\boldsymbol{\rho}-\mathbf{d})^2/2
-i\mathbf{q}\cdot\boldsymbol{\rho}\right\}=2\pi\exp(-q^2/2-i\mathbf{d}
\cdot\mathbf{q}).
\vspace{3mm}
\end{equation}
Then, following the procedure in Section \ref{SS:over4} once again,
we obtain
\vspace{3mm}
\begin{align}
\langle\left\{n\right\}|\hat{V}_{1h}|\left\{n'\right\}\rangle &=
-\frac{C^2A}{3}\prod_{j=1}^3 i^{|n_j-n'_j|}\int d\mathbf{f}
\int d\mathbf{q}\;V_{ah}(q)e^{-q^2/2-i\mathbf{d}\cdot\mathbf{q}}
\\ \notag 
&\times\exp\left\{i\phi_{+}(n_1-n_1^{\prime})
+i\phi_{\mathbf{f}}[(L-L')-(n_1-n_1^{\prime})]\right\} \\
&\times\mathcal{M}_{n_1 n'_1}(f_{+})\mathcal{M}_{n_2 n'_2}(f)
\mathcal{M}_{n_3 n'_3}(f), \notag
\end{align}
where $\mathcal{M}_{mm'}$ are as defined in Eq.~\eqref{curlym4}, 
$\mathbf{f}_{+}=\mathbf{f}+\mathbf{q}$, and $\phi_{+}$
is the phase of $\mathbf{f}_{+}$.
As before, $L=n_1+n_2+n_3$.

We now eliminate $\phi_{\mathbf{f}}$ by substituting 
$\phi=\phi_{\mathbf{f}}-\phi_{\mathbf{q}}$ and
$\psi=\phi_{+}-\phi_{\mathbf{f}}$. Choosing the $x$-axis along
$\mathbf{d}$, the integration over $\phi_{\mathbf{q}}$ is then
\begin{equation}
\int_0^{2\pi}d\phi_{\mathbf{q}}\;\exp\left\{i\phi_{\mathbf{q}}
(L-L^{\prime})-idq\cos\phi_{\mathbf{q}}\right\}
=2\pi{(-i)}^{|L-L^{\prime}|}J_{|L-L^{\prime}|}(dq),
\vspace{3mm}
\end{equation}
which gives for the matrix elements:
\begin{align}\label{ahole4}
\langle\left\{n\right\}|\hat{V}_{1h}|\left\{n'\right\}\rangle &=
-\frac{C^2A(2\pi)^2}{3}i^{|L-L'|}\prod_{j=1}^3 i^{|n_j-n'_j|}
\int_0^{\infty}dq\;e^{-q^2/2-qh}J_{|L-L'|}(dq) \notag \\ 
&\times\int_0^{\infty}df\: f\int_0^{2\pi}d\phi\:\exp\left\{i\phi(L-L^{\prime})
+i\psi(n_1-n_1^{\prime})\right\} \notag \\
&\times\mathcal{M}_{n_1 n'_1}(f_{+})\mathcal{M}_{n_2 n'_2}(f)
\mathcal{M}_{n_3 n'_3}(f),
\end{align}
where the new variables $\psi$ and $f_{+}$ can be expressed in terms of 
the variables of integration as
\begin{equation}
e^{i\psi}=\left(f+qe^{-i\phi}\right)/f_{+},\qquad
f_{+}^2=f^2+q^2+2fq\cos\phi.
\end{equation}
For $d=0$, the Bessel function $J_{|L-L'|}(dq)$ entering the integrand
of Eq.~\eqref{ahole4} is non-zero only if \mbox{$L=L^{\prime}$}.
This indicates that for the case of zero in-plane momentum \mbox{$(k=d=0)$},
the anyon-hole interaction matrix has the same block diagonal
structure as the anyon-anyon and overlap matrices.
Therefore, states with different angular momentum $L_z$ decouple.

Eq.~\eqref{ahole4} may be simplified further using the explicit form
of $\mathcal{M}_{mm'}(f)$ in Eq.~\eqref{chyp} and subsequent results.
This yields the following expression for the anyon-hole interaction 
matrix elements for basis functions in terms of symmetric polynomials:
\begin{equation}\label{Qpoly4}
\langle L,M|\hat{V}_{ah}|L',M'\rangle =
-\int_0^{\infty}\exp(-3q^2/2-qh)J_{|L-L'|}(dq)Q_{LM,L'M'}(q)\;dq,
\end{equation}
where $Q_{LM,L'M'}(q)$ is a polynomial in $q$ (the lowest order
polynomial is \mbox{$Q_{00,00}=1$}). 
This expression was previously derived in \cite{Portnoi1996}.
Note that for zero 2DEG-hole separation $(h=0)$, Eq.~\eqref{Qpoly4} further 
reduces to an expression in terms of elementary functions.  
For the case of $k=d=0$ and $L=L^{\prime}=0$, the anyon-hole interaction 
matrix element is then
\begin{equation}\label{ahah4}
\langle 0|\hat{V}_{ah}|0\rangle=-\sqrt{\frac{\pi}{6}}.
\end{equation}

\subsection{Six-particle anyon exciton}

\subsubsection{Formulation}

We now introduce an anyon exciton consisting of a hole and five 
anyons with charge $-e/5$. 
We also make a boson approximation so that the statistical factor 
$\alpha=0$. Our justification for this step is as follows.
It was shown in \cite{Portnoi1996} that for large values 
of $h$ (which is required for the AEM to be valid)
the statistical factor $\alpha$ becomes unimportant, and the
results for $\alpha=0$ were very similar to those for $\alpha=\pm 1/3$.
We would therefore also expect this to be true for $\alpha=\pm 1/N$, 
as this is even closer to zero. From now on we consider a boson
approximation $(\alpha=0)$ for anyon statistics.

The explicit form of the exciton basis functions is
\begin{multline}\label{basisfuncs}
\Psi_{L,M,\mathbf{k}}\left(\mathbf{R},\boldsymbol{\rho},\{\bar{\zeta_i}\}
\right)=\exp\left\{i\mathbf{k}\cdot\mathbf{R}
+i\hat{\mathbf{z}}\cdot\left[\mathbf{R}\times\boldsymbol{\rho}
\right]/2-(\boldsymbol{\rho}-\mathbf{d})^2/4\right\} \\
\times P_{L,M}({\bar{\zeta}}_1,\ldots,{\bar{\zeta}}_5)\prod_{j=1}^5
\exp\left\{-|\zeta_j|^2/20\right\},
\end{multline}
where $M$ enumerates all linearly-independent symmetric polynomials
of degree $L$.

To construct the symmetric polynomials $P_{L,M}$ we need to consider only
the elementary symmetric polynomials $\sigma_2$, $\sigma_3$, $\sigma_4$
and $\sigma_5$. Note that $\sigma_1=0$ because of constraint
\eqref{constraint}. From Eq.~\eqref{genprod}, we find that the 
number of possible ways of constructing a polynomial of degree $L$ 
is therefore the coefficient of $x^L$ in the expansion
\vspace{3mm}
\begin{multline}
\frac{1}{1-x^2}\cdot\frac{1}{1-x^3}\cdot\frac{1}{1-x^4}\cdot\frac{1}{1-x^5}
=1+x^2+x^3+2x^4+2x^5 \\
+3x^6+3x^7+5x^8+5x^9+7x^{10}+\ldots\;.
\vspace{3mm}
\end{multline}

The possible ways of constructing the first twelve polynomials are shown 
explicitly in Table~\ref{Table2}.
It is evident that the number and complexity of the polynomials
increases rapidly with the degree $L$.
\begin{table}
\caption{Possible ways of constructing a symmetric polynomial $P_L$
from the elementary symmetric polynomials $\sigma_2$, $\sigma_3$, 
$\sigma_4$ and $\sigma_5$.}
\vspace{0.5cm}
\begin{center}
\begin{tabular}{|c|c|l|} \hline
Order, $L$ & No. of polynomials & Structure \\
\hline
0 & 1 & 1 \\
\hline
1 & 0 & - \\
\hline
2 & 1 & $\sigma_2$ \\
\hline
3 & 1 & $\sigma_3$ \\
\hline
4 & 2 & $\sigma_2^2$, $\sigma_4$ \\
\hline
5 & 2 & $\sigma_2\sigma_3$, $\sigma_5$ \\
\hline
6 & 3 & $\sigma_2^3$, $\sigma_2\sigma_4$, $\sigma_3^2$ \\
\hline
7 & 3 & $\sigma_2^2\sigma_3$, $\sigma_2\sigma_5$, $\sigma_3\sigma_4$ \\
\hline
8 & 5 & $\sigma_2^4$, $\sigma_2^2\sigma_4$, $\sigma_2\sigma_3^2$,
$\sigma_3\sigma_5$, $\sigma_4^2$ \\
\hline
9 & 5 & $\sigma_2^3\sigma_3$, $\sigma_2^2\sigma_5$,
$\sigma_2\sigma_3\sigma_4$, $\sigma_3^3$, $\sigma_4\sigma_5$ \\
\hline
10 & 7 & $\sigma_2^5$, $\sigma_2^3\sigma_4$, $\sigma_2^2\sigma_3^2$,
$\sigma_2\sigma_3\sigma_5$, $\sigma_2\sigma_4^2$, $\sigma_3^2\sigma_4$, 
$\sigma_5^2$ \\
\hline
11 & 7 & $\sigma_2^4\sigma_3$, $\sigma_2^3\sigma_5$, 
$\sigma_2^2\sigma_3\sigma_4$, $\sigma_2\sigma_3^3$,
$\sigma_2\sigma_4\sigma_5$, \\
\hspace{2cm} & \hspace{4cm} & $\sigma_3^2\sigma_5$, $\sigma_3\sigma_4^2$ \\
\hline
12 & 10 & $\sigma_2^6$, $\sigma_2^4\sigma_4$, $\sigma_2^3\sigma_3^2$, 
$\sigma_2^2\sigma_3\sigma_5$, $\sigma_2^2\sigma_4^2$,
$\sigma_2\sigma_3^2\sigma_4$, \\
\hspace{2cm} & \hspace{4cm} & $\sigma_2\sigma_5^2$, $\sigma_3^4$,
$\sigma_3\sigma_4\sigma_5$, $\sigma_4^3$ \\
\hline
\end{tabular}
\label{Table2}
\end{center}
\vspace{1cm}
\end{table}

Following the method in Section \ref{SS:formulation} once again, we 
expand a general exciton wavefunction for given $\mathbf{k}$ 
in terms of the complete set of basis functions \eqref{basisfuncs}:
\begin{equation}
\Phi=\sum_i\chi_i\Psi_i.
\end{equation}
As mentioned in Section \ref{SS:form4}, basis functions with different 
values of $\mathbf{k}$ and $L$ are orthogonal, although basis functions 
with the same value of $L$ but different $M$ are not necessarily orthogonal, 
so we introduce the overlap matrix $\hat{B}$ of scalar products into the 
Schr\"{o}dinger equation as follows:
\begin{equation}\label{mateqn}
\hat{H}\boldsymbol{\chi}=\varepsilon\hat{B}\boldsymbol{\chi}.
\end{equation}
Note that $\hat{B}$ is block diagonal, and in the case of a six-particle 
exciton (see Table~\ref{Table2}), the block corresponding to $L=12$
for example will be of size $10\times 10$. 
As the Hamiltonian matrix $\hat{H}$ and the overlap matrix $\hat{B}$ are 
diagonal in $\mathbf{k}$, we only need to consider the matrix elements 
in Eq.~\eqref{mateqn} that are diagonal in $\mathbf{k}$.
We use the following functions in monomials in the matrix element 
calculations, from which the basis functions in terms of symmetric
polynomials can be obtained as a linear combination:
\begin{multline}\label{mono}
\Psi_{\{n\},\mathbf{k}}\left(\mathbf{R},\boldsymbol{\rho},\{\bar{\zeta_i}\}
\right)=C\exp\left\{i\mathbf{k}\cdot\mathbf{R}
+i\hat{\mathbf{z}}\cdot\left[\mathbf{R}\times\boldsymbol{\rho}
\right]/2-(\boldsymbol{\rho}-\mathbf{d})^2/4\right\} \\
\times {\bar{\zeta}}_1^{n_1}{\bar{\zeta}}_2^{n_2}
{\bar{\zeta}}_3^{n_3}{\bar{\zeta}}_4^{n_4}{\bar{\zeta}}_5^{n_5}
\prod_{j=1}^5\exp\left\{-|\zeta_j|^2/20\right\},
\end{multline}
where $\{n\}$ denotes the set of quantum numbers $n_1$ to $n_5$, for which
it should be noted that \mbox{$n_1+n_2+n_3+n_4+n_5=L$}, and the constant 
$C$ will be defined in Section \ref{SS:overlap}.

Constraint \eqref{constraint} is taken into account via the
following transformation:
\vspace{3mm}
\begin{equation}
\delta\left(\boldsymbol{\xi}_1+\boldsymbol{\xi}_2+\boldsymbol{\xi}_3
+\boldsymbol{\xi}_4+\boldsymbol{\xi}_5\right)
=\int\frac{d\mathbf{f}}{(2\pi)^2}\,\exp\left\{i\mathbf{f}\cdot
\left(\boldsymbol{\xi}_1+\boldsymbol{\xi}_2+\boldsymbol{\xi}_3
+\boldsymbol{\xi}_4+\boldsymbol{\xi}_5\right)\right\}.
\vspace{3mm}
\end{equation}

\subsubsection{Overlap matrix elements}\label{SS:overlap}

To calculate the overlap matrix $\hat{B}$, let us consider
the scalar product of two monomial functions:
\begin{multline}
\langle\left\{n\right\}|\left\{n'\right\}\rangle=\int d\mathbf{R}
\int d\boldsymbol{\rho}\int\frac{d\mathbf{f}}{(2\pi)^2}
\int d\boldsymbol{\xi}_1 d\boldsymbol{\xi}_2 d\boldsymbol{\xi}_3 
d\boldsymbol{\xi}_4 d\boldsymbol{\xi}_5 \\
\times\bar{\Psi}_{{\{n\},\mathbf{k}}}\Psi_{\{n'\},\mathbf{k}}
\exp\left\{i\mathbf{f}\cdot
\left(\boldsymbol{\xi}_1+\boldsymbol{\xi}_2+\boldsymbol{\xi}_3
+\boldsymbol{\xi}_4+\boldsymbol{\xi}_5\right)\right\}.
\end{multline}
The integration over $\mathbf{R}$ and $\boldsymbol{\rho}$ gives
a factor of $2\pi A$, 
and this leaves the following:
\vspace{3mm}
\begin{equation}
\langle\left\{n\right\}|\left\{n'\right\}\rangle=C^2(2\pi A)
\int\frac{d\mathbf{f}}{(2\pi)^2}\;\prod_{j=1}^5 M_{n_j n'_j},
\vspace{3mm}
\end{equation}
where
\begin{equation}
M_{m m^\prime}(\mathbf{f})=\int d\boldsymbol{\xi}\:
\zeta^m{\bar{\zeta}}^{m'}e^{-\xi^2/10+i\mathbf{f}\cdot
\boldsymbol{\xi}}.
\vspace{3mm}
\end{equation}
If we now make the substitution $\zeta=\xi e^{i\phi_{\boldsymbol{\xi}}}$,
and let $\phi$ be the angle between $\mathbf{f}$ and $\boldsymbol{\xi}$, 
i.e.
\begin{equation}
\phi=\phi_{\boldsymbol{\xi}}-\phi_\mathbf{f},
\end{equation}
where $\phi_\mathbf{f}$ is the phase of $\mathbf{f}$, we obtain
\begin{multline}
M_{m m^\prime}(\mathbf{f})=\exp\left\{i(m-m')\phi_{\mathbf{f}}\right\}
\int_0^{2\pi}d\phi\int_0^{\infty}d\xi\:\xi^{1+m+m'} \\ 
\times\exp\left\{i(m-m')\phi+if\xi\cos\phi-\xi^2/10\right\}.
\end{multline}
We can now make use of Bessel's integral
\vspace{3mm}
\begin{equation}
\int_0^{2\pi}d\phi\: e^{\pm i(m-m')\phi+if\xi\cos\phi}
=2\pi i^{|m-m'|}J_{|m-m'|}(f\xi),
\vspace{3mm}
\end{equation}
to reduce our expression to the form
\begin{equation}
M_{m m^\prime}(\mathbf{f})=2\pi i^{|m-m'|}e^{i(m-m')
\phi_{\mathbf{f}}}\mathcal{M}_{mm'}(f),
\end{equation}
where
\begin{equation}\label{curlym}
\mathcal{M}_{mm'}(f)=\int_0^{\infty}d\xi\:\xi^{1+m+m'}
e^{-\xi^2/10}J_{|m-m'|}(f\xi).
\vspace{3mm}
\end{equation}
This can also be expressed in terms of a 
confluent hypergeometric function $\Phi(\beta,\gamma;z)$ as
\begin{multline}\label{confhyp}
\mathcal{M}_{mm'}(f)=\frac{2^{(m+m')/2}5^{(m+m')/2+1}
\Gamma\left(\max\{m,m'\}+1\right)}{|m-m'|!}t^{|m-m'|/2} \\ 
\times\Phi\left(\max\{m,m'\}+1,|m-m'|+1;-t\right),
\end{multline}
where $t=5f^2/2$ and $\max\{m,m'\}$ is the largest of the integers
$m$ and $m'$. Eq.~\eqref{confhyp} can be simplified further if we apply 
the Kummer transformation:
\begin{equation}
\Phi(\beta,\gamma;-t)=e^{-t}\Phi(\gamma-\beta,\gamma;t),
\end{equation} 
and note that $(\gamma-\beta)$ is always a negative integer:
\begin{equation}
\gamma-\beta=|m-m'|-\max\{m,m'\}.
\end{equation}
This means that $\Phi(\gamma-\beta,\gamma;t)$ reduces to a polynomial,
and $\Phi(\beta,\gamma;-t)$ is then just a polynomial multiplied by $e^{-t}$.

If we now perform the integration over $\phi_{\mathbf{f}}$ we obtain the
final form of the overlap matrix elements
\begin{equation}\label{monom}
\langle\left\{n\right\}|\left\{n'\right\}\rangle=C^2A(2\pi)^5\:\delta_{LL'}
\int_0^{\infty}df\: f\prod_{j=1}^5 i^{|n_j-n'_j|}\mathcal{M}_{n_j n'_j}(f),
\end{equation}
where $L=n_1+n_2+n_3+n_4+n_5$.
Note that nothing depends on $\mathbf{k}$ in this expression for the
matrix elements.
It is convenient to define the constant $C$ so that for $L=L^{\prime}=0$
the matrix element $\langle\left\{n\right\}|\left\{n'\right\}\rangle=
\langle 0|0\rangle=1$.
This yields
\begin{equation}\label{csquared}
C^2=\frac{1}{125(2\pi)^5A}.
\end{equation}

In the above formulation we have only considered matrix elements in
terms of monomial functions. As mentioned earlier, matrix elements in
terms of symmetric polynomials can be easily constructed as a linear 
combination of the monomial matrix elements \eqref{monom}.

\subsubsection{Anyon-anyon interaction}\label{SS:anyonanyon}

The anyon-anyon interaction has the form
\begin{equation}\label{Vaa}
\hat{V}_{aa}=\frac{1}{25}\sum_{j<l}\;\frac{1}{|\boldsymbol{\xi}_j
-\boldsymbol{\xi}_l|}.
\end{equation}
We shall only calculate the matrix elements for
the first term $\hat{V}_{12}$ as the others follow by analogy.
We begin by taking the Fourier transform of Eq.~\eqref{Vaa}:
\vspace{3mm}
\begin{equation}
\hat{V}_{12}\left({\boldsymbol{\xi}}_1,{\boldsymbol{\xi}}_2\right)=
\frac{1}{25}\int\frac{d\mathbf{q}}{2\pi q}
\exp\left\{i\mathbf{q}\cdot\left({\boldsymbol{\xi}}_1
-{\boldsymbol{\xi}}_2\right)\right\}.
\vspace{3mm}
\end{equation}
This gives a matrix element in terms of monomials
\begin{align}
\langle\left\{n\right\}|\hat{V}_{12}|\left\{n'\right\}\rangle &=
\frac{1}{25}\int d\mathbf{R}
\int d\boldsymbol{\rho}\int\frac{d\mathbf{f}}{(2\pi)^2}
\int\frac{d\mathbf{q}}{2\pi q} \notag \\
&\times\int d\boldsymbol{\xi}_1 d\boldsymbol{\xi}_2 d\boldsymbol{\xi}_3
d\boldsymbol{\xi}_4 d\boldsymbol{\xi}_5\:
\bar{\Psi}_{{\{n\},\mathbf{k}}}\Psi_{\{n'\},\mathbf{k}} \notag \\
&\times\exp\left\{i\mathbf{f}\cdot
\left(\boldsymbol{\xi}_1+\boldsymbol{\xi}_2+\boldsymbol{\xi}_3
+\boldsymbol{\xi}_4+\boldsymbol{\xi}_5\right)\right\}
\exp\left\{i\mathbf{q}\cdot\left({\boldsymbol{\xi}}_1
-{\boldsymbol{\xi}}_2\right)\right\}.
\end{align}
Following the procedure in Section \ref{SS:overlap} yields
\vspace{3mm}
\begin{align}
\langle\left\{n\right\}|\hat{V}_{12}|\left\{n'\right\}\rangle &=
\frac{C^2A(2\pi)^3}{25}\prod_{j=1}^5 i^{|n_j-n'_j|}
\int d\mathbf{f}\int d\mathbf{q} \notag \\
&\times\exp\left\{i\phi_{+}(n_1-n_1^{\prime})
+i\phi_{-}(n_2-n_2^{\prime})\right\} \notag \\
&\times\exp\left\{i\phi_{\mathbf{f}}\right[(n_3-n_3^{\prime})
+(n_4-n_4^{\prime})+(n_5-n_5^{\prime})\left]\right\} \notag \\
&\times\mathcal{M}_{n_1 n'_1}(f_{+})\mathcal{M}_{n_2 n'_2}(f_{-})
\prod_{j=3}^5\mathcal{M}_{n_j n'_j}(f),
\vspace{3mm}
\end{align}
where $\mathcal{M}_{mm'}$ are as defined in Eq.~\eqref{curlym}, 
$\mathbf{f}_{\pm}=\mathbf{f}\pm\mathbf{q}$, and $\phi_{\pm}$
are the phases of $\mathbf{f}_{\pm}$.

We now seek to eliminate $\phi_{\mathbf{f}}$ by the change of variables
$\phi=\phi_{\mathbf{f}}-\phi_{\mathbf{q}}$ and $\psi_{\pm}=
\phi_{\pm}-\phi_{\mathbf{f}}$. After substitution and integration 
over $\phi_{\mathbf{q}}$ we obtain
\vspace{3mm}
\begin{align}\label{aafinal}
\langle\left\{n\right\}|\hat{V}_{12}|\left\{n'\right\}\rangle & =
\frac{C^2(2\pi)^4A}{25}\:\delta_{LL'}\prod_{j=1}^5 i^{|n_j-n'_j|}
\int_0^{\infty}df\: f
\int_0^{\infty}dq\int_0^{2\pi}d\phi \notag \\
&\times\exp\left\{i\phi(L-L^{\prime})\right\}
\exp\left\{i\psi_{+}(n_1-n_1^{\prime})+i\psi_{-}(n_2-n_2^{\prime})
\right\} \notag \\
&\times\mathcal{M}_{n_1 n'_1}(f_{+})\mathcal{M}_{n_2 n'_2}(f_{-})
\mathcal{M}_{n_3 n'_3}(f)\mathcal{M}_{n_4 n'_4}(f)\mathcal{M}_{n_5 n'_5}(f),
\vspace{3mm}
\end{align}
where $\psi_{\pm}$ and $f_{\pm}$ can be expressed in terms of the 
variables of integration as
\begin{equation}
e^{i\psi_{\pm}}=\left(f\pm qe^{-i\phi}\right)/f_{\pm},\qquad
f_{\pm}^2=f^2+q^2\pm 2fq\cos\phi.
\end{equation}
Eq.~\eqref{aafinal} shows that the anyon-anyon interaction
matrix has the same 
block diagonal structure as the overlap matrix.
For the case of $L=L^{\prime}=0$, 
using Eq.~\eqref{csquared} for $C^2$,
the anyon-anyon interaction matrix element reduces to
\begin{equation}\label{aaaa}
\langle 0|\hat{V}_{aa}|0\rangle=\frac{1}{5}\sqrt{\frac{\pi}{5}}.
\end{equation}

\subsubsection{Anyon-hole interaction}\label{SS:anyonhole}

The anyon hole interaction takes the form
\begin{equation}
\hat{V}_{ah}=-\frac{1}{5}\left\{\frac{1}{\left|\mathbf{r}_{1h}\right|}
+\frac{1}{\left|\mathbf{r}_{2h}\right|}+\frac{1}{\left|\mathbf{r}_{3h}
\right|}+\frac{1}{\left|\mathbf{r}_{4h}\right|}
+\frac{1}{\left|\mathbf{r}_{5h}\right|}\right\},
\end{equation}
where
\begin{equation}
\mathbf{r}_{jh}=\boldsymbol{\xi}_j-\boldsymbol{\rho}+h\hat{\mathbf{z}}.
\vspace{3mm}
\end{equation}
Considering only $\hat{V}_{1h}$, we take the Fourier transform:
\vspace{3mm}
\begin{equation}
\hat{V}_{1h}\left(\boldsymbol{\rho},{\boldsymbol{\xi}}_1\right)=
-\frac{1}{5}\int\frac{d\mathbf{q}}{(2\pi)^2}V_{ah}(q)
\exp\left\{i\mathbf{q}\cdot\left({\boldsymbol{\xi}}_1-\boldsymbol{\rho}
\right)\right\},
\vspace{3mm}
\end{equation}
where
\begin{equation}
V_{ah}(q)=\frac{2\pi}{q}e^{-qh}.
\vspace{3mm}
\end{equation}
The matrix elements are then
\begin{align}
\langle\left\{n\right\}|\hat{V}_{1h}|\left\{n'\right\}\rangle &=
-\frac{1}{5}\int d\mathbf{R}
\int d\boldsymbol{\rho}\int\frac{d\mathbf{f}}{(2\pi)^2}
\int\frac{d\mathbf{q}}{(2\pi)^2}V_{ah}(q) \notag \\
&\times\int d\boldsymbol{\xi}_1 d\boldsymbol{\xi}_2 d\boldsymbol{\xi}_3 
d\boldsymbol{\xi}_4 d\boldsymbol{\xi}_5\:
\bar{\Psi}_{{\{n\},\mathbf{k}}}\Psi_{\{n'\},\mathbf{k}} \notag \\
&\times\exp\left\{i\mathbf{f}\cdot
\left(\boldsymbol{\xi}_1+\boldsymbol{\xi}_2+\boldsymbol{\xi}_3
+\boldsymbol{\xi}_4+\boldsymbol{\xi}_5\right)\right\}
\exp\left\{i\mathbf{q}\cdot\left({\boldsymbol{\xi}}_1-\boldsymbol{\rho}
\right)\right\}.
\end{align}
Integration over $\mathbf{R}$ and $\boldsymbol{\rho}$ gives
$2\pi A\exp\left(-q^2/2-i\mathbf{d}\cdot\mathbf{q}\right)$.
Then, following the procedure in Section \ref{SS:overlap} once again,
we obtain
\begin{align}
\langle\left\{n\right\}|\hat{V}_{1h}|\left\{n'\right\}\rangle &=
-\frac{C^2(2\pi)^2A}{5}\prod_{j=1}^5 i^{|n_j-n'_j|}
\int d\mathbf{f}\int d\mathbf{q}\;V_{ah}(q)
e^{-q^2/2-i\mathbf{d}\cdot\mathbf{q}} \notag \\
&\times\exp\left\{i\phi_{+}(n_1-n_1^{\prime})
+i\phi_{\mathbf{f}}\left[(L-L^{\prime})-(n_1-n_1^{\prime})
\right]\right\} \notag \\
&\times\mathcal{M}_{n_1 n'_1}(f_{+})\prod_{j=2}^5
\mathcal{M}_{n_j n^{\prime}_j}(f),
\end{align}
where $\mathcal{M}_{mm'}$ are as defined in Eq.~\eqref{curlym}, 
$\mathbf{f}_{+}=\mathbf{f}+\mathbf{q}$, and $\phi_{+}$
is the phase of $\mathbf{f}_{+}$.
As before, $L=n_1+n_2+n_3+n_4+n_5$.

We now eliminate $\phi_{\mathbf{f}}$ by substituting 
$\phi=\phi_{\mathbf{f}}-\phi_{\mathbf{q}}$ and
$\psi=\phi_{+}-\phi_{\mathbf{f}}$. Choosing $\mathbf{d}$ along
the $x$-axis, the integration over $\phi_{\mathbf{q}}$ is then
\vspace{3mm}
\begin{equation}
\int_0^{2\pi}d\phi_{\mathbf{q}}\;\exp\left\{i\phi_{\mathbf{q}}
(L-L^{\prime})-idq\cos\phi_{\mathbf{q}}\right\}
=2\pi{(-i)}^{|L-L^{\prime}|}J_{|L-L^{\prime}|}(dq),
\vspace{3mm}
\end{equation}
which gives for the matrix elements:
\begin{align}\label{ahole}
\langle\left\{n\right\}|\hat{V}_{1h}|\left\{n'\right\}\rangle &=
-\frac{C^2(2\pi)^4A}{5}{(-i)}^{|L-L'|}\prod_{j=1}^5 i^{|n_j-n'_j|} \notag \\
&\times\int_0^{\infty}df\: f
\int_0^{\infty}dq\:e^{-q^2/2-qh}J_{|L-L'|}(dq) \notag \\
&\times\int_0^{2\pi}d\phi\;\exp\left\{i\phi(L-L^{\prime})
+i\psi(n_1-n_1^{\prime})\right\} \notag \\
&\times\mathcal{M}_{n_1 n'_1}(f_{+})\prod_{j=2}^5
\mathcal{M}_{n_j n^{\prime}_j}(f),
\end{align}
where $\psi$ and $f_{+}$ can be expressed in terms of the variables of
integration as
\vspace{3mm}
\begin{equation}
e^{i\psi}=\left(f+qe^{-i\phi}\right)/f_{+},\qquad
f_{+}^2=f^2+q^2+2fq\cos\phi.
\vspace{3mm}
\end{equation}

Eq.~\eqref{ahole} may be simplified further, and this yields the
following expression for the anyon-hole interaction matrix elements
for basis functions in terms of symmetric polynomials:
\vspace{3mm}\begin{equation}\label{Qpoly}
\langle L,M|\hat{V}_{ah}|L',M'\rangle =
-\int_0^{\infty}\exp(-5q^2/2-qh)J_{|L-L'|}(dq)Q_{LM,L'M'}(q)\;dq,
\vspace{3mm}
\end{equation}
where $Q_{LM,L'M'}(q)$ is a polynomial in $q$ (the lowest order
polynomial is $Q_{00,00}=1$).
For $h=0$, Eq.~\eqref{Qpoly} further reduces to an
expression in terms of elementary functions.  For the case of 
$L=L^{\prime}=0$, the anyon-hole interaction matrix element is then
\begin{equation}\label{ahah}
\langle 0|\hat{V}_{ah}|0\rangle=-\sqrt{\frac{\pi}{10}}.
\end{equation}

\subsection{Exact results for $\mathbf{k}=0$, $L=0$}

All the above results are simplified significantly for the state
with $\mathbf{k}=0$ and $L=0$. For such a state, the $(N+1)$-particle
wavefunction \eqref{finalbasis} with $\alpha=0$ reduces to the form
\begin{equation}
\Psi\left(\mathbf{R},\boldsymbol{\rho},\{\zeta_i\}\right)=
C_N\exp\left\{i\hat{\mathbf{z}}\cdot\left[\mathbf{R}\times\boldsymbol{\rho}
\right]/2-\rho^2/4\right\}\exp\left\{-\sum_{p=1}^N|\zeta_p|^2/4N\right\},
\vspace{3mm}
\end{equation} 
where the normalisation constant $C_N$ can be determined from the integral
\begin{multline}
C_N^2\int d\mathbf{R}\int\frac{d\mathbf{f}}{{(2\pi)}^2}
\int d\boldsymbol{\rho}\;e^{-\rho^2/2}\int d\boldsymbol{\xi}_1
\cdots d\boldsymbol{\xi}_N \\
\times\exp\left\{\sum_{p=1}^N\left(
i\boldsymbol{\xi}_p\cdot\mathbf{f}-\xi_p^2/2N\right)\right\}
=\frac{C_N^2 A}{N^2}(2\pi N)^N=1.
\end{multline}

\vspace{3mm}
For the case of $\mathbf{k}=0$ and $L=0$, the interaction matrix elements
of a $(N+1)$-particle exciton are straightforward to evaluate.
The anyon-anyon matrix element is
\vspace{3mm}
\begin{align}\label{anyany}
\langle 0|\hat{V}_{aa}|0\rangle &=\frac{N(N-1)}{2A(2\pi N)^N}
\int d\mathbf{R}\int\frac{d\mathbf{q}}{2\pi q}\int d\boldsymbol{\rho}
\;e^{-\rho^2/2}\int d\boldsymbol{\xi}_1\cdots d\boldsymbol{\xi}_N \notag \\
&\times\int\frac{d\mathbf{f}}{(2\pi)^2}\;
\exp\left\{\sum_{p=1}^N\left(
i\boldsymbol{\xi}_p\cdot\mathbf{f}-\xi_p^2/2N\right)+i\mathbf{q}\cdot
\left(\boldsymbol{\xi}_1-\boldsymbol{\xi}_2\right)\right\} \notag \\
&=\frac{(N-1)}{4N}\sqrt{\frac{\pi}{N}}.
\vspace{3mm}
\end{align}
For non-zero inter-plane separation $h$, the anyon-hole
matrix element can be reduced to
\vspace{3mm}
\begin{align}\label{erfc}
\langle 0|\hat{V}_{ah}|0\rangle &=-\frac{N^2}{A(2\pi N)^N}
\int d\mathbf{R}\int\frac{d\mathbf{q}}{2\pi q}\int d\boldsymbol{\rho}
\;e^{-\rho^2/2}\int d\boldsymbol{\xi}_1
\cdots d\boldsymbol{\xi}_N \notag \\
&\times\int\frac{d\mathbf{f}}{(2\pi)^2}\exp\left\{\sum_{p=1}^N\left(
i\boldsymbol{\xi}_p\cdot\mathbf{f}-\xi_p^2/2N\right)+i\mathbf{q}\cdot
\left(\boldsymbol{\xi}_1-\boldsymbol{\rho}\right)-qh \right\} \notag \\
&=-\int_0^{\infty}dq\;e^{-Nq^2/2-qh}
=-\sqrt{\frac{\pi}{2N}}e^{h^2/2N}\mbox{erfc}\left( h/\sqrt{2N}\right),
\vspace{3mm}
\end{align}
where $\mbox{erfc}(x)$ is the complementary error function.
(Note that for $h=0$, Eq.~\eqref{erfc} can be simplified further to a 
numerical value.)
Using the asymptotic expansion of $\mbox{erfc}(x)$ it can be easily
seen from Eq.~\eqref{erfc} that $\langle 0|\hat{V}_{ah}|0\rangle
\rightarrow -1/h$ as $h\rightarrow\infty$, as expected.

Eqs.~\eqref{anyany} and \eqref{erfc} also allow us to calculate the
critical inter-plane separation $h_c$ at which the $\mathbf{k}=0$,
$L=0$ state becomes unbound, i.e. when
\vspace{3mm}
\begin{equation}
\sqrt{\frac{\pi}{2N}}\;e^{x_c^2}\;\mbox{erfc}(x_c)=
\frac{(N-1)\sqrt{2}}{4N}\sqrt{\frac{\pi}{N}},
\vspace{3mm}
\end{equation}
where $x_c=h_c/\sqrt{2N}$. So, for $N=3$ the critical separation 
$h_c\approx 5.39\:l$, for $N=5$ we find that $h_c\approx 5.59\:l$, 
and for $N\gg 1$ we have $h_c\approx 1.32\sqrt{2N}\:l$. 
Notably, these critical 
separations are well inside the region for which the AEM is applicable.
It should be emphasised that the state with $L=0$ is not the ground
state for the anyon exciton at large separation $h$.
For example, for a four-particle exciton \cite{Portnoi1996},
the ground states for large separation satisfy a superselection rule
$L=3m$, where $m$ is an integer, and when $h\rightarrow\infty$
the ground state energy tends to its classical value, 
$\varepsilon_{cl}=-(2/3)^{3/2}/h$.

The model is applicable to real physical situations at large layer 
separations ($h>l$), although it remains solvable for
all values of $h$, including $h=0$. Moreover, it has been shown 
in \cite{Portnoi1996} that
the ground state of the four-particle problem ($N=3$) at $h=0$ is
the state with $\mathbf{k}=0$ and $L=0$. We expect the same to be
true for $N\geqslant 5$, since the anyon-hole attraction will always overcome
the anyon-anyon repulsion at small inter-particle separations. 
It can be shown (in a similar way to that in
\cite{Portnoi1996}) that the smallest average inter-particle 
separation corresponds to the $L=0$ case. Therefore, the case of
$\mathbf{k}=0$, $L=0$ is of special interest since it predicts the
anyon exciton binding energies in the limit $h\rightarrow 0$.

We are now in a position to write down a general expression for the
binding energy of a $(N+1)$-particle exciton at zero 2DEG-hole
separation ($h=0$):
\vspace{3mm}
\begin{equation}\label{bind}
E_b=-\left(\langle 0|\hat{V}_{ah}|0\rangle
+\langle 0|\hat{V}_{aa}|0\rangle\right)
=\left[1-\frac{(N-1)}{2\sqrt{2}N}\right]
\sqrt{\frac{\pi}{2N}}.
\vspace{3mm}
\end{equation}
Eq.~\eqref{bind} has been written in this particular form to emphasise
the key result that for any value of $N$ there always exists at least
one bound state of a neutral $(N+1)$-particle anyon exciton.  
For $N=1$, Eq.~\eqref{bind} yields the value \mbox{$\sqrt{\pi/2}$},
which corresponds to that obtained in \cite{Lerner1980} for a
standard diamagnetic exciton. For a four-particle exciton $(N=3)$,
\mbox{$E_b=\left(1-\sqrt{2}/6\right)\sqrt{\pi/6}$}, 
which agrees with the result of \cite{Portnoi1996}. 
Finally, for $N=5$ we have a binding energy of
\mbox{$(1-\sqrt{2}/5)\sqrt{\pi/10}$}, which can also be obtained from
Eqs.~\eqref{aaaa} and \eqref{ahah}.

\subsection{Conclusions}

The anyon exciton model has been generalised to the case of a neutral
exciton consisting of a valence hole and an arbitrary number $N$ of anyons, 
and several important mathematical results have been obtained. 
A complete set of exciton basis functions has been obtained and
these functions have been fully classified using a result from the
theory of partitions.
We have derived expressions for the overlap and interaction matrix 
elements of a six-particle system $(N=5)$, which describes an exciton
against the background of an IQL with filling factor $\nu=2/5$.
In the particular case of $\mathbf{k}=0$ and $L=0$, we have found an 
expression for the binding energy of a $(N+1)$-particle exciton 
at zero anyon-hole separation, which agrees with known results for a 
standard diamagnetic exciton and four-particle anyon exciton.
We have also shown that the $(N+1)$-particle exciton remains bound for
2DEG-hole separations exceeding several magnetic lengths, when the anyon
exciton model becomes applicable to real physical systems.

\section{Appendices}
\subsection{Solution of real-space Schr\"{o}dinger equation
in two dimensions}\label{S:radial}

We apply the method of separation of variables to Eq.~\eqref{wannier},
making the substitution
\begin{equation}
\Psi(\boldsymbol{\rho})=R(\rho)\Phi(\phi).
\end{equation}
Introducing a separation constant $m^2$, we can obtain the angular
equation
\begin{equation}
\frac{d^2\Phi}{d\phi^2}+m^2\Phi=0,
\end{equation}
with the solution
\begin{equation}
\Phi(\phi)=\frac{1}{\sqrt{2\pi}}e^{im\phi}.
\vspace{3mm}
\end{equation}

The corresponding radial equation (with $E=-q_0^2$) is
\vspace{3mm}
\begin{equation}\label{radialeqn}
\frac{d^2
R}{d\rho^2}+\frac{1}{\rho}\frac{dR}{d\rho}+\left(\frac{2}{\rho}-q_0^2-
\frac{m^2}{\rho^2}\right)R=0.
\vspace{3mm}
\end{equation}
We make the substitution
\begin{equation}
R(\rho)=C\rho^{|m|}e^{-q_{0}\rho}w(\rho),
\end{equation}
where $C$ is a normalisation constant. This leads to the equation
\begin{equation}
\rho\frac{d^2w}{d\rho^2}+(2|m|+1-2q_0\rho)\frac{dw}{d\rho}+(2-2|m|q_0-q_0)w=0.
\end{equation}
Making a final change of variables $\beta=2q_0\rho$, we obtain
\vspace{3mm}
\begin{equation}\label{hyper}
\beta\frac{d^2w}{d\beta^2}+(2|m|+1-\beta)\frac{dw}{d\beta}+
\left(\frac{1}{q_0}-|m|-\frac{1}{2}\right)w=0.
\vspace{3mm}
\end{equation}
This is the confluent hypergeometric equation \cite{Gradshteyn2000}, which has 
two linearly independent solutions. If we choose the solution which is 
regular at the origin, then this becomes a polynomial of finite degree if
$q_0=(n+1/2)^{-1}$, with $n=0,1,2,\ldots$  Eq.~\eqref{hyper} then
becomes the associated Laguerre equation \cite{Arfken1985}, 
the solutions of which are the associated Laguerre polynomials
\begin{equation}
w=L_{n-|m|}^{2|m|}(\beta)=L_{n-|m|}^{2|m|}(2q_0\rho).
\end{equation}

We can now write the real-space wavefunction in the form
\begin{equation}
\Psi_{nm}(\boldsymbol{\rho})=\frac{C}{2\pi}\rho^{|m|}e^{-q_0\rho}
L_{n-|m|}^{2|m|}(2q_0\rho)e^{im\phi_\rho},
\end{equation}
where the reason for the subscript on $\phi$ is explained in
Section \ref{SS:ints}.

To normalise this wavefunction we need to make use of the integral
\cite{Arfken1985}:
\begin{multline}
\int_0^{\infty}e^{-2q_0\rho}(2q_0\rho)^{2|m|+1}L_{n-|m|}^{2|m|}(2q_0\rho)
L_{n-|m|}^{2|m|}(2q_0\rho)\,d(2q_0\rho) \\
=\frac{(n+|m|)!}{(n-|m|)!}(2n+1).
\end{multline}
The normalised wavefunctions are therefore:
\vspace{3mm}
\begin{equation}
\Psi_{nm}(\boldsymbol{\rho})=\sqrt{\frac{q_0^3(n-|m|)!)}{\pi(n+|m|)!}}
(2q_0\rho)^{|m|}e^{-q_0\rho}L_{n-|m|}^{2|m|}(2q_0\rho)e^{im\phi_\rho},
\vspace{3mm}
\end{equation}
satisfying the following orthogonality condition:
\begin{equation}
\int\Psi_{n_1m_1}^*(\boldsymbol{\rho})\Psi_{n_2m_2}(\boldsymbol{\rho})\,d
\boldsymbol{\rho}=\delta_{n_1n_2}\delta_{m_1m_2}.
\end{equation}

\subsection{Derivation of expression for $\hat{\mathbf{A}}^2$}
\label{S:asquared}

From Eq.~\eqref{dimless} we have
\begin{align}\label{asq}
\hat{\mathbf{A}}^2&=\left[(\hat{\mathbf{q}}\times\hat{\mathbf{L}}_z-
\hat{\mathbf{L}}_z\times\hat{\mathbf{q}})-\frac{2}{\rho}
\boldsymbol{\rho}\right]^2 \\
&=[2(\hat{\mathbf{q}}\times\hat{\mathbf{L}}_z)-i\hat{\mathbf{q}}]^2-
\frac{2}{\rho}\boldsymbol{\rho}\cdot[2(\hat{\mathbf{q}}
\times\hat{\mathbf{L}}_z)-i\hat{\mathbf{q}}]-\frac{2}{\rho}[2(\hat{\mathbf{q}}
\times\hat{\mathbf{L}}_z)-i\hat{\mathbf{q}}]\cdot\boldsymbol{\rho}+4.
\notag
\end{align}
We further expand as follows:
\begin{align}\label{exp1}
[2(\hat{\mathbf{q}}\times\hat{\mathbf{L}}_z)-i\hat{\mathbf{q}}]^2&=
4(\hat{\mathbf{q}}\times\hat{\mathbf{L}}_z)^2-2i\hat{\mathbf{q}}
\cdot(\hat{\mathbf{q}}\times\hat{\mathbf{L}}_z)-2i(\hat{\mathbf{q}}
\times\hat{\mathbf{L}}_z)\cdot\hat{\mathbf{q}}-\hat{\mathbf{q}}^2 \\
&=4\hat{\mathbf{q}}^2\hat{\mathbf{L}}_z^2+2\hat{\mathbf{q}}^2-
\hat{\mathbf{q}}^2
=\hat{\mathbf{q}}^2(4\hat{\mathbf{L}}_z^2+1),
\notag
\end{align}
and
\begin{equation}\label{exp2}
-\frac{2}{\rho}\boldsymbol{\rho}\cdot[2(\hat{\mathbf{q}}\times
\hat{\mathbf{L}}_z)-i\hat{\mathbf{q}}]-\frac{2}{\rho}[2(\hat{\mathbf{q}}
\times\hat{\mathbf{L}}_z)-i\hat{\mathbf{q}}]\cdot\boldsymbol{\rho}
=-\frac{2}{\rho}(4\hat{\mathbf{L}}_z^2+1).
\vspace{3mm}
\end{equation}
Substituting Eqs.~\eqref{exp1} and \eqref{exp2} into Eq.~\eqref{asq} gives
\begin{equation}
\hat{\mathbf{A}}^2=\hat{\mathbf{q}}^2(4\hat{\mathbf{L}}_z^2+1)-
\frac{2}{\rho}(4\hat{\mathbf{L}}_z^2+1)+4,
\end{equation}
which, from Eq.~\eqref{ham2}, is just
\begin{equation}
\hat{\mathbf{A}}^2=\hat{H}(4\hat{\mathbf{L}}_z^2+1)+4.
\end{equation}

\subsection{Variable phase method}\label{S:vphase}

In this Section, which is based on the formulation in \cite{Portnoi1999}, 
we derive the basic equations of the variable 
phase approach in two dimensions from the radial Schr\"{o}dinger equation. 
This derivation is similar to that in three dimensions 
\cite{Calogero1967}.

\subsubsection{Scattering phase shifts}

The relative in-plane motion of two interacting particles with masses 
$M_a$ and $M_b$ and energy of relative motion $E$ can be 
considered as the motion of a particle with mass 
$\mu_{ab}=M_a M_b/(M_a+M_b)$ and energy $E$, moving in an external 
central potential $V(\rho)$. 
This motion is described by a wavefunction satisfying the stationary 
Schr\"{o}dinger equation  
\begin{equation}\label{vpm1}               
\hat{H}_{rel}\psi=-\frac{\hbar^2}{2\mu_{ab}}\left(\frac{1}{\rho}
\frac{\partial}{\partial\rho}\rho\frac{\partial}{\partial\rho}
+\frac{1}{\rho^2}\frac{\partial^2}{\partial\varphi^2}
\right)\psi+V(\rho)\psi=E\psi.
\vspace{3mm}
\end{equation}
Owing to the axial symmetry of the potential $V(\rho)$, we can separate 
variables in the expression for the wavefunction
\begin{equation}\label{vpm2}
\psi_m(\rho,\varphi)=R_m(\rho)e^{i m\varphi},\qquad m=0,\pm1,\pm2,\ldots\; .
\end{equation}
The  equation for the radial function $R_m(\rho)$ reads
\begin{equation}\label{vpm3}
R''_m+\frac{1}{\rho}R'_m+
\left(k^2-U(\rho)-\frac{m^2}{\rho^2}\right)R_m=0,
\end{equation}
where $k^2=2\mu_{ab}E/\hbar^2$ and $U(\rho)=2\mu_{ab}V(\rho)/\hbar^2$.
In what follows we consider $m\geq0$ only, 
as $R_{-m}(\rho)=R_{m}(\rho)$.

We assume that the interaction potential vanishes at infinity
(the precise decay rate will be discussed later).
Then, at large distances the radial function satisfies the free 
Bessel equation, whose general solution is 
\begin{align}\label{vpm4}
R_m(\rho)&=A_m[J_m(k\rho)\cos{\delta_m}-N_m(k\rho)\sin{\delta_m}] 
\\ \notag 
&\stackrel{\rho\rightarrow\infty}{\longrightarrow} 
A_m\left(\frac{2}{\pi k \rho}\right)^{1/2} 
\cos(k\rho - (2m+1)\pi/4 + \delta_m),  
\end{align} 
where $\delta_m$ is the scattering phase shift \cite{Stern1967,
Landau1977}, and $J_m(k\rho)$ and $N_m(k\rho)$ are the Bessel and 
Neumann functions, respectively.

In the variable phase approach, $A_m$ and $\delta_m$ are considered 
not as constants, but as functions of the distance $\rho$. 
The amplitude function $A_m(\rho)$ and the phase function 
$\delta_m(\rho)$ are introduced by the equation
\begin{equation}\label{vpm5}
R_m(\rho)=A_m(\rho)[J_m(k\rho)\cos{\delta_m(\rho)}
-N_m(k\rho)\sin{\delta_m(\rho)}], 
\end{equation}
with an additional condition, which we are free to choose as
\begin{equation}\label{vpm6}
R'_m(\rho)=A_m(\rho)[J'_m(k\rho)\cos{\delta_m(\rho)} 
-N'_m(k\rho)\sin{\delta_m(\rho)}], 
\end{equation}
where the prime indicates differentiation with respect to $\rho$.
The phase function $\delta_m(\rho)$ has a natural physical 
interpretation as the phase shift produced by a potential 
cut-off at a distance $\rho$.

Differentiating Eq.~\eqref{vpm6} and substituting the resulting 
expression, together with Eqs.~\eqref{vpm5} and \eqref{vpm6}, 
into Eq.~\eqref{vpm3}, we get
\begin{align}\label{vpm7}
A'_m(\rho)&[J'_m(k\rho)\cos{\delta_m(\rho)}
-N'_m(k\rho)\sin{\delta_m(\rho)}]
\\ \notag
&-\delta'_m(\rho)A_m(\rho)[J'_m(k\rho)\sin{\delta_m(\rho)} 
+N'_m(k\rho)\cos{\delta_m(\rho)}]
\\ \notag
&=U(\rho)A_m(\rho)[J_m(k\rho)\cos{\delta_m(\rho)}
-N_m(k\rho)\sin{\delta_m(\rho)}]. 
\end{align}
To obtain Eq.~\eqref{vpm7} we used the fact that the functions 
$J_m(k\rho)$ and $N_m(k\rho)$ satisfy the free Bessel equation
\begin{equation}
F''_m+\frac{1}{\rho}F'_m+\left(k^2-\frac{m^2}{\rho^2}\right)F_m=0.
\end{equation}

Equating the derivative of Eq.~\eqref{vpm5} to Eq.~\eqref{vpm6} 
implies the following condition on the derivatives 
of the amplitude and the phase functions:
\begin{multline}\label{vpm8}
A'_m(\rho)[J_m(k\rho)\cos{\delta_m(\rho)}
-N_m(k\rho)\sin{\delta_m(\rho)}] \\
=\delta'_m(\rho)A_m(\rho)[J_m(k\rho)\sin{\delta_m(\rho)}
+N_m(k\rho)\cos{\delta_m(\rho)}]. 
\end{multline}
Substituting $A'_m(\rho)$, obtained from Eq.~\eqref{vpm8}, into 
Eq.~\eqref{vpm7} yields
\begin{multline}\label{vpm9}
-\delta'_{m}(\rho)[J_m(k\rho)N'_m(k\rho)-N_m(k\rho)J'_m(k\rho)] \\
=U(\rho)[J_m(k\rho)\cos{\delta_m(\rho)}-N_m(k\rho)\sin{\delta_m(\rho)}]^2. 
\end{multline}
Eq.~\eqref{vpm9} can be simplified further, using the Wronskian 
of the Bessel functions
\vspace{3mm}
\begin{equation}
W\{J_m(x), N_m(x)\}=J_m(x)\frac{d}{dx}N_m(x)-
N_m(x)\frac{d}{dx}J_m(x)=\frac{2}{\pi x},
\vspace{3mm}
\end{equation}
and thus becomes
\vspace{3mm}
\begin{equation}\label{vpm10}
\frac{d}{d\rho}\,\delta_m(\rho)=-\frac{\pi}{2}\,\rho\,U(\rho)
[J_m(k\rho)\cos{\delta_m(\rho)}-N_m(k\rho)\sin{\delta_m(\rho)}]^2. 
\vspace{3mm}
\end{equation}
This \emph{phase equation}, Eq.~\eqref{vpm10}, is a first-order
nonlinear differential equation of the Ricatti type, which must 
be solved with the initial condition 
\begin{equation}\label{vpm11}
\delta_m(0)=0,  
\end{equation} 
thus  ensuring  that the radial function does not diverge at $\rho=0$.
The total scattering phase shift $\delta_m$ can be obtained as a 
large-distance limit of the phase function $\delta_m(\rho)$:
\begin{equation}\label{vpm12}
\delta_m=\lim_{\rho\rightarrow\infty}\delta_m(\rho).
\end{equation}
For numerical convenience, instead of the initial condition 
Eq.~\eqref{vpm11}, the small-$\rho$ expansion is used:
\begin{equation}\label{vpm13}
\delta_m(\rho)\approx-\frac{\pi k^{2m}}{2^{2m+1}(m!)^2} 
\int_{0}^{\rho}U(\rho^\prime){\rho^\prime}^{2m+1}d\rho^\prime, 
\quad\rho\rightarrow 0. 
\end{equation}

From Eq.~\eqref{vpm10} and the asymptotic expansions of the Bessel 
functions, one can see that the variable phase method is applicable 
only if the scattering potential $U(\rho)$ satisfies the 
necessary conditions
\vspace{3mm}
\begin{equation}\label{vpm14}
\int_{\rho}^{\infty}U(\rho^\prime)d\rho^\prime\rightarrow 0, 
\quad\rho\rightarrow\infty,  
\end{equation} 
and 
\begin{equation}\label{vpm15}
{\rho}^{2}U(\rho)\rightarrow 0,\quad\rho\rightarrow 0.
\vspace{3mm}
\end{equation}
The Stern-Howard potential $V_s(\rho)$, defined by 
Eq.~\eqref{Stern}, behaves like $\rho^{-1}$ at small distances 
and like $\rho^{-3}$ at large distances. Such behaviour allows 
the application of the variable phase method to this potential.

\subsubsection{Bound-state energies}

For states with negative energy of the relative motion (bound states),
the wavenumber $k$ is imaginary $(k=i\kappa)$,        
and we introduce the function $\eta_m (\rho,\kappa)$ which 
vanishes at the origin and satisfies the nonlinear equation  
\vspace{3mm}
\begin{equation}\label{vpmb1}
\frac{d}{d\rho}\eta_m(\rho,\kappa)=-\frac{\pi}{2}\rho U(\rho)
\left[ I_m(\kappa\rho)\cos{\eta_m(\rho,\kappa)}+ 
\frac{2}{\pi} K_m(\kappa\rho)\sin{\eta_m(\rho,\kappa)}\right]^2,
\vspace{3mm}
\end{equation}
where $I_m(\kappa\rho)$ and $K_m(\kappa\rho)$ are modified Bessel 
functions of the first and  second kind, respectively. 
Eq.~\eqref{vpmb1} is derived in the same fashion as 
Eq.~\eqref{vpm10}.
The functions $I_m(\kappa\rho)$ and $K_m(\kappa\rho)$ represent two 
linearly-independent solutions of the free radial-wave Schr\"{o}dinger 
equation for the negative value of energy $E=-\hbar^2\kappa^2/2\mu_{ab}$, 
and $\cot{\eta_m}$ characterises the weights of the diverging 
[$I_m(\kappa\rho)$] and converging [$K_m(\kappa\rho)$] 
solutions as $\rho\rightarrow\infty$. 
For the bound state the diverging solution vanishes, 
implying the asymptotic condition 
\begin{equation}\label{vpmb2}
\eta_m(\rho\rightarrow\infty,\kappa_\nu)=\left(\nu-\frac{1}{2}\right)\pi,
\qquad\nu=1,2,3,\ldots\; . 
\vspace{3mm}
\end{equation} 
Here, $\nu$ numerates the bound states for a given $m$ and $(\nu-1)$ is the 
number of non-zero nodes of the radial wavefunction. For numerical solution 
of Eq.~\eqref{vpmb1}, instead of the boundary condition $\eta_m(0,\kappa)=0$, 
an asymptotic initial condition (analogous to the condition Eq.~\eqref{vpm13} 
for the phase function $\delta_m(\rho)$) is used.

\bibliographystyle{unsrt}

\begin{thebibliography}{100}
 

\bibitem{Stern1967}
F.~Stern and W.~E. Howard.
\newblock {\em Phys. Rev.}, 163:816, 1967.

\bibitem{Portnoi1997}
M.~E. Portnoi and I.~Galbraith.
\newblock {\em Solid State Commun.}, 103:325, 1997.

\bibitem{Portnoi1999}
M.~E. Portnoi and I.~Galbraith.
\newblock {\em Phys. Rev. B}, 60:5570, 1999.

\bibitem{Tanguy2001}
C.~Tanguy.
\newblock {\em cond-mat/0106184}, 2001.

\bibitem{Rashba1982}
E.~I. Rashba and M.~Sturge, editors.
\newblock {\em Excitons}.
\newblock North-Holland, Amsterdam, 1982.

\bibitem{Haug1998}
H.~Haug and S.~W. Koch.
\newblock {\em Quantum Theory of the Optical and Electronic Properties of
  Semiconductors}.
\newblock World Scientific, Singapore, 3rd edition, 1998.

\bibitem{Noether1918}
E.~Noether.
\newblock {\em Nachrichten von der Gesselschaft der Wissenschaften zu \\
  G{\"o}ttingen}, pages 235--258, 1918.

\bibitem{Landau1977}
L.~D. Landau and E.~M. Lifshitz.
\newblock {\em Quantum Mechanics (Non-relativistic Theory)}.
\newblock Pergamon Press, Oxford, 3rd edition, 1977.

\bibitem{Kalnins1976}
E.~G. Kalnins, W.~Miller, and P.~Winternitz.
\newblock {\em SIAM J. Appl. Math.}, 30:630, 1976.

\bibitem{Goldstein1980}
H.~Goldstein.
\newblock {\em Classical Mechanics}.
\newblock Addison-Wesley, New York, 2nd edition, 1980.

\bibitem{Bertrand1873}
J.~Bertrand.
\newblock {\em Comptes Rendus}, 77:849, 1873.

\bibitem{Pauli1926}
W.~Pauli.
\newblock {\em Z. Phys.}, 36:336, 1926.

\bibitem{Fock1935}
V.~A. Fock.
\newblock {\em Z. Phys.}, 98:145, 1935.

\bibitem{Bargmann1936}
V.~Bargmann.
\newblock {\em Z. Phys.}, 99:576, 1936.

\bibitem{Alliluev1958}
S.~P. Alliluev.
\newblock {\em Sov. Phys.--JETP}, 6:156, 1958.

\bibitem{Shibuya1965}
T.~Shibuya and C.~E. Wulfman.
\newblock {\em Am. J. Phys.}, 33:570, 1965.

\bibitem{Flugge1952}
S.~Fl{\"{u}}gge and H.~Marschall.
\newblock {\em Rechenmethoden der Quantentheorie}.
\newblock Springer-Verlag, Berlin, 2nd edition, 1952.
\newblock Problem 24.

\bibitem{Zaslow1967}
B.~Zaslow and M.~E. Zandler.
\newblock {\em Am. J. Phys.}, 35:1118, 1967.

\bibitem{Bander1966a}
M.~Bander and C.~Itzykson.
\newblock {\em Rev. Mod. Phys.}, 38:330, 1966.

\bibitem{Bander1966b}
M.~Bander and C.~Itzykson.
\newblock {\em Rev. Mod. Phys.}, 38:346, 1966.

\bibitem{Yang1991a}
X.~L. Yang, M.~Lieber, and F.~T. Chan.
\newblock {\em Am. J. Phys.}, 59:231, 1991.

\bibitem{Yang1991b}
X.~L. Yang, S.~H. Guo, F.~T. Chan, K.~W. Wong, and W.~Y. Ching.
\newblock {\em Phys. Rev. A}, 43:1186, 1991.

\bibitem{Guo1991}
S.~H. Guo, X.~L. Yang, F.~T. Chan, K.~W. Wong, and W.~Y. Ching.
\newblock {\em Phys. Rev. A}, 43:1197, 1991.

\bibitem{Aquilanti1997}
V.~Aquilanti, S.~Cavalli, and C.~Coletti.
\newblock {\em Chem. Phys.}, 214:1, 1997.

\bibitem{Nouri1999}
S.~Nouri.
\newblock {\em J. Math. Phys.}, 40:1294, 1999.

\bibitem{Daboul1993}
J.~Daboul, P.~Slodowy, and C.~Daboul.
\newblock {\em Phys. Lett. B}, 317:321, 1993.

\bibitem{Daboul1998}
C.~Daboul and J.~Daboul.
\newblock {\em Phys. Lett. B}, 425:135, 1998.

\bibitem{Dahl1997}
J.~P. Dahl.
\newblock {\em J. Phys. A}, 30:6831, 1997.

\bibitem{Kamath2002}
S.~G. Kamath.
\newblock {\em J. Math. Phys.}, 43:318, 2002.

\bibitem{Muljarov1999}
E.~A. Muljarov, A.~L. Yablonskii, S.~G. Tikhodeev, A.~E. Bulatov, and J.~L.
  Birman.
\newblock {\em Phys. Rev. B}, 59:4600, 1999.

\bibitem{Muljarov2000}
E.~A. Muljarov, A.~L. Yablonskii, S.~G. Tikhodeev, A.~E. Bulatov, and J.~L.
  Birman.
\newblock {\em J. Math. Phys.}, 41:6026, 2000.

\bibitem{Schindlmayr1997}
A.~Schindlmayr.
\newblock {\em Eur. J. Phys.}, 18:374, 1997.

\bibitem{Gradshteyn2000}
I.S. Gradshteyn and I.M. Ryzhik.
\newblock {\em Table of Integrals, Series, and Products}.
\newblock Academic Press, San Diego, 6th edition, 2000.

\bibitem{Bateman1992}
D.~S. Bateman, C.~Boyd, and B.~Dutta-Roy.
\newblock {\em Am. J. Phys.}, 60:833, 1992.

\bibitem{Racah1951}
G.~Racah.
\newblock {\em Group Theory and Spectroscopy}.
\newblock Princeton Lectures, Princeton, 1951.

\bibitem{Torres1998}
G.~F.~Torres del Castillo and J.~L.~Calvario Acocal.
\newblock {\em Rev. Mex. F{\'i}s.}, 44:344, 1998.

\bibitem{Parfitt2002a}
D.~G.~W. Parfitt and M.~E. Portnoi.
\newblock {\em J. Math. Phys.}, 43:4681, 2002.

\bibitem{Parfitt2003a}
D.~G.~W. Parfitt and M.~E. Portnoi.
\newblock {\em Phys. Status Solidi A}, 195:596, 2003.

\bibitem{Mathews1970}
J.~Mathews and R.L. Walker.
\newblock {\em Mathematical Methods of Physics}.
\newblock Benjamin/Cummings, California, 2nd edition, 1970.

\bibitem{Arfken1985}
G.~Arfken.
\newblock {\em Mathematical Methods for Physicists}.
\newblock Academic Press, London, 3rd edition, 1985.

\bibitem{Berry1973}
M.~V. Berry and A.~M.~Ozorio de~Almeida.
\newblock {\em J. Phys. A}, 6:1451, 1973.

\bibitem{Yi1994}
H.~S. Yi, H.~R. Lee, and K.~S. Sohn.
\newblock {\em Phys. Rev. A}, 49:3277, 1994.

\bibitem{Parfitt2003b}
D.~G.~W. Parfitt and M.~E. Portnoi.
\newblock {\em Physica E}, 17:212, 2003.

\bibitem{Goldberg1988}
B.~B. Goldberg, D.~Heiman, A.~Pinczuk, C.~W. Tu, A.~C. Gossard, and J.~H.
  English.
\newblock {\em Surf. Sci.}, 196:209, 1988.

\bibitem{Heiman1988}
D.~Heiman, B.~B. Goldberg, A.~Pinczuk, C.~W. Tu, A.~C. Gossard, and J.~H.
  English.
\newblock {\em Phys. Rev. Lett.}, 61:605, 1988.

\bibitem{Turberfield1990}
A.~J. Turberfield, S.~R. Haynes, P.~A. Wright, R.~A. Ford, R.~G. Clark, J.~F.
  Ryan, J.~J. Harris, and C.~T. Foxon.
\newblock {\em Phys. Rev. Lett.}, 65:637, 1990.

\bibitem{Portnoi1996}
M.~E. Portnoi and E.~I. Rashba.
\newblock {\em Phys. Rev. B}, 54:13791, 1996.

\bibitem{Plentz1996}
F.~Plentz, D.~Heiman, A.~Pinczuk, L.~N. Pfeiffer, and K.~W. West.
\newblock {\em Surf. Sci.}, 361/362:30, 1996.

\bibitem{Ashkinadze2002a}
B.~M. Ashkinadze, V.~Voznyy, E.~Cohen, A.~Ron, and V.~Umansky.
\newblock {\em Phys. Rev. B}, 65:073311, 2002.

\bibitem{Klitzing1980}
K.~von Klitzing, G.~Dorda, and M.~Pepper.
\newblock {\em Phys. Rev. Lett.}, 45:494, 1980.

\bibitem{Tsui1982}
D.~C. Tsui, H.~L. St{\"o}rmer, and A.~C. Gossard.
\newblock {\em Phys. Rev. Lett.}, 48:1559, 1982.

\bibitem{Prange1987}
R.~E. Prange and S.~M. Girvin.
\newblock {\em The Quantum Hall Effect}.
\newblock Springer-Verlag, New York, 1987.

\bibitem{Chakraborty1995}
T.~Chakraborty and P.~Pietil{\"{a}}inen.
\newblock {\em The Quantum Hall Effects: Integer and Fractional}.
\newblock Springer, Berlin, 2nd edition, 1995.

\bibitem{Laughlin1983b}
R.~B. Laughlin.
\newblock {\em Phys. Rev. Lett.}, 50:1395, 1983.

\bibitem{Chang1983}
A.~M. Chang, M.~A. Paalanen, D.~C. Tsui, H.~L. St{\"o}rmer, and J.~C.~M. Hwang.
\newblock {\em Phys. Rev. B}, 28:6133, 1983.

\bibitem{Kukushkin1986}
I.~V. Kukushkin and V.~B. Timofeev.
\newblock {\em JETP Lett.}, 44:228, 1986.

\bibitem{Buhmann1990}
H.~Buhmann, W.~Joss, K.~von Klitzing, I.~V. Kukushkin, G.~Martinez, A.~S.
  Plaut, K.~Ploog, and V.~B. Timofeev.
\newblock {\em Phys.Rev. Lett.}, 65:1056, 1990.

\bibitem{Clark1988}
R.~G. Clark, J.~R. Mallett, S.~R. Haynes, J.~J. Harris, and C.~T. Foxon.
\newblock {\em Phys. Rev. Lett.}, 60:1747, 1988.

\bibitem{Simmons1989}
J.~A. Simmons, H.~P. Wei, L.~W. Engel, D.~C. Tsui, and M.~Shayegan.
\newblock {\em Phys. Rev. Lett.}, 63:1731, 1989.

\bibitem{Dorozhkin1995}
S.~I. Dorozhkin, R.~J. Haug, K.~von Klitzing, and K.~Ploog.
\newblock {\em Phys. Rev. B}, 51:14729, 1995.

\bibitem{Haldane1983}
F.~D.~M. Haldane.
\newblock {\em Phys. Rev. Lett.}, 51:605, 1983.

\bibitem{Haldane1985}
F.~D.~M. Haldane and E.~H. Rezayi.
\newblock {\em Phys. Rev. Lett.}, 54:237, 1985.

\bibitem{Fano1986}
G.~Fano, F.~Ortolani, and E.~Colombo.
\newblock {\em Phys. Rev. B}, 34:2670, 1986.

\bibitem{Halperin1984}
B.~I. Halperin.
\newblock {\em Phys. Rev. Lett.}, 52:1583, 1984.

\bibitem{Wilczek1990}
F.~Wilczek.
\newblock {\em Fractional Statistics and Anyon Superconductivity}.
\newblock World Scientific, Singapore, 1990.

\bibitem{Wilczek1991}
F.~Wilczek.
\newblock {\em Sci. Am.}, 264:24, May 1991.

\bibitem{Girvin1985}
S.~M. Girvin, A.~H. MacDonald, and P.~M. Platzman.
\newblock {\em Phys. Rev. Lett.}, 54:581, 1985.

\bibitem{Girvin1986}
S.~M. Girvin, A.~H. MacDonald, and P.~M. Platzman.
\newblock {\em Phys. Rev. B}, 33:2481, 1986.

\bibitem{Pinczuk1993}
A.~Pinczuk, B.~S. Dennis, L.~N. Pfeiffer, and K.~West.
\newblock {\em Phys. Rev. Lett.}, 70:3983, 1993.

\bibitem{Kivelson1992}
S.~Kivelson, D.-H. Lee, and S.-C. Zhang.
\newblock {\em Phys. Rev. B}, 46:2223, 1992.

\bibitem{Jain1989}
J.~K. Jain.
\newblock {\em Phys. Rev. Lett.}, 63:199, 1989.

\bibitem{Jain1990}
J.~K. Jain.
\newblock {\em Phys. Rev. B}, 41:7653, 1990.

\bibitem{Dyakonov2001}
M.~I. Dyakonov.
\newblock {\em cond-mat/0209206}, 2002.

\bibitem{Picciotto1997}
R.~de~Picciotto, M.~Reznikov, M.~Heiblum, V.~Umansky, G.~Bunin, and D.~Mahalu.
\newblock {\em Nature}, 389:162, September 1997.

\bibitem{Saminadayar1997}
L.~Saminadayar, D.~C. Glattli, Y.~Jin, and B.~Etienne.
\newblock {\em Phys. Rev. Lett.}, 79:2526, 1997.

\bibitem{Collins1997}
G.~P. Collins.
\newblock {\em Physics Today}, page~17, November 1997.

\bibitem{Goldman1995}
V.~J. Goldman and B.~Su.
\newblock {\em Science}, 267:1010, February 1995.

\bibitem{MacDonald1992}
A.~H. MacDonald, E.~H. Rezayi, and D.~Keller.
\newblock {\em Phys. Rev. Lett.}, 68:1939, 1992.

\bibitem{Apalkov1991b}
V.~M. Apalkov and E.~I. Rashba.
\newblock {\em JETP Lett.}, 54:155, 1991.

\bibitem{Kohn1961}
W.~Kohn.
\newblock {\em Phys. Rev.}, 123:1242, 1961.

\bibitem{Apalkov1991a}
V.~M. Apalkov and E.~I. Rashba.
\newblock {\em JETP Lett.}, 53:442, 1991.

\bibitem{Plentz1997}
F.~Plentz, D.~Heiman, A.~Pinczuk, L.~N. Pfeiffer, and K.~W. West.
\newblock {\em Solid State Commun.}, 101:103, 1997.

\bibitem{Yusa2001}
G.~Yusa, H.~Shtrikman, and I.~Bar-Joseph.
\newblock {\em Phys. Rev. Lett.}, 87:216402--1, 2001.

\bibitem{Yusa2002}
G.~Yusa, H.~Shtrikman, and I.~Bar-Joseph.
\newblock {\em Physica E}, 12:49, 2002.

\bibitem{Ashkinadze2002b}
B.~M. Ashkinadze, V.~Voznyy, E.~Cohen, and A.~Ron.
\newblock In {\em Proceedings of the 26th International Conference on the
  Physics of Semiconductors}, Edinburgh, UK, 2002.

\bibitem{Apalkov1992a}
V.~M. Apalkov and E.~I. Rashba.
\newblock {\em JETP Lett.}, 55:37, 1992.

\bibitem{Apalkov1992b}
V.~M. Apalkov and E.~I. Rashba.
\newblock {\em Phys. Rev. B}, 46:1628, 1992.

\bibitem{Apalkov1995a}
V.~M. Apalkov and E.~I. Rashba.
\newblock {\em Solid State Commun.}, 93:193, 1995.

\bibitem{Apalkov1995b}
V.~M. Apalkov and E.~I. Rashba.
\newblock {\em Solid State Commun.}, 95:421, 1995.

\bibitem{Apalkov1995c}
V.~M. Apalkov, F.~G. Pikus, and E.~I. Rashba.
\newblock {\em Phys. Rev. B}, 52:6111, 1995.

\bibitem{Zang1995}
J.~Zang and J.~L. Birman.
\newblock {\em Phys. Rev. B}, 51:5574, 1995.

\bibitem{Rashba1993}
E.~I. Rashba and M.~E. Portnoi.
\newblock {\em Phys. Rev. Lett.}, 70:3315, 1993.

\bibitem{Portnoi1994}
M.~E. Portnoi and E.~I. Rashba.
\newblock {\em J. Lumin.}, 60/61:782, 1994.

\bibitem{Portnoi1995a}
M.~E. Portnoi and E.~I. Rashba.
\newblock {\em Il Nuovo Cimento}, 17D:1669, 1995.

\bibitem{Portnoi1995b}
M.~E. Portnoi and E.~I. Rashba.
\newblock {\em Mod. Phys. Lett. B}, 9:123, 1995.

\bibitem{Wu1984a}
Y.~S. Wu.
\newblock {\em Phys. Rev. Lett.}, 52:2103, 1984.

\bibitem{Wu1984b}
Y.~S. Wu.
\newblock {\em Phys. Rev. Lett.}, 53:111, 1984.

\bibitem{Parfitt2003c}
D.~G.~W. Parfitt and M.~E. Portnoi.
\newblock {\em Phys. Rev. B}, 68:035306, 2003.

\bibitem{Bolton1994}
F.~Bolton.
\newblock {\em Phys. Rev. Lett.}, 73:158, 1994.

\bibitem{Gorkov1968}
L.~P. Gor'kov and I.~E. Dzyaloshinskii.
\newblock {\em Sov. Phys.--JETP}, 26:449, 1968.

\bibitem{Lerner1980}
I.~V. Lerner and Y.~E. Lozovik.
\newblock {\em Sov. Phys.--JETP}, 51:588, 1980.

\bibitem{Birkhoff1977}
G.~Birkhoff and S.~MacLane.
\newblock {\em A Survey of Modern Algebra}.
\newblock Macmillan, New York, 4th edition, 1977.

\bibitem{Slomson1991}
A.~Slomson.
\newblock {\em An Introduction to Combinatorics}.
\newblock Chapman and Hall, London, 1991.

\bibitem{Hardy1918}
G.~H. Hardy and S.~Ramanujan.
\newblock {\em Proc. Lond. Math. Soc.}, 2:75, 1918.

\bibitem{Calogero1967}
F.~Calogero.
\newblock {\em Variable Phase Approach to Potential Scattering}.
\newblock Academic Press, New York, 1967.

\end{thebibliography}

\end{document}